\documentclass[twocolumn,twocolappendix]{aastex63}

\usepackage{xspace}

\newcommand{\Ha}{H$\alpha$\xspace}
\newcommand{\Hb}{H$\beta$\xspace}
\newcommand{\NII}{[N{\sc ii}]\xspace}
\newcommand{\SII}{[S{\sc ii}]\xspace}
\newcommand{\OIII}{[O{\sc iii}]\xspace}
\newcommand{\OII}{[O{\sc ii}]\xspace}
\newcommand{\gleam}{\textsc{gleam}\xspace}

\received{November 2, 2020}
\revised{January 25, 2021}
\accepted{February 16, 2021}

\submitjournal{ApJ}

\graphicspath{{./}{figures/}}

\usepackage{pgffor}
\usepackage[caption=false]{subfig}

\usepackage{amssymb}	
\usepackage{amstext}    
\usepackage{graphicx}

\begin{document}

\title{ENISALA: II. Distinct Star Formation and Active Galactic Nucleus Activity in Merging and Relaxed Galaxy Clusters}

\correspondingauthor{Andra Stroe}
\email{andra.stroe@cfa.harvard.edu}

\author[0000-0001-8322-4162]{Andra Stroe}
\altaffiliation{Clay Fellow}
\affiliation{Center for Astrophysics \text{\textbar} Harvard \& Smithsonian, 60 Garden St., Cambridge, MA 02138, USA}

\author[0000-0001-8823-4845]{David Sobral}
\affiliation{Department of Physics, Lancaster University, Lancaster LA1 4YB, UK}

\begin{abstract}
  The growth of galaxy clusters is energetic and may trigger and/or quench star formation and black hole activity. The ENISALA\footnote{\label{enisala}The project is named as a tribute to the storied Enisala citadel (Dobrogea, Romania). Enisala (`new settlement', in Turkish and Romanian) sits on top of a windswept hill, at the crossroads of the Danube Delta and the Pontus Euxinus sea (`hospitable sea', Black Sea), forever shaped by forces of nature. It stands as a metaphor for the ever-evolving galaxy cluster environment and its profound influence on galaxy and black hole evolution. ENISALA can also be understood to stand for `ENvironmental Influence on Star formation and AGN through Line Astrophysics'.} project is a collection of multiwavelength observations aimed at understanding how large-scale structure drives galaxy and black hole evolution. Here, we introduce optical spectroscopy of over 800 H$\alpha$ emission-line galaxies, selected in 14 $z\sim0.15-0.31$ galaxy clusters, spanning a range of masses and dynamical states. We investigate the nature of the emission lines in relation to the host galaxy properties, its location within the cluster, and the properties of the parent cluster. We uncover remarkable differences between mergers and relaxed clusters. The majority of H$\alpha$ emission-line galaxies in merging cluster fields are located within 3\,Mpc of their center. A large fraction of these line-emitters in merging clusters are powered by star formation irrespective of cluster-centric radius, while the rest are powered by active galactic nuclei. Star-forming galaxies are rare within 3\,Mpc of relaxed clusters and active galactic nuclei are most abundant at their outskirts ($\sim1.5-3$\,Mpc). We discover a population of star-forming galaxies with large equivalent widths and blue UV--optical colors, found exclusively in the merging clusters in our sample. The widespread emission-line activity in merging clusters is likely supported by triggered activity in recently-accreted, gas-rich galaxies. By contrast, our observations for relaxed clusters match established models, in which black hole activity is enhanced at the virial radius and star-formation is quenched within the infall region. We conclude that emission-line galaxies experience distinct evolutionary paths in merging and relaxed clusters.
\end{abstract}


\section{Introduction} \label{sec:intro}

Large-scale structure plays a critical role in the evolution of galaxies, accelerating the growth of star-forming spirals into passive ellipticals. Galaxies in evolved, relaxed galaxy clusters have systematically redder colors, with more evolved, elliptical and lenticular morphologies, lower star formation rates (SFR), and lower gas fractions compared to field counterparts \citep[e.g.][]{Gunn1972, Kenney2004, Dressler1980, Goto2003, Melnick1977, Balogh2004, Chung2009a}. One of the main pathways for galaxy evolution in clusters is the interaction between infalling galaxies and the dense, hot intracluster medium (ICM). The ICM can remove interstellar gas from the galaxy altogether or prevent the accretion of new gas reservoirs, thus depriving galaxies of the fuel needed for future star formation (SF) episodes \citep{Gunn1972, Larson1980, Bekki2002}. Before falling into the cluster, galaxies can experience pre-processing in filaments \citep{Darvish2017, Paulino2018, Connor2018}. Especially at the outskirts of relaxed clusters, high-speed close encounters between pairs of galaxies can lead to a truncation of the halo and a shut down of SF \citep{Moore1996}. Environmental effects are very efficient at quenching galaxies as even a single passage through the cluster can render galaxies significantly redder than the infalling population \citep[e.g.][]{Pimbblet2011, Muriel2014}.

\begin{deluxetable*}{ccccDcc}[htb!]
  \tablecaption{List of galaxy clusters with spectroscopic follow-up of \Ha cluster member candidates.\label{tab:clusters}}
  \tablewidth{0pt}
  \tablehead{
    \colhead{Cluster} & \colhead{Nickname} & \colhead{R.A.}  & \colhead{Decl.} &\multicolumn2c{$z$} & \colhead{L$_\mathrm{X-ray}$}             & \colhead{State} \\
    & & \colhead{$hh\,mm\,ss$} & \colhead{$^{\circ}\,'\,''$} &     &        & \colhead{($10^{44}$\,erg\,s$^{-1}$)} &                   }
  \decimalcolnumbers
  \startdata
  Abell 1689        & A1689         & $13\,11\,29$  & $-01\,20\,17$ & 0.183  & 14 & relaxed \\
  Abell 963	        &	A963          & $10\,17\,13$	&	$+39\,01\,31$	&	0.206	 & \phn6	& relaxed	\\
  Abell 2390	      &	A2390         & $21\,53\,35$	&	$+17\,41\,12$	&	0.228	 & 13	&	relaxed	\\
  Zwicky 2089	      &	Z2089         & $09\,00\,36$	&	$+20\,53\,39$	&	0.2343 & \phn7	&	relaxed	\\
  RX J2129+0005     & RXJ2129       & $21\,29\,38$  & $+00\,05\,39$ & 0.235  & 12 & relaxed \\
  RX J0437.1+0043 	&	RXJ0437       & $04\,37\,10$	&	$+00\,43\,38$	&	0.285	 & \phn9  & relaxed	\\ \hline
  Abell 2254        &	A2254         & $17\,17\,40$	&	$+19\,42\,51$	&	0.178	 & \phn5	& merger (+ turbulence)	\\
  CIZA J2242.8+5301 & Sausage     &	$22\,42\,50$  &	$+53\,06\,30$	&	0.188  & \phn7 	&	merger (+ shocks)	\\
  Abell 115         & A115          & $00\,55\,59$	& $+26\,22\,41$	&	0.1971 & \phn9  & merger (+ shocks)	\\
  Abell 2163        & A2163         & $16\,15\,34$  & $-06\,07\,26$ & 0.203  & 38 & merger (+ turbulence) \\
  Abell 773         & A773          &	$09\,17\,59$	&	$+51\,42\,23$	&	0.217	 & \phn6  & merger (+ turbulence)	\\
  1RXS J0603.3+4214 & Toothbrush  & $06\,03\,30$  & $+42\,17\,30$	&	0.225	 & \phn8  &	merger (+ shocks and turbulence)	\\
  Abell 2219        & A2219         &	$16\,40\,21$	&	$+46\,42\,21$	&	0.2256 & 12	& merger (+ turbulence) 	\\
  Abell 2744        & A2744         &	$00\,14\,18$	&	$-30\,23\,22$	&	0.308	 & 13 & merger (+ shocks and turbulence)	\\
  \enddata
\end{deluxetable*}

Through invaluable work spanning the entire electromagnetic spectrum, complemented by simulations and theoretical developments, galaxy evolution studies from the perspective of large scale structure have established the critical role relaxed cluster environments have in shaping the evolutionary tracks of their member galaxies. However, the bulk of research carried over the last 50 years has mostly focused on contrasting field and relaxed cluster environments \citep{Dressler1984, Boselli2006}. Matter in the Universe is distributed in a non-uniform matter, spanning voids, filaments, sheets, groups, and large clusters. Therefore, a large fraction of galaxies do not reside in either a highly-evolved massive local cluster or in average density field environments. To distinguish between galaxy evolution models, we need to fill in the missing information on the effect of intermediate environments such as galaxy groups, filaments, disturbed clusters, and their outskirts \citep[e.g.][]{Kodama2001, Darvish2017}.

The most transformational events in the lifetime of a galaxy cluster are collisions and mergers with other clusters. Mergers between clusters are the most energetic events since the Big Bang and significantly alter the evolutionary pathways of the participating clusters. Simulations and observations find that mergers inject non-thermal components, in the form of magnetic fields, relativistic particles, bulk motion, cluster-wide turbulence, and weak shocks. Mergers can also produce Mpc-wide strong shocks, which heat the ICM, accelerate particles to relativistic speeds and provide $5-25$\% of pressure support, with the highest impact at cluster outskirts \citep[e.g.][]{Eckert2019, Biffi2016}. Effects of turbulence can be measured through the broadening of X-ray emission lines or radio observations revealing synchrotron emission associated with relativistic electrons. Shock waves can be detected as X-ray discontinuities or as arc-like patches of diffuse radio emission from shock-accelerated particles \citep{Markevitch2007, Vazza2009, vanWeeren2019}.

Cosmological simulations unveiled that mergers were common at $z>1$ and are part of every local cluster's history \citep[e.g.][]{Cohn2005, Boylan-Kolchin2009}. Traditionally, it was assumed that mergers are rare at redshifts $z<1$. However, in their pioneering study, \citet{Jones1999} discovered that 40\% of X-ray selected galaxy clusters at $z<0.2$ have significant substructure. The large fraction of merging and interacting galaxy clusters has been confirmed in many subsequent studies, which have found that $55-70$\% of mass-selected and/or X-ray volume-limited samples are not relaxed \citep{Andrade-Santos2017, Rossetti2017, Lovisari2017, Chon2017}.

Thus, a large fraction of SF in cluster galaxies actually happens in merging, disturbed structures. If we are to understand the evolution of galaxies from the field to relaxed clusters, it is crucial to understand the intermediary evolution stages. What happens to galaxies when their parent cluster is undergoing a massive merger? Studies have found striking differences between the SF properties, morphologies, active galactic nucleus (AGN), and gas reservoir properties of galaxies in merging clusters compared to those in relaxed clusters. Some disturbed galaxy clusters have a higher density of \Ha-bright galaxies \citep{Stroe2014, Stroe2015a, Stroe2017}, a higher fraction of star-forming \citep[e.g.][]{Cohen2014, Yoon2020} and blue galaxies \citep[e.g.][]{Wang1997, Cortese2004, Hou2012, Cava2017} and are more gas-rich than counterparts in relaxed clusters \citep{Stroe2015, Jaffe2012, Jaffe2016, Cairns2019}. There is also compelling evidence from multiwavelength data that cluster mergers trigger AGN activity \citep{Miller2003, Owen1999, Sobral2015, Hwang2009}. Not only is the number of star-forming galaxies increased in merging clusters compared to relaxed clusters, but the morphological and spectroscopic properties of the star formers also different between merging clusters, relaxed environments, and the field. For example, \citet{Yoon2019} found that cluster-cluster interactions at $z<0.06$ trigger the formation of bars in galaxies at all stellar masses studied ($10^{10-11.5}$\,M$_\odot$), an effect attributed to strong asymmetric perturbations induced by the rapidly changing tidal field in merging galaxy clusters. \citet{Mulroy2017} find that galaxy colors are standardized by a cluster-wide process, such as shock waves, in merging clusters. In a spectroscopic study, \citet{Sobral2015} find that active galaxies in merging clusters are metal-rich, have evidence for outflows, and have very low electron densities, which indicate high supernova rates (SNR) and sustained SF for timescales of 500\,Myr. In addition to emission-line galaxies, some merging clusters contain significant populations of post starburst (E+A) galaxies in merging clusters, showing a possible correlation with cluster merger timescale \citep[e.g.][]{Ma2010, Pranger2014}. By contrast, \citet{Deshev2017} found a depletion of star-forming galaxies in the core of the massive merging Abell 520 cluster, with the bulk of SF happening along infalling filaments. \citet{Chung2009} and \citet{Shim2010} found that mid-infrared colors, a measure of specific SFR (sSFR), did not vary significantly across the shock fronts in the Bullet and the Abell 2255 clusters. Simulations predict that the increased pressure caused by the merger-induced traveling shock waves can cause a temporary burst of SF, but ultimately lead to a fast consumption of gas and a shut down on SF \citep{Fujita1999, Roediger2014}. Further, \citet{Ebeling2019} propose a scenario that brings some of these results into agreement, in which galaxies in the core of merging clusters experience increased quenching, while late-type galaxies falling into the merging system along filaments can experience a burst of SF trigger by a cluster-wide, merger-induced shock wave.

The drivers of SF and black hole (BH) activity in cluster galaxies seem to be closely linked to the merger history of the host cluster and its associated filaments and infalling groups, opening a new window into how galaxies evolve. Despite the growing amount of evidence to their profound influence, some results are in disagreement, and the exact mechanisms through which cluster mergers drive galaxy evolution are still poorly understood.

\begin{figure*}[ht!]
  \includegraphics[width=0.5\textwidth]{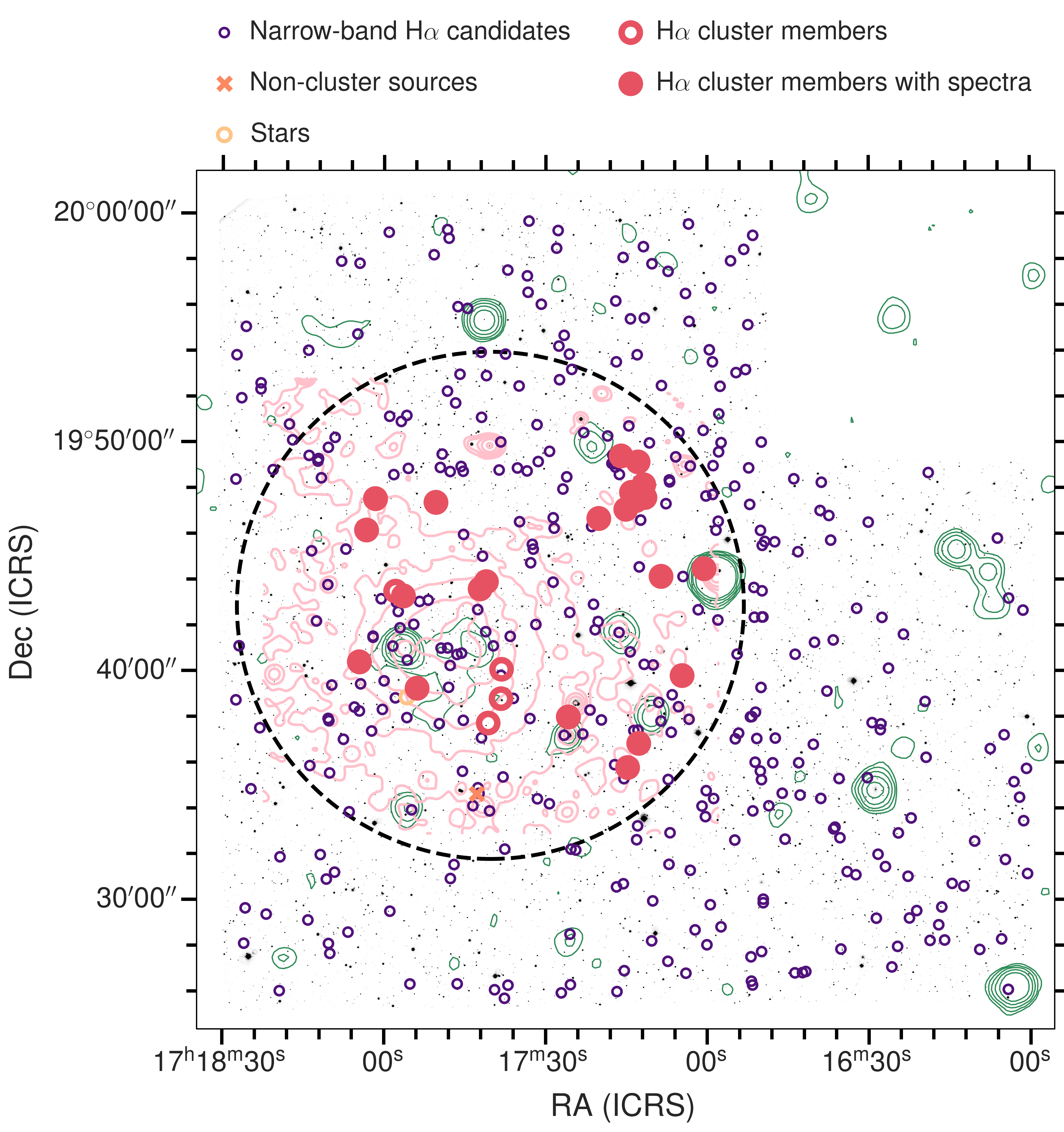}
  \includegraphics[width=0.5\textwidth]{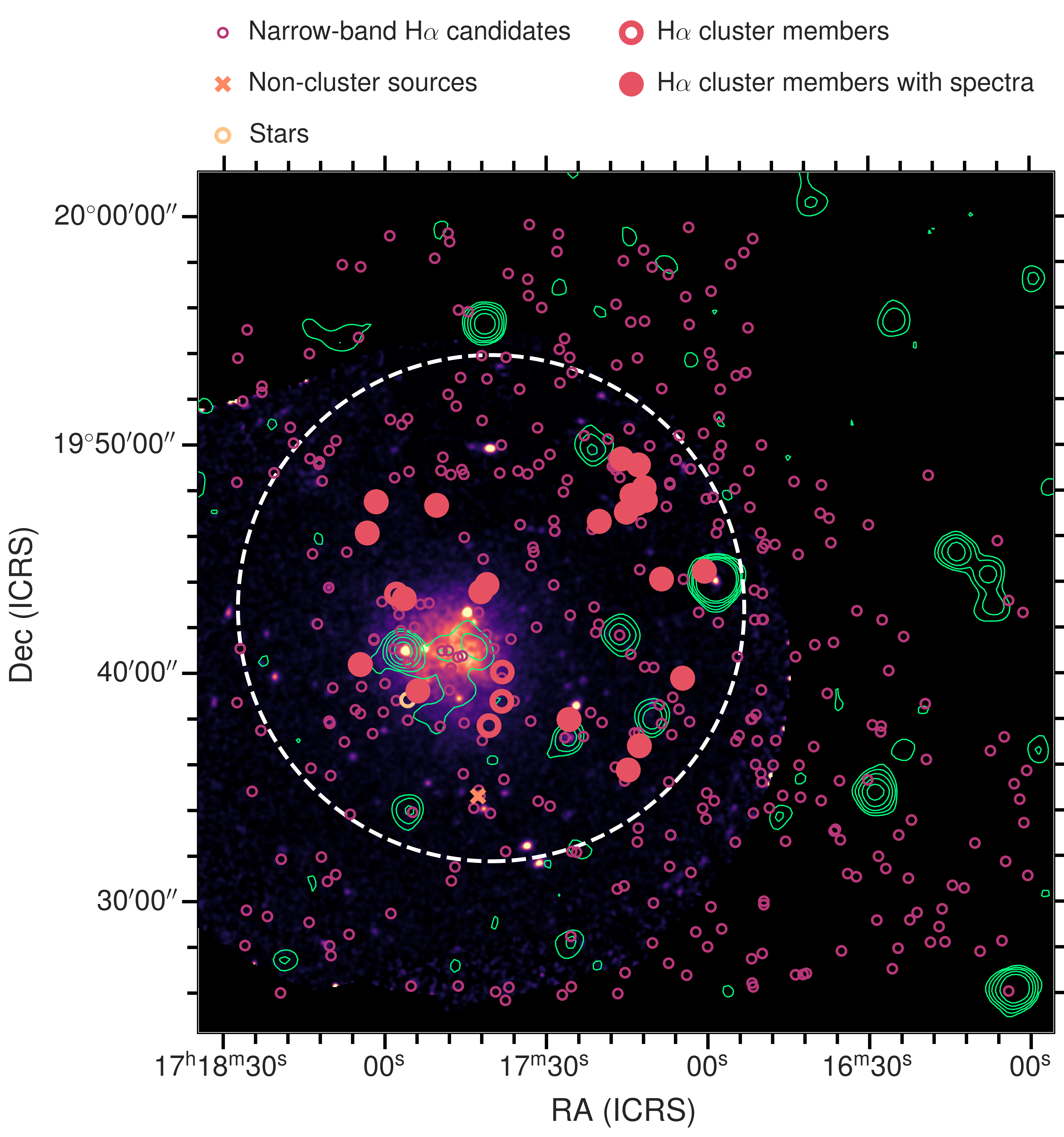}
  \caption{A multiwavelength view of Abell 2254, a massive $1.5-3\times10^{15}$\,M$_\odot$ cluster undergoing a major merger that injects turbulence in the ICM. Symbols mark the positions of \Ha candidates selected through our NB survey (small empty circles), those confirmed to be \Ha at the cluster redshift (large circles), and emitters confirmed to be at other redshifts (crosses). Filled symbols show the confirmed \Ha candidates with spectroscopy available through our project. Left: Our NB image targeting \Ha at the cluster redshift is shown in grayscale, with X-ray emission from \textit{XMM-Newton} tracing the ICM in pink contours and radio observations from NVSS in green contours, highlighting locations of radio galaxies and the diffuse emission associated with the ICM. Right: The same X-ray image is shown in the background with radio contours overlaid.}
  \label{fig:composite}
\end{figure*}

\section{\textbf{ENISALA} - an emission line spectroscopic survey of merging and relaxed galaxy clusters}
We commenced the \textbf{ENISALA} project, an ambitious multiwavelength photometric and spectroscopic observing campaign to unveil the evolutionary pathways of galaxies in merging clusters and their large scale structure, along with a comparison sample of relaxed clusters. In the first paper from the series \citep{Stroe2017}, we presented the results from the first systematic survey of SF activity in a statistically-significant set of $19$ $0.15<z<0.31$ clusters samples a range of masses, luminosities, and dynamical states. We employed custom-made narrow-band (NB) filters to select star-forming and active galaxies through their \Ha emission, over the entire 3D volume of their host clusters. Our method results in a very simple selection function, which uniformly selected star-forming galaxies in and around clusters down to a well-understood star-formation rate (SFR) limit. We found striking differences between relaxed and merging clusters, with merging environments having over two times more \Ha galaxies compared to merging clusters, especially those hosting large scale shocks, which are overdense by a factor of 4 \citep{Stroe2017}. Unlike relaxed, low-redshift clusters, merging clusters are surprisingly similar to high-redshift (proto-)clusters that are rich in massive, star-forming galaxies \citep[e.g.][]{Hatch2011, Koyama2013, Cooke2014, Suzuki2019, Darvish2020}.

Our spectroscopic observing campaign follows up a representative sample of \Ha galaxies selected through our NB observations in 14 merging and relaxed clusters and their immediate cosmic web (see Table~\ref{tab:clusters} and Figure~\ref{fig:composite}). The main drivers of our spectroscopic observations are to confirm the cluster membership and constrain the star-formation, ionization, metallicity, and electron density properties of the galaxies. Our sample contains over 800 galaxies with measurements of at least one main optical emission line and over 300 galaxies with measurements of enough lines to classify them securely in a \citet{Baldwin1981} (BPT) diagram.

In this paper (Paper II) of the series, we will give a general introduction to our ENISALA spectroscopic survey, including the survey strategy, data acquisition, reduction, and the initial spectroscopic products, such as redshifts, line measurements, and ratios. We also discuss the first results from the ENISALA spectroscopic survey, focusing on the SF properties and ionization sources of the \Ha-selected galaxies, while future papers will focus on properties such as metallicity, ionization potential, temperatures, and electron density.

The paper is structured as follows: Section~\ref{sec:data} presents the parent NB sample and the observing strategy for the follow-up spectroscopy, while in Section~\ref{sec:redshifts} we measure redshifts from the spectra and present the \textsc{redshifts} package. Section~\ref{sec:selection} presents a validation of the NB selection and the distribution of follow-up sources with respect to the parent sample. Section~\ref{sec:measurements} describes how measurements are derived from spectroscopy. We present the final sample used in this paper in Section~\ref{sec:sample}. In Section~\ref{sec:results}, we discuss velocity width and equivalent width (EW) properties of the sample as a function of the host galaxy color, ionization source, and location within the cluster. Section~\ref{sec:discussion} aims to embed our results into the overall picture of environmentally driven evolution by focusing on viable galaxy and BH evolutionary pathways in the context of galaxy cluster growth. We present our conclusions and outlook to the future in Section~\ref{sec:discussion}. We assume a $\Lambda$CDM cosmology, with $H_{0}=70$\,km\,s$^{-1}$\,Mpc$^{-1}$, $\Omega_M=0.3$ and $\Omega_\Lambda=0.7$. We report AB magnitudes throughout. The clusters in the sample range from 0.178 to 0.308 in redshift, which corresponds to physical scales of $3.01-4.55$\,kpc\,arcsec$^{-1}$.

\begin{deluxetable}{cccDc}[htb!]
  \tablecaption{Details of the spectroscopic observations. For each cluster, we list the telescope used for taking the data, the number of independent pointing/source setups, the spectral resolution of the instrument, and the total exposure time for each pointing.\label{tab:obs}}
  \tablewidth{0pt}
  \tablehead{
    \colhead{Cluster} & \colhead{Instrument} & \colhead{Pointings} & \multicolumn2c{$\Delta  \lambda$} & \colhead{Exp. Time}\\
    &                      &  & \multicolumn2c{(\AA)} & (hours) }
  \decimalcolnumbers
  \startdata
  A1689             & Hectospec/MMT & 6 & 6.   & $0.5-0.75$ \\
  A963              & Hectospec/MMT & 6 & 6.   & $0.5-1.0$ \\
  A2390             & VLT/VIMOS     & 1 & 12.5 & 2. \\
  & Hectospec/MMT & 4 & 6.   & $0.375-1.0$ \\
  Z2089             & Hectospec/MMT & 6 & 6.   & $0.5-1.25$ \\
  RXJ2129           & Hectospec/MMT & 4 & 6.   & $0.5-0.75$ \\
  RXJ0437           & VLT/VIMOS     & 1 & 12.5 & 2. \\ \hline
  A2254             & VLT/VIMOS     & 1 & 12.5 & 2. \\
  Sausage           & WHT/AF2       & 2 & 4.4  & 1.5, 2.5 \\
  & Keck/DEIMOS   & 4 & 1.   & 0.75 \\
  A115              & Hectospec/MMT & 5 & 6.   & $0.5-0.59$ \\
  A2163             & VLT/VIMOS     & 1 & 12.5 & 1. \\
  A773              & WHT/AF2       & 1 & 8.1  & 2.25 \\
  Toothbrush        & WHT/AF2       & 1 & 8.1  & 3. \\
  & Keck/DEIMOS   & 4 & 1.   & $0.83-1.0$ \\
  A2219             & Hectospec/MMT & 4 & 6.   & $0.5-1.0$ \\
  A2744             & VLT/VIMOS     & 1 & 12.5 & 2. \\
  \enddata
\end{deluxetable}

\section{Sample, Observing Strategy and Data Reduction}\label{sec:data}

\subsection{Parent NB \Ha sample}

In \citet{Stroe2017}, using NB filters, we selected over 3000 \Ha emitting candidates in the fields around 19 galaxy clusters at $0.15<z<0.31$. The sample included relaxed and merging clusters, with cluster-wide shocks and turbulence, and spanned masses of $5-35\times10^{14}$\,M$_\odot$. The observations cover a field of view (FOV) of $0.5$\,deg$^2$ centered on each cluster, or about a maximum $3-5$\,Mpc from the cluster center, and $>2$ times the velocity dispersion of the cluster in redshift space, for a total volume of $1.3\times10^3$\,Mpc$^3$. The limiting dust-uncorrected \Ha luminosities are $10^{40.2-41.3}$\,erg\,s$^{-1}$. The observations reach equivalent dust-corrected SFRs at the level of $0.03-0.3$ of the typical SFRs at $z\sim0.2$.

\begin{figure*}[ht!]
  {\subfloat[SF dominated spectrum taken with WHT/AF2.]{%
      \begin{minipage}{0.08\textwidth}
        \includegraphics[width=\textwidth]{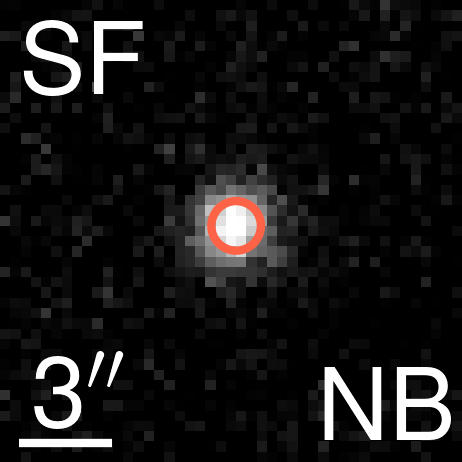}\\
        \includegraphics[width=\textwidth]{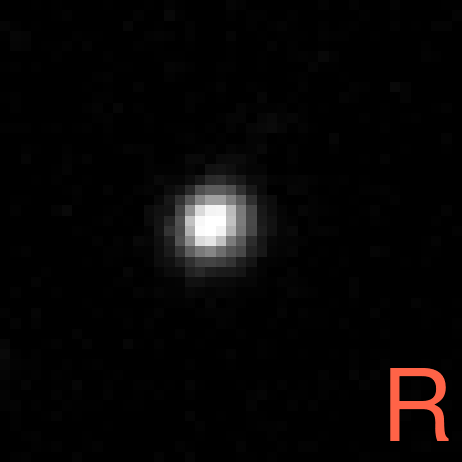}\\
        \includegraphics[width=\textwidth]{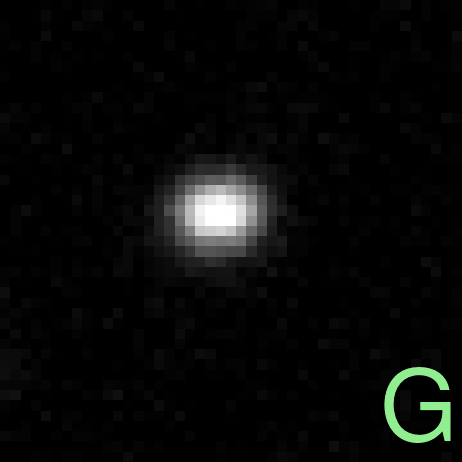}\\
        \includegraphics[width=\textwidth]{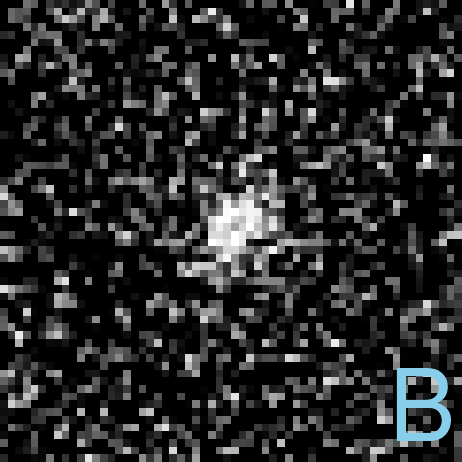}
      \end{minipage}%
      \hspace{5pt}
      \begin{minipage}{0.92\textwidth}
        \includegraphics[height=0.4\textwidth]{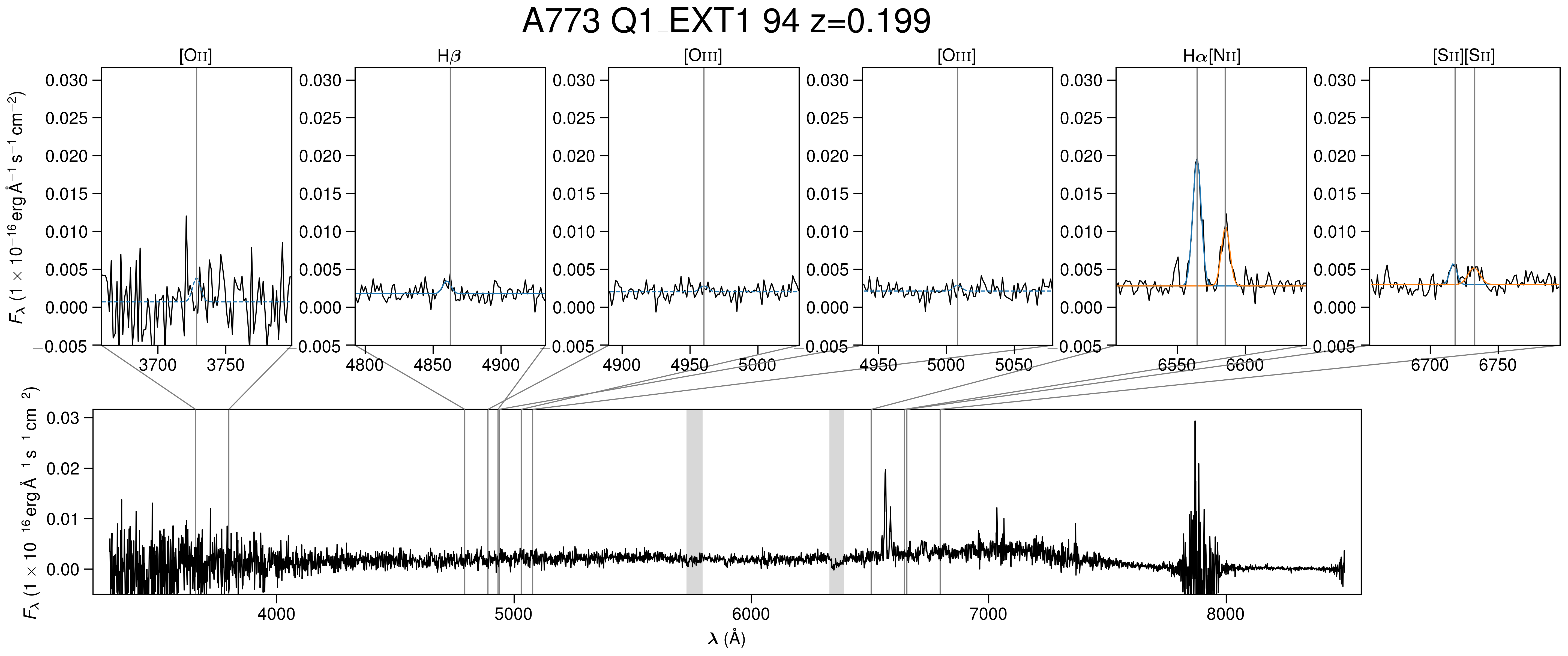}
      \end{minipage}%
    }\\}%
  {\subfloat[AGN dominated spectrum taken with VLT/VIMOS.]{%
      \begin{minipage}{0.08\textwidth}
        \includegraphics[width=\textwidth]{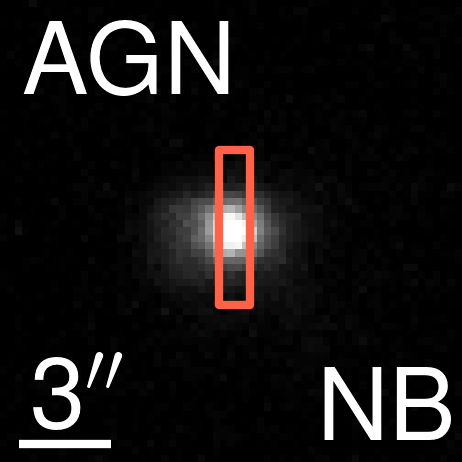}\\
        \includegraphics[width=\textwidth]{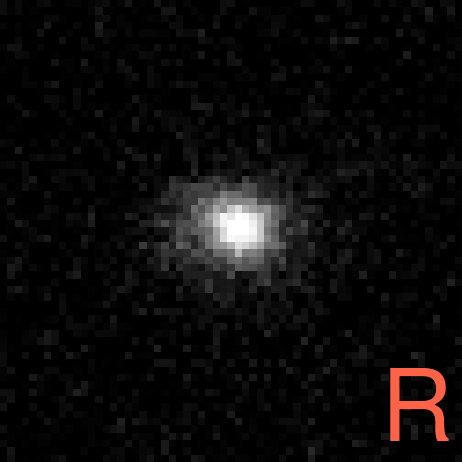}\\
        \includegraphics[width=\textwidth]{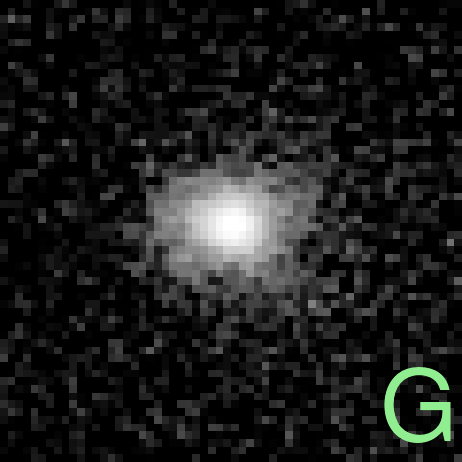}\\
        \includegraphics[width=\textwidth]{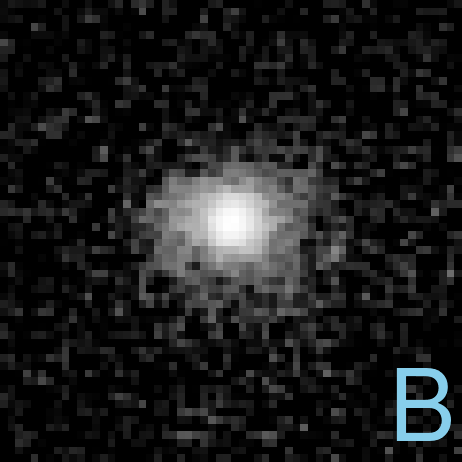}
      \end{minipage}%
      \hspace{5pt}
      \begin{minipage}{0.92\textwidth}
        \includegraphics[height=0.4\textwidth]{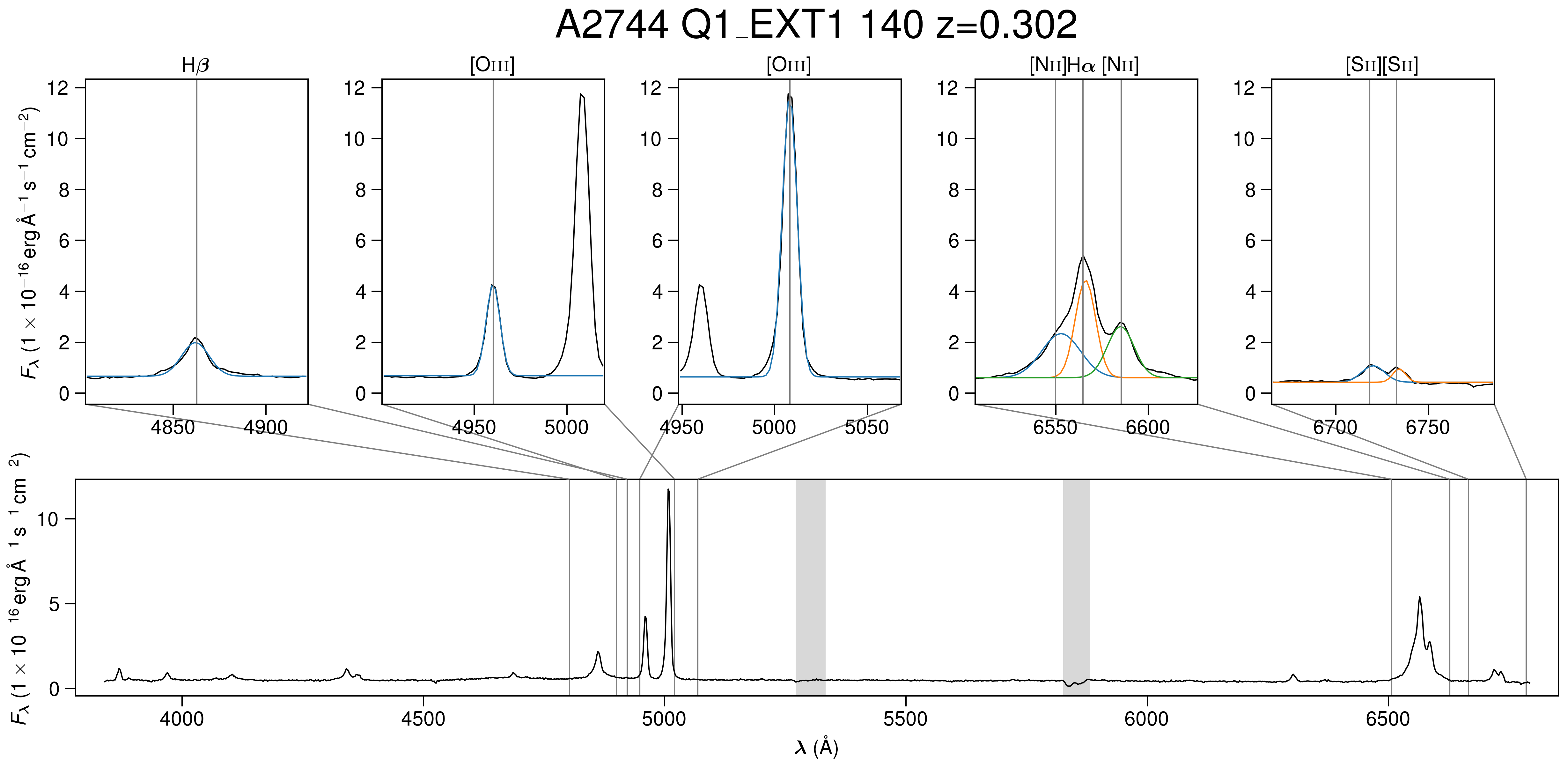}
      \end{minipage}%
    }\\}%
  \caption{Examples of cluster member spectra. We show $15''\times15''$ cutouts in the NB, red ($i$ band), green ($r$ band), and blue ($g$ band) light. We show illustrations of the fiber and slit sizes used for each instrument.}
  \label{fig:spectra}
\end{figure*}
\vspace{20pt}

\begin{figure*}[ht!]
  \ContinuedFloat
  {\subfloat[Composite spectrum taken with Keck/DEIMOS.]{%
      \begin{minipage}{0.08\textwidth}
        \includegraphics[width=\textwidth]{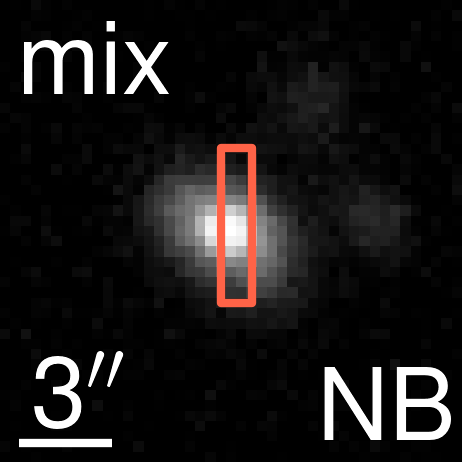}\\
        \includegraphics[width=\textwidth]{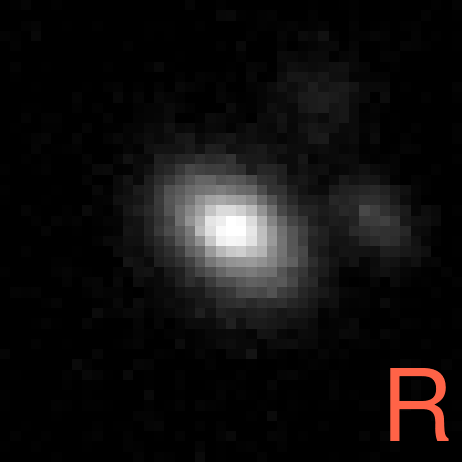}\\
        \includegraphics[width=\textwidth]{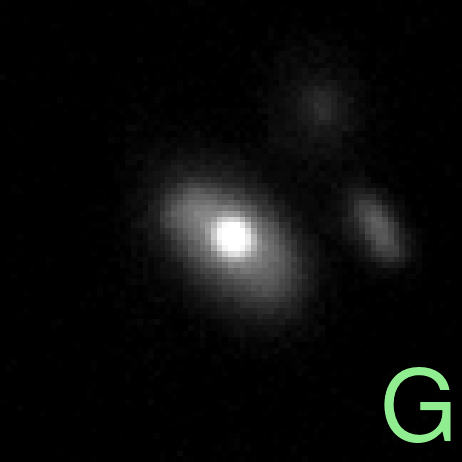}\\
        \includegraphics[width=\textwidth]{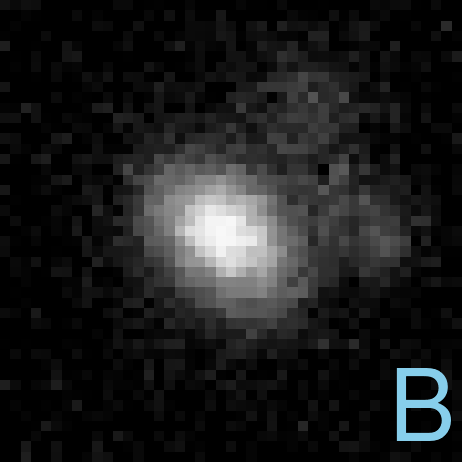}
      \end{minipage}%
      \hspace{5pt}
      \begin{minipage}{0.92\textwidth}
        \includegraphics[height=0.4\textwidth]{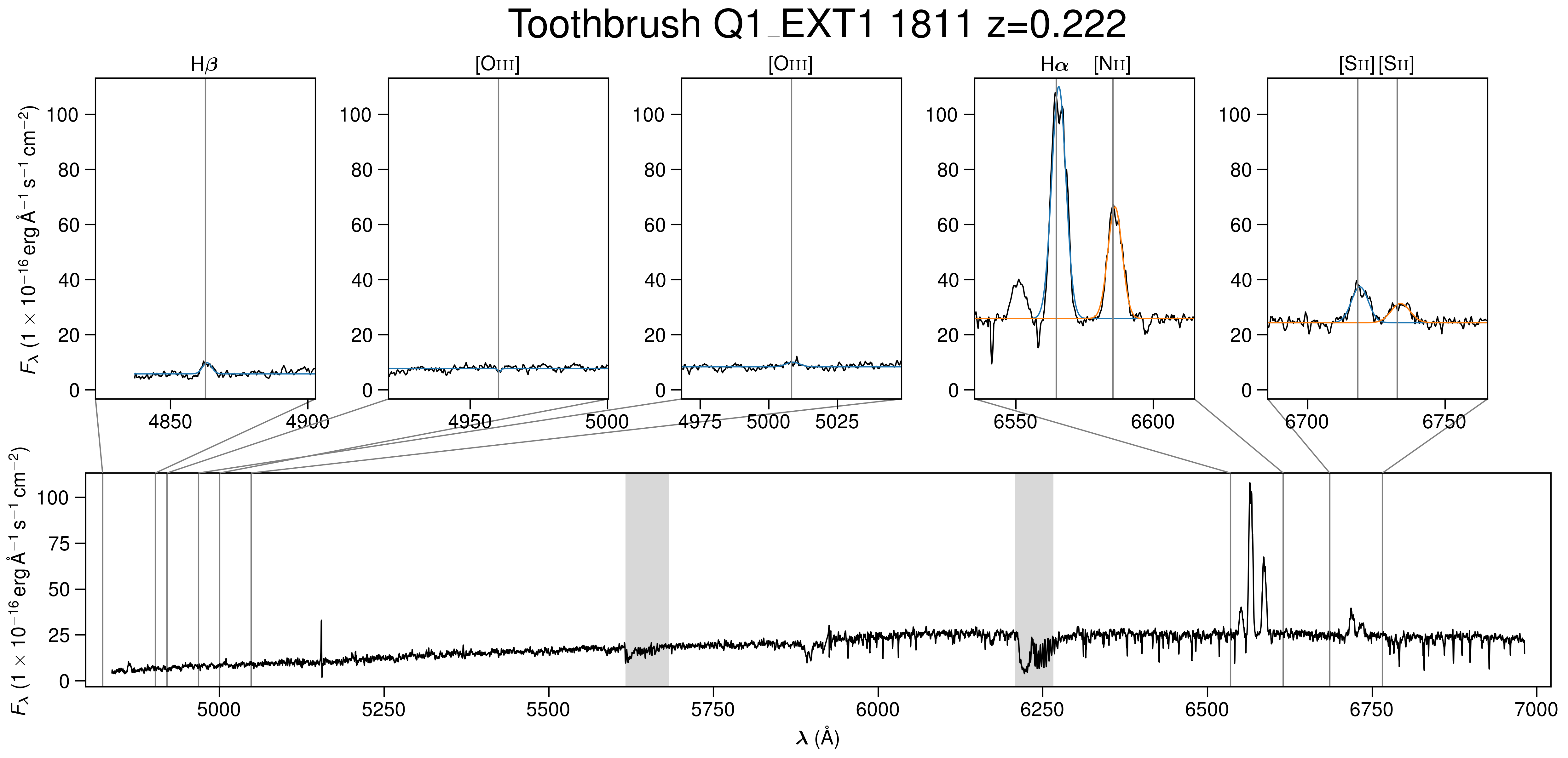}
      \end{minipage}%
    }\\}%
  {\subfloat[A composite spectrum taken with Hectospec/MMT.]{%
      \begin{minipage}{0.08\textwidth}
        \includegraphics[width=\textwidth]{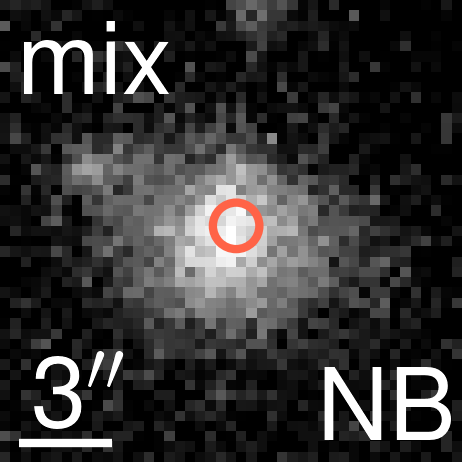}\\
        \includegraphics[width=\textwidth]{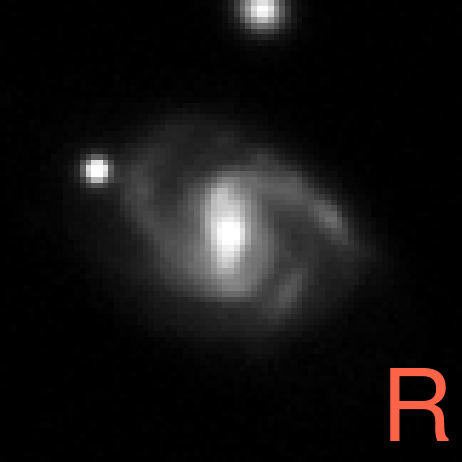}\\
        \includegraphics[width=\textwidth]{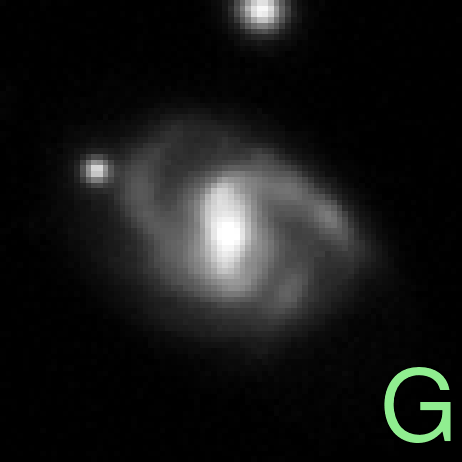}\\
        \includegraphics[width=\textwidth]{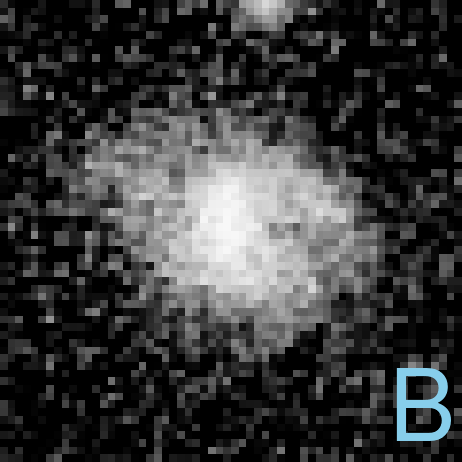}
      \end{minipage}%
      \hspace{5pt}
      \begin{minipage}{0.92\textwidth}
        \includegraphics[height=0.4\textwidth]{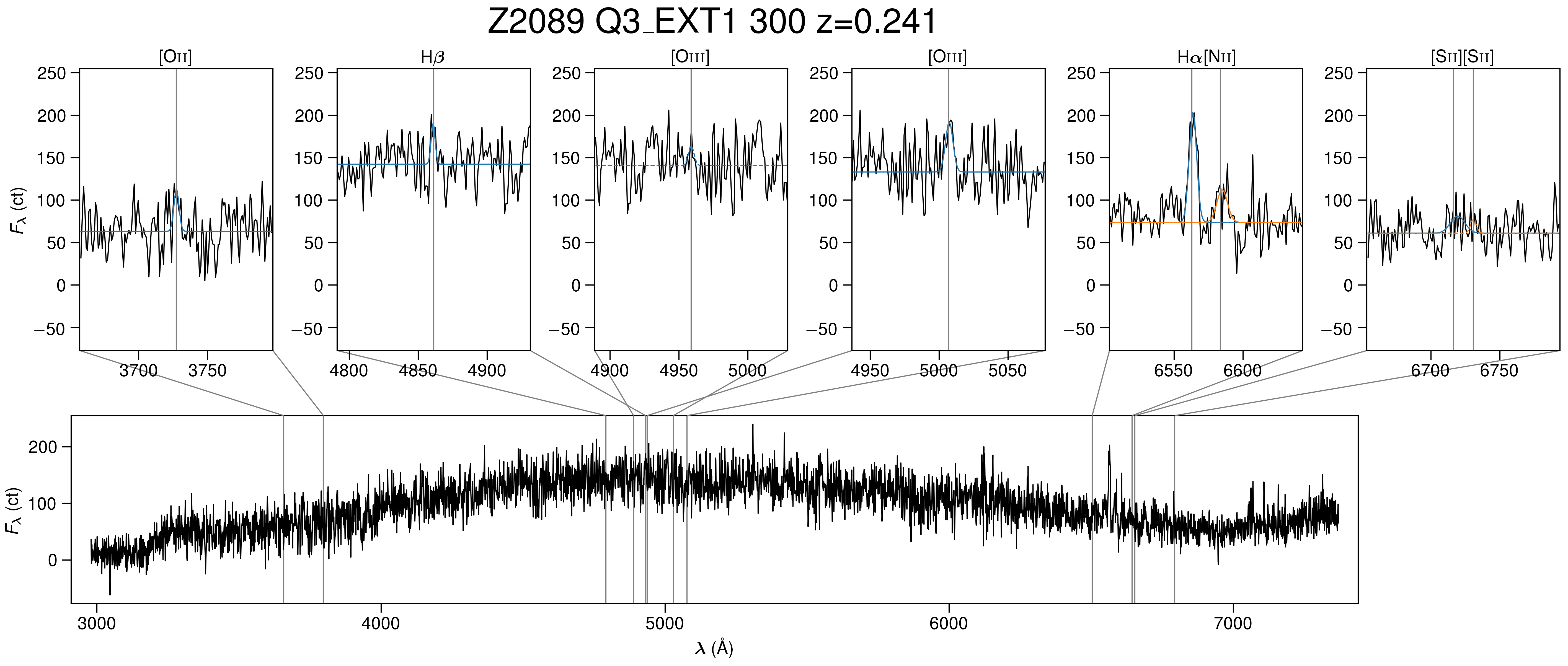}
      \end{minipage}%
    }\\}%
  \caption{Continued.}
\end{figure*}

\subsection{Target Selection, Observations, and Data Reduction}

We carried out spectroscopic follow-up observations of a subset of the emitters in 14 galaxy clusters, of which six are relaxed, and eight are undergoing mergers (see Table~\ref{tab:clusters} for details on the cluster properties). We obtained new multi-object spectroscopy (MOS) using three different instruments (VLT/VIMOS, WHT/AF2, Keck/DEIMOS) described below. We also leveraged publicly available spectroscopy obtained through the Arizona Cluster Redshift Survey (ACReS\footnote{\url{http://herschel.as.arizona.edu/acres/acres.html}}). In designing our observations, the main drivers were to cover and detect all main rest-frame optical emission lines, including spectral coverage of \Ha, \NII ($\lambda\lambda\,6550,\,6585$\,\AA), \Hb, \OIII ($\lambda\,5007$\,\AA) and \SII ($\lambda\lambda\,6718,\,6733$\,\AA). The resolution was required to be sufficient to separate \Ha and \NII\footnote{Unless specified, when used alone, \NII will refer to the $\lambda$\,6585\,\AA{} component.}, and the \SII doublet, which enables us to accurately deblend lines, precisely measure redshifts, and, in some cases, resolve some lines in velocity. While not designed with absorption in mind, the Mg and Na absorption lines, for example, are covered. We designed masks and fiber configurations to prioritize NB \Ha candidates with bright NB \Ha luminosities and ensure the detection of emission lines with a small telescope time investment. Any remaining space was allocated to other likely cluster members. Detecting continuum emission for individual sources was not part of the main aims of our survey, but it is nevertheless detected in brighter targets (with typical $i$-band magnitudes brighter than 21.5\,mag). The FOV of all the MOS instruments was smaller than the coverage of our NB observations, which generally resulted in a denser sampling at low cluster-centric radii and a sparser sampling at larger radii when compared to our NB selection.

Figure~\ref{fig:spectra} displays representative examples of \Ha NB candidates confirmed to be SF galaxies and AGN-dominated sources at the cluster redshift, together with NB, red, green, and blue postage stamps of the galaxy. Figure~\ref{fig:composite} shows the distribution of candidate sources and those confirmed to be \Ha cluster members for the merging cluster A2254.

In this paper, we rely on the spectroscopic observations mainly for measuring redshifts to confirm cluster membership, for measuring line ratios to classify galaxies as star-forming or AGN-dominated, and for obtaining line EWs. Further properties available from the data, such as metallicities, ionization parameters, electron densities, will be discussed in a forthcoming paper.

\subsubsection{VLT/VIMOS}\label{sec:vimos}

We employed the multi-slit capabilities of the VIMOS\footnote{\url{https://www.eso.org/sci/facilities/paranal/decommissioned/vimos.html}} instrument mounted on the UT3 telescope at Paranal Observatory. Slits can be distributed over the 4 VIMOS detectors, for a total FOV of $4\times7'\times8'$. With the MR grating in combination with the GG475 blocking filter with $1''$ slits, our observations covered the $500-1000$\,\AA{} range at a resolution of 12.5\,\AA. Five clusters (see Table~\ref{tab:obs} for details) were observed in a single slit configuration, for an average of 2\,h on target, in $0.6-1.2''$ seeing and thin cloud conditions during August 2017. The data were reduced using the VIMOS pipeline implemented in EsoReflex \citep{2013A&A...559A..96F}. The reduction first creates a set of final combined biases, darks, flat fields, and arcs. Each science exposure is debiased, flat-fielded, and corrected for spatial curvature and cosmic rays and hot pixels are flagged using a bad pixel table. The sky contribution is subtracted using a local sky model, and a set of sky lines are used to obtain the wavelength solution, calibrated to vacuum wavelengths. We align and combine the individual exposures pertaining to a common pointing/setup into a single image, on which we perform the detection and extraction of objects. A standard star was observed for each observing block and calibrated in the same fashion as the science data. The standard star is used to estimate the response curve. Finally, the response curve is applied to the science data to flux-calibrate the extracted spectra.

\subsubsection{WHT/AF2}

We observed three clusters with the AF2\footnote{\url{http://www.ing.iac.es/astronomy/instruments/af2/}} instrument mounted of the William Herschel Telescope in La Palma, Spain. AF2 has configurable $1.6''$ fibers, which can be deployed over a FOV of about $30'\times30'$. Observations of the Sausage cluster were taken in two separate fiber configurations in July 2014, using the R600R grating and reaching a resolution of 4.4\,\AA{} and were presented in \citet{Sobral2015}. The rest of the data on two clusters were taken in January-February 2017. The data were taken in good seeing conditions of $0.8-1.0''$ for most of the run. We used the R316R grating, covering the $4000-10000$\,\AA{} range at 8.1\,\AA{} resolution. Compared to the VIMOS observations, the observations extend bluer and they cover the \OII ($\lambda$\,3727\,\AA{}), albeit with lower sensitivity and at a much-increased noise level. As \citet{Sobral2015} contains details of the AF2 data reduction, we provide a summary here. After correcting the fiber traces for flexure, the data were corrected for bias, flat-fielded, and the sky was subtracted. We extracted 1D spectra for each source, which were wavelength calibrated and flux calibrated.

\subsubsection{Keck/DEIMOS}

Two clusters in this sample were observed with the DEIMOS\footnote{\url{https://www2.keck.hawaii.edu/inst/deimos/}} instrument on the Keck telescope as part of the Merging Cluster Collaboration (MC$^2$) spectroscopic follow-up efforts \citep{Golovich2019}. While the primary targets for MC$^2$ were candidate passive cluster members, in this paper, we focus on the minority of \Ha candidates included in the selection. The Sausage cluster data have been presented in detail in \citet{Sobral2015}, while the entire Keck sample is presented in detail in \citet{Golovich2019}. We refer the reader to those two papers for full details on the target selection and data reduction. In short, each cluster was covered with four slit masks for a total of about 0.75\,h, under excellent seeing conditions of about $0.7''$. With the 1200 line\,mm grating, a resolution of 1\,\AA{} is achieved over the $5400-8000$\,\AA{} range, although this can vary by $\pm500$\,\AA{} from source to source, which means some very red emission lines can be missed. The data are reduced using the DEEP2 version of the SPEC2D package \citep{Newman2013}, which performs the same set of steps as described for VIMOS (see Section~\ref{sec:vimos}).

\subsubsection{Hectospec/MMT (through ACReS)}

We leverage publicly available spectroscopy for seven galaxy clusters through the Arizona Cluster Redshift Survey \citep[ACReS,][]{Newman2013}. ACReS followed up thousands of galaxies in the field of galaxy clusters, down to a limiting K-band magnitude. Unlike our other datasets, which included a clear selection for emission-line galaxies, the ACReS dataset, therefore, includes a multitude of stars and galaxies, of which some are in the cluster, and some are foreground and background galaxies. In this paper, we use only galaxies with significant emission line detections. ACReS was conducted with the Hectospec\footnote{https://www.cfa.harvard.edu/mmti/hectospec.html} instrument \citep{Fabricant2005} on the MMT telescope in Arizona, which can deploy 300 $1.5''$ fibers over a FOV of 1$^\circ$, in combination with the 270 line grating, to cover the $3650-9200$\,\AA{} at a resolution of 6\,\AA. The MMT observations cover \OII, and all the other lines covered in our other observations. The MMT observations are reduced through the Smithsonian Astrophysical Observatory Optical/Infrared Telescope Data Center using the Hectospec pipeline\footnote{\url{http://tdc-www.harvard.edu/instruments/hectospec/pipeline.html}}. Similar steps are conducted to the other data sets. Taking advantage of the stable fiber response, we correct the response curve by deriving the pixel response and fiber throughput from the flat fields. Note that, while a relative flux calibration is applied, bringing wavelengths on the same relative flux scale within each spectrum, no absolute flux calibration with a standard star is performed in the Hectospec pipeline. By comparing the seven pairs of galaxies with both an MMT and a VIMOS spectrum, we conclude this has minimal effects on our results (see also Section~\ref{sec:redshifts} and \ref{sec:BPT}). The wavelength calibration is done to air wavelengths, unlike the rest of the data, but this is inconsequential for the emission line measurements presented in our paper.

\subsection{Ancillary Imagining and Photometry}

We use ancillary imaging and photometry to assess our selection of targets and interpret our results. We particularly rely on the NB observations centered on \Ha at the cluster redshift, the associated $i$ broad-band observations and derived products, such as the NB and $i$ magnitudes and NB-derived \Ha luminosities, as presented in \citet{Stroe2017}.

\begin{figure}[ht!]
  \includegraphics[width=\linewidth]{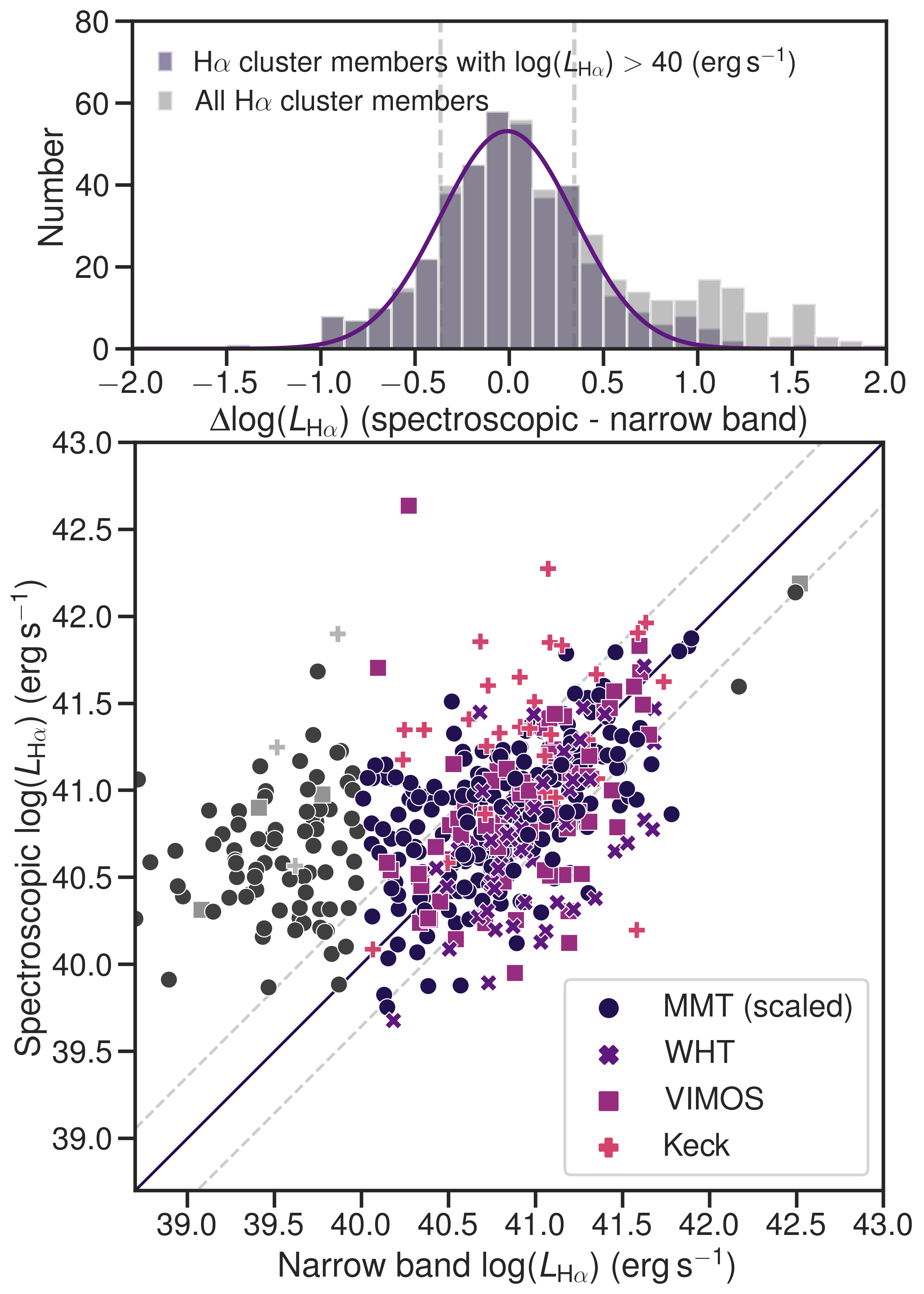}
  \caption{The \Ha luminosity measured from spectroscopy correlates well with the NB \Ha luminosity. The spectroscopic luminosities follow a Gaussian distribution ($\sigma=0.36$) around the expected 1:1 relation in the luminosity range used in our science analysis ($10^{40}-10^{42}$ erg\,s).
  }
  \label{fig:lum_lum}
\end{figure}

Throughout the paper, we choose to use \Ha luminosities derived from the NB photometry and not from the spectroscopy because of two main reasons. Firstly, the \Ha luminosities were derived from the photometry using large 5$''$ apertures, which encompass all the emission even for the largest galaxies at $z\sim0.15-0.3$. The spectroscopic follow-up included fibers and slits of varied size and shape, which can lead to slightly different estimates of the \Ha luminosity. Secondly, the Hectospec/MMT observations were not absolute flux calibrated. Overall, the \Ha luminosities derived from spectroscopy match those derived from the NB (see Figure~\ref{fig:lum_lum}), especially for \Ha luminosity ranges of most interest in the paper ($10^{40}-10^{42}$\,erg\,s$^{-1}$). After applying a single scaling factor to the MMT luminosities (which lack flux calibration), we find that spectroscopic \Ha luminosities have a Gaussian spread with a standard deviation of $0.36$\,dex around the expected 1:1 relation with the NB measurements. Hence, the NB-derived \Ha luminosities provide us with a clean, simple way to compare all the data, including the MMT observations. Note that all NB luminosities are corrected for \NII contamination and for dust extinction within the host galaxy, as described in \citet{Stroe2017}. For a detailed description of the process of measuring properties, including \Ha luminosities from spectroscopy, see Section~\ref{sec:measurements}.

We also employ our own $g$ and $r$ data \citep{Stroe2017}, and when not available, imaging and photometry from the Sloan Digital Sky Survey\footnote{\url{https://www.sdss.org/dr12/}} \citep[SDSS Data Release 12,][]{Alam2015} or the Pan-STARRS\footnote{\url{https://panstarrs.stsci.edu/}} survey \citep[PS1,][]{Flewelling2016}. In building our figures, we also make use of publicly available X-ray observations through the \textit{XMM-Newton}\footnote{\url{http://nxsa.esac.esa.int/nxsa-web}} and \textit{Chandra}\footnote{\url{https://cxc.harvard.edu/cda/}} archives and 1.4\,GHz radio data from the NRAO VLA Sky Survey\footnote{\url{https://www.cv.nrao.edu/nvss/}} (NVSS).

\begin{figure}[t!]
  \includegraphics[width=\linewidth]{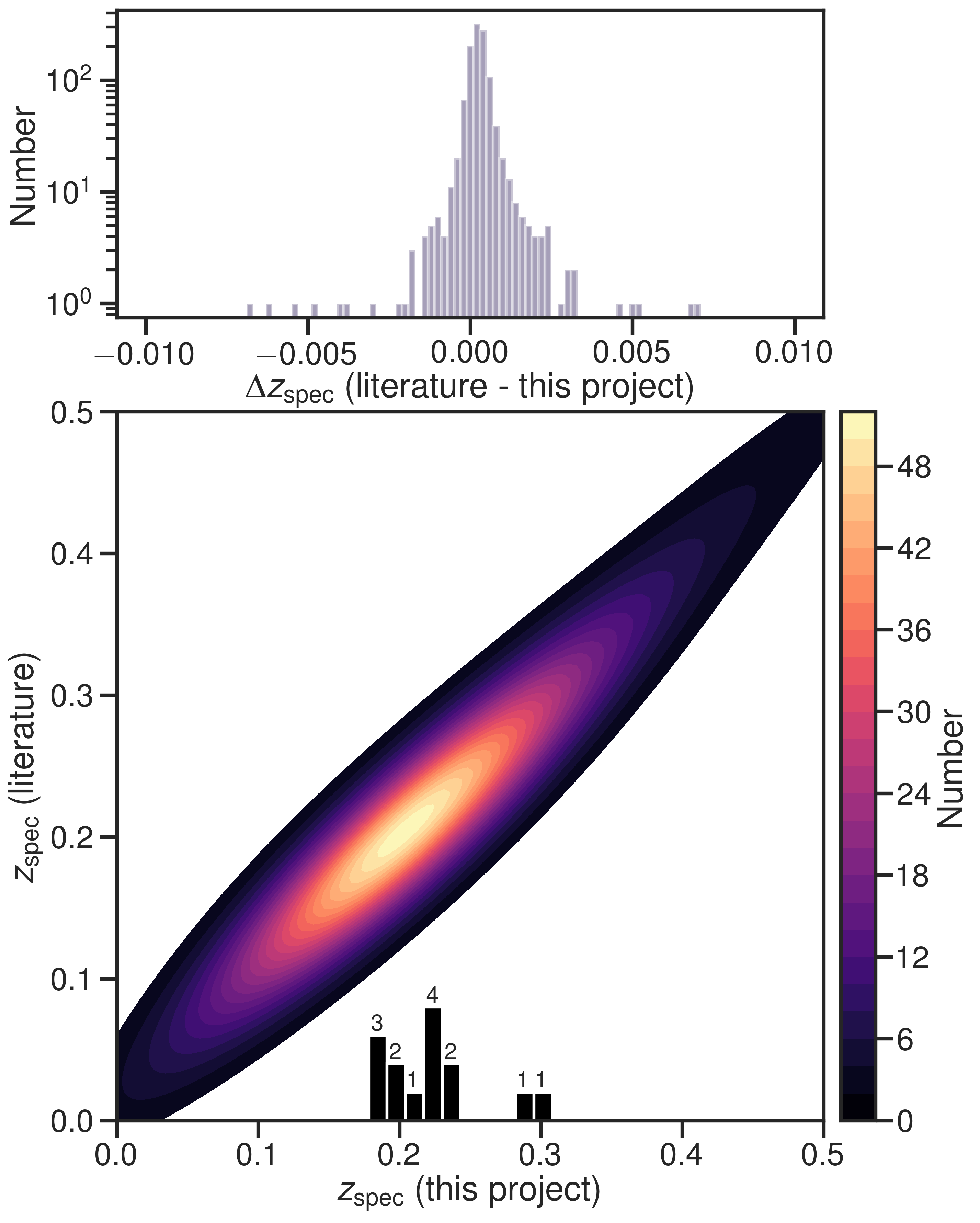}
  \caption{The accuracy of our redshifts, compared to values from the literature. All available spectroscopy is included, not only \Ha candidates. Bottom: Our redshifts match well with those from the literature. Of note is the high density of sources around the cluster redshifts, whose distribution is shown in the histogram. Top: The distribution of shifts between the literature redshifts and our redshifts (note the logarithmic scale). The bulk of the sources have matching redshifts within 0.0005.}
  \label{fig:redshift_accuracy}
\end{figure}

\begin{figure*}[htb!]
  \includegraphics[width=\textwidth]{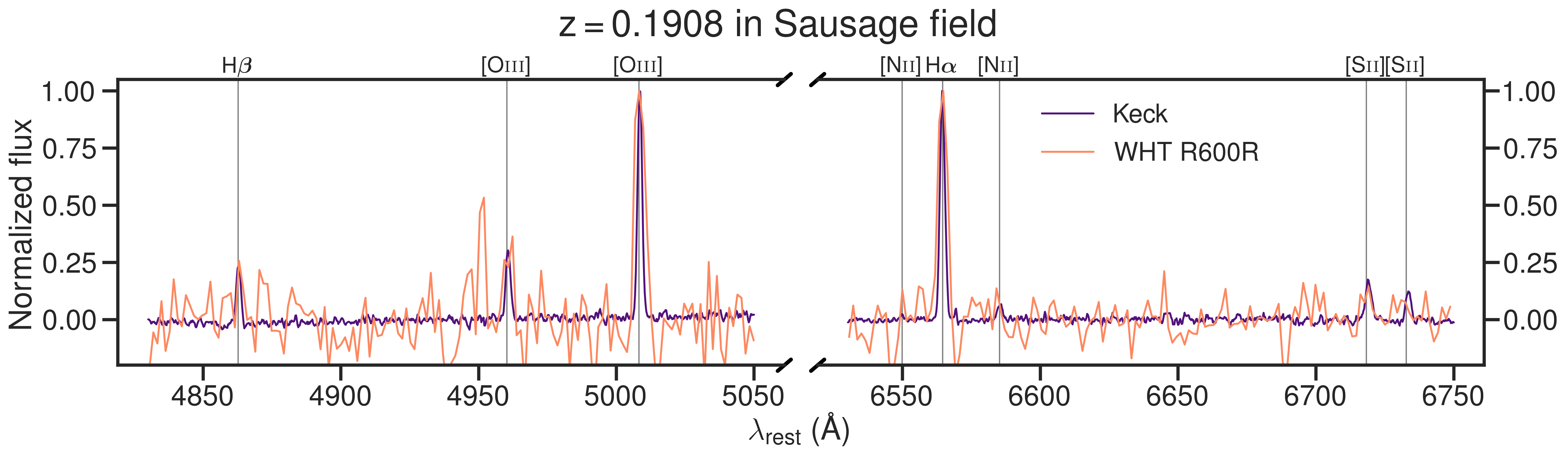}\\
  \includegraphics[width=\textwidth]{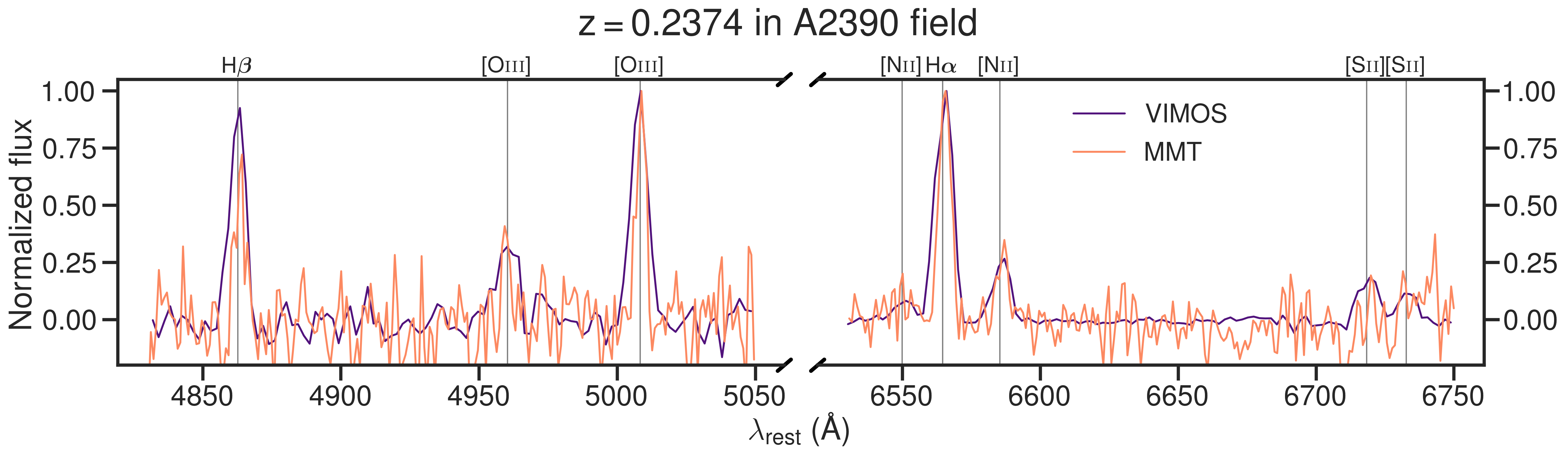}
  \caption{\Ha cluster members confirmed with spectroscopy with two independent observations from different telescopes.
    The redshifts and emission line properties are in excellent agreement, despite the different telescope sizes (4, 6.5, 8, and 10-m), exposure times (0.5-2\,h), observing conditions (dark time versus gray time with thin  clouds), and instrument properties (1$''$ slits vs. 1.5$''$ fibers, $1-12.5$\,\AA{} spectral resolution). For display purposes, we subtracted the median and normalized the data to the maximum value in each panel.}
  \label{fig:pairs}
\end{figure*}

\subsection{\textsc{redshifts} Package and Ancillary Redshifts}

While the overwhelming majority of the sources are expected to be \Ha emitters at the cluster redshift, the NB filters are also sensitive to higher redshift emitters as \OII, \OIII and \Hb at $z > 0.5$ and to M-class stars which can mimic emission lines in the NB because of their strong spectral features. Therefore, we also collect spectroscopic redshifts from the literature, which enable us to further study the reliability of our NB selection of \Ha candidates and to conduct a supplemental check of the redshifts derived from our spectroscopy.

To aid in this endeavor, we introduce the \textsc{redshifts} package \citep{stroe_redshifts}. \textsc{redshifts} is a Python package that collects all unique spectroscopic redshifts from the VizieR\footnote{\url{https://vizier.u-strasbg.fr/}} and the NASA Extragalactic Database (NED\footnote{\url{https://ned.ipac.caltech.edu/}}) online databases. With \textsc{redshifts}, the user can perform a flexible search within a radius of a given set of (R.A., Decl.) coordinates. \textsc{redshifts} leverages Astroquery\footnote{\url{https://astroquery.readthedocs.io/en/latest/}} to use column names and descriptions (including UCD keywords) to identify columns containing spectroscopic redshifts or radial velocities. \textsc{redshifts} weeds out photometric redshifts and duplicates and returns a unique list of `best' spectroscopic redshift measurements in FITS table format. \textsc{redshifts}\footnote{\url{https://github.com/multiwavelength/redshifts}} uses a configuration file, written in YAML\footnote{\url{https://yaml.org/}}, a popular human-readable markup language, to specify the search radius, column names, and any VizieR `banned' catalogs. The search is performed from the specified (R.A., Decl.) position out to the radius from the configuration file. The uncertainty is used to evaluate whether redshifts could be photometric instead of spectroscopic. The `banned' catalogs encompass any VizieR catalogs one does not want included in the search, for example, because they were found to mix spectroscopic and photometric redshifts in one column. One of the limitations of the package is that it relies on the original authors to use the UCD and other column names correctly. For wide-area searches (i.e. large radius), NED and VizieR sometimes time out. Additionally, the search requires a stable internet connection and can fail if the connection is interrupted during the search. For details on installation, usage, and full functionality, we refer the reader to \citet{stroe_redshifts} and \url{https://github.com/multiwavelength/redshifts}.

For the ENISALA project, we used \textsc{redshifts} to collect available spectroscopic redshifts out to 0.7$^\circ$ from the cluster core, covering the FOVs of our NB observations in their entirety.

\section{Redshift Measurements} \label{sec:redshifts}

We derive redshifts through visual inspection, primarily from emission line features, focusing on \Ha, \NII, \Hb, \OIII and \SII, and secondarily from absorption features, such as CaHK and G band absorption where covered, in the 1D spectra. In line with their respective calibrations, vacuum wavelengths are used as a reference for the VIMOS, Keck, and WHT samples, while air wavelengths are used for the MMT sample. For completeness and self-consistency, we derived redshifts for all 1D spectra available to us, including, for example, any passive galaxies selected as fillers in the VIMOS observations and all galaxies in the MMT fields, many of which are passive cluster galaxies and non-cluster members.

We cross-match the positions of sources in our sample with the publicly available redshifts measurements in the literature, using a 1$''$ tolerance. Figure~\ref{fig:redshift_accuracy} shows the distribution of redshifts in our sample, as well as a comparison to the literature values. We note that, in terms of source numbers, the sample presented in Figure~\ref{fig:redshift_accuracy} is dominated by the ACReS spectroscopy, whose broad, magnitude limited sample, results in a highly complete cluster coverage, but also a significant contribution from foreground and background galaxies, and to a lesser degree, stars. However, the bottom panel of the figure highlights the high density of sources concentrated around the cluster redshifts. The top panel shows a slight systematic median shift of $0.00026$ between our redshifts and those in the literature, dominated by the large number of non-cluster MMT observations we included in the plot. The shift is consistent with a wavelength error of just 1\,\AA{}, expected given the resolution (6\,\AA{}) of the Hectospec/MMT instrument. The bulk of the source show excellent agreement, with about 75\% of sources with redshift measurements within $<0.0005$ of those in the literature.

By virtue of chance, there are 16 galaxies with observations from two telescopes. Based on these pairs of galaxies, we estimate that the quality of the calibration and redshift estimation is excellent. We show two examples in Figure~\ref{fig:pairs}.

\begin{figure}[t!]
  \includegraphics[width=\linewidth]{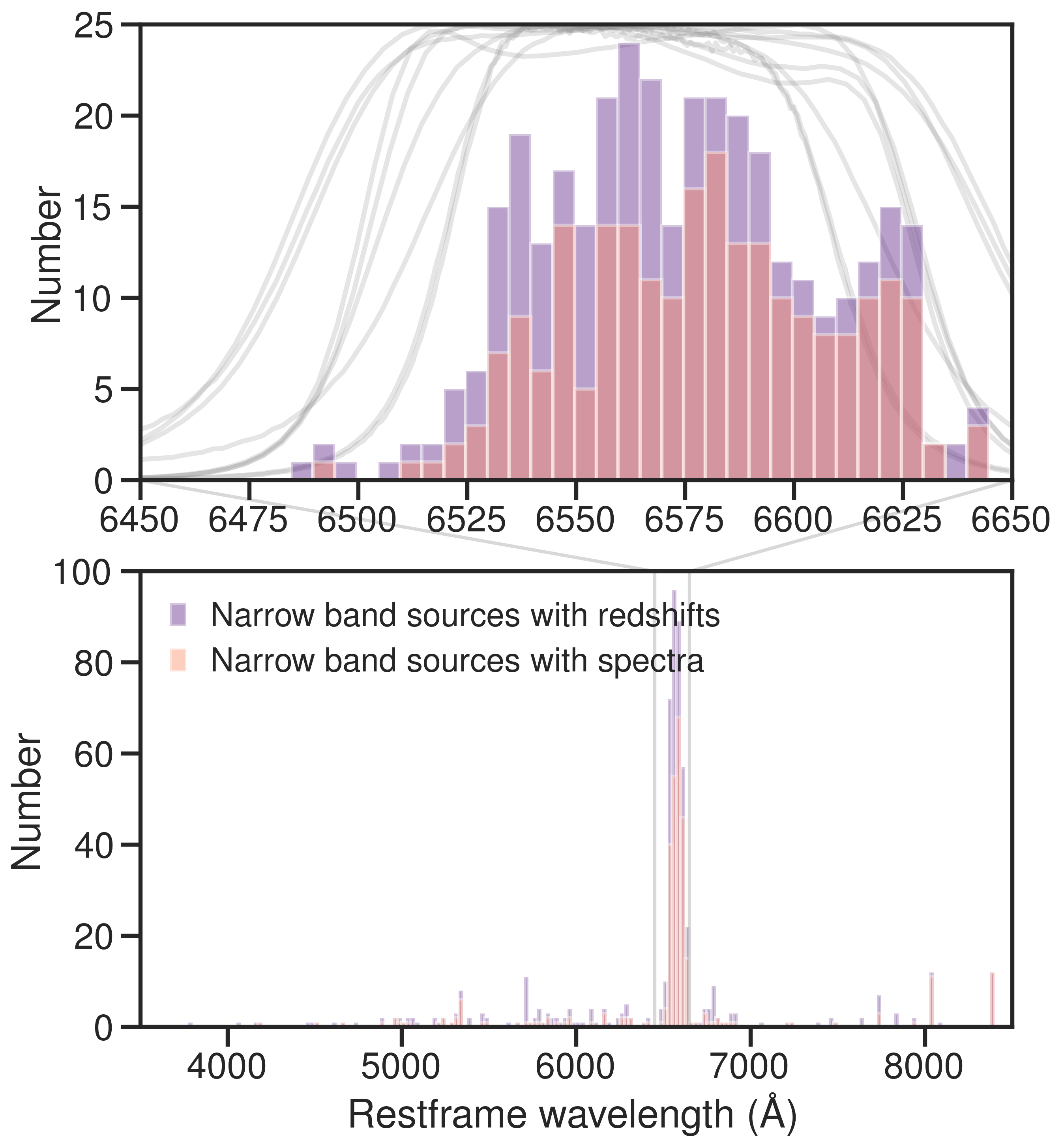}
  \caption{Distribution of rest-frame wavelengths for the spectroscopically confirmed NB candidates. The top panel shows a zoomed-in view on the wavelength range around \Ha. We also show the transmission profiles of the nine NB filters used to select the \Ha candidates, converted to the \Ha rest-frame (thin gray lines). We confirm the majority of NB candidates ($>$70\%) as \Ha emitters at the cluster redshifts, with the rest confirmed as either stars or galaxies at other redshifts.}
  \label{fig:redshift_distrib}
\end{figure}

\section{Selection function} \label{sec:selection}

\subsection{How good was the NB selection?}

With spectroscopy in hand, both from our work and the literature, we can further quantify the robustness of our NB selection for \Ha cluster candidates. Figure~\ref{fig:redshift_distrib} shows the distribution of redshifts for the NB sources. When we include spectroscopy from the literature, 65\% of \Ha candidates are confirmed to be at the cluster redshift. When focusing on our sample of sources for which we have direct access to spectroscopy, the fraction rises to 71\% confirmed as \Ha candidates. This discrepancy can be explained as follows: when compared to our spectroscopic campaign that prioritized bright emitters, the selection function of spectroscopy from the literature is biased towards lower magnitudes and \Ha luminosities. The NB selection purity for \Ha is expected to drop as the contamination from interlopers increases at fainter fluxes. This is because luminosity functions of higher redshift emission lines are steeper at fainter fluxes contributing more at faint fluxes than at brighter fluxes \citep[e.g.][]{Khostovan2015}. The selection was particularly poor for RXJ0437 and RXJ2129, where none of the followed-up sources is confirmed to be a cluster member. The high fraction of M-type stars in RXJ0437 (63\% of candidates followed up) is most likely due to the combination of central wavelength ($\sim$8370\,\AA) and width (210\,\AA) of the NB filter used for the NB selection in this cluster, which is the widest NB filter used in our survey and which traces one of the more prominent spectral features in M stars. When removing contaminating stars from the sample, we confirm that $\sim$78\% of galaxies are \Ha emitters. Overall, the spectroscopic follow-up purity in 7/13 clusters is over 80\%.

\subsection{Distribution with cluster-centric distance}

\begin{figure}[t!]
  \includegraphics[width=\linewidth]{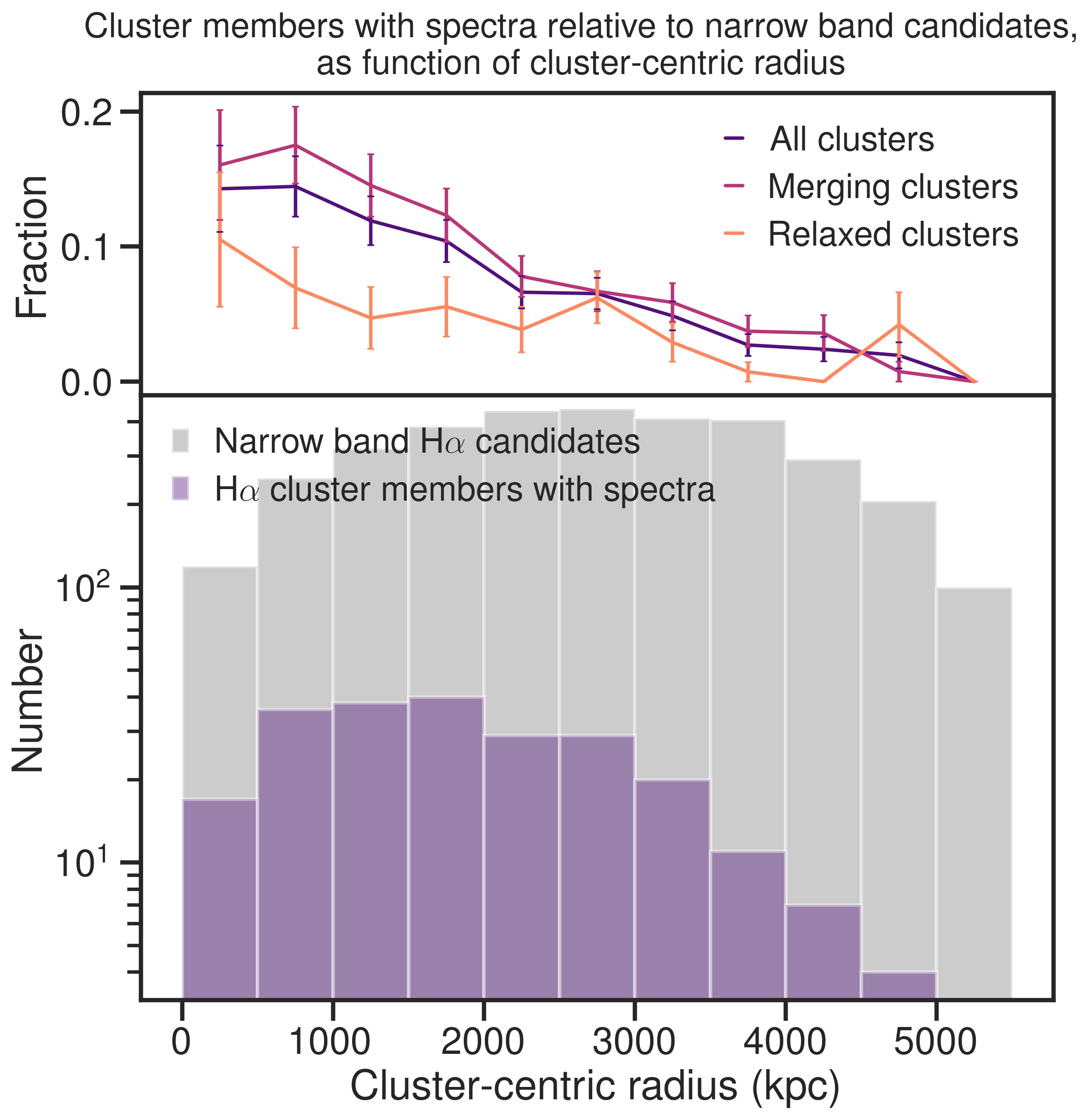}
  \caption{Distribution of NB candidates and those with follow-up spectroscopy with cluster-centric distance. The spectroscopic follow-up for both relaxed and merging clusters more densely samples the parent population towards the cluster core, as expected.}
  \label{fig:radial_distrib}
\end{figure}

Figure~\ref{fig:radial_distrib} shows the distribution of the spectroscopic and parent samples with cluster-centric radius. The distribution of the underlying NB candidate sample is driven by both physical factors (e.g. cluster influence on the SF activity in infalling galaxies), as well as the FOV of the NB observations, which results in less coverage at large radii. In the figure, the bins are equally spaced in radius, which translates to increasing areas at larger cluster-centric radii, which partially explains the large number of sources at $2000-4000$\,kpc distances. We followed up NB candidates at similar rates for merging and relaxed clusters, with higher rates at low cluster-centric distances. For both merging and relaxed clusters, we followed-up sources within 2000\,kpc at about twice the rate of those at larger distances.

\subsection{Distribution with NB magnitude and NB \Ha luminosity}

\begin{figure*}[t!]
  \includegraphics[width=0.5\linewidth]{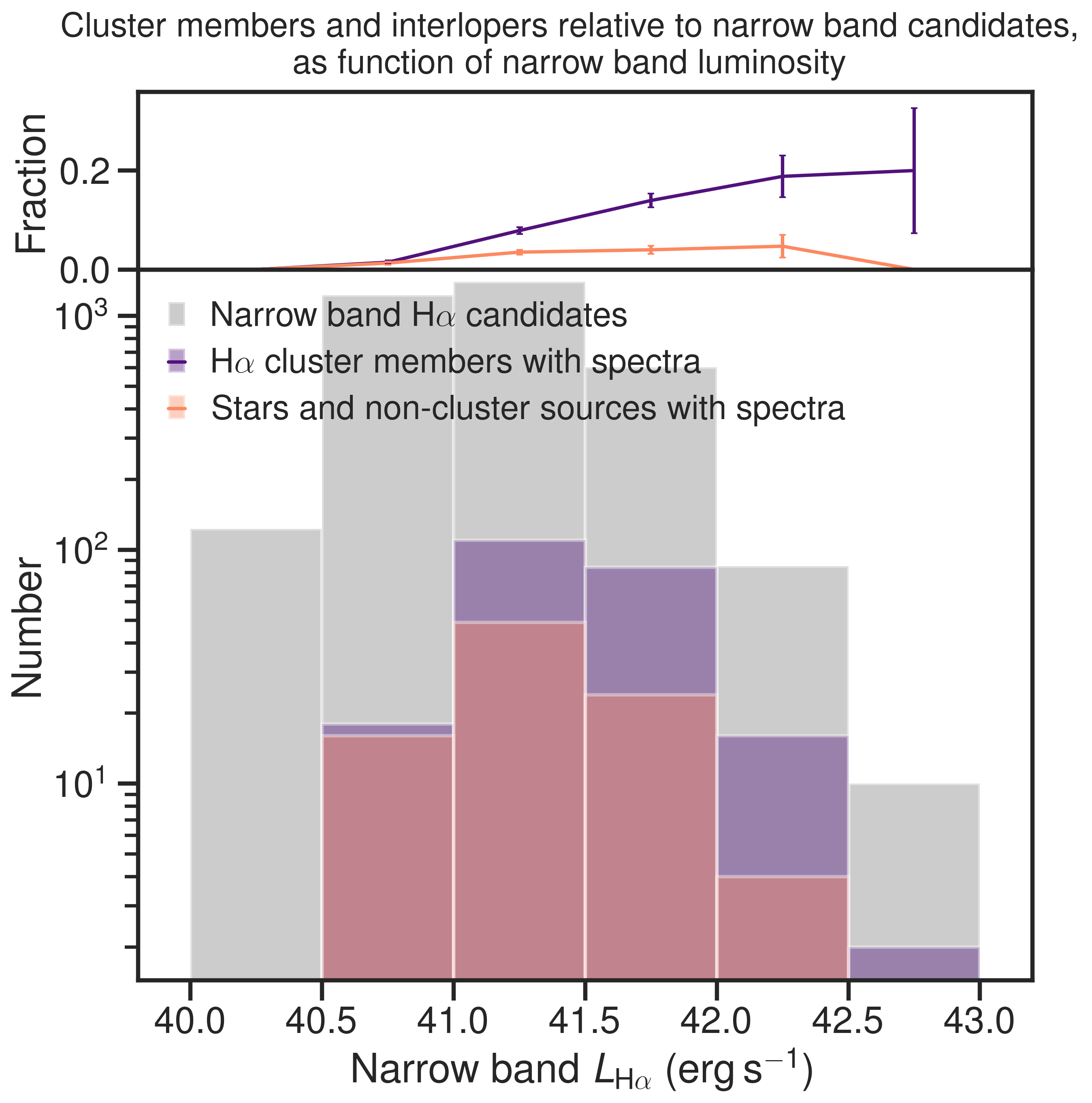}
  \includegraphics[width=0.5\linewidth]{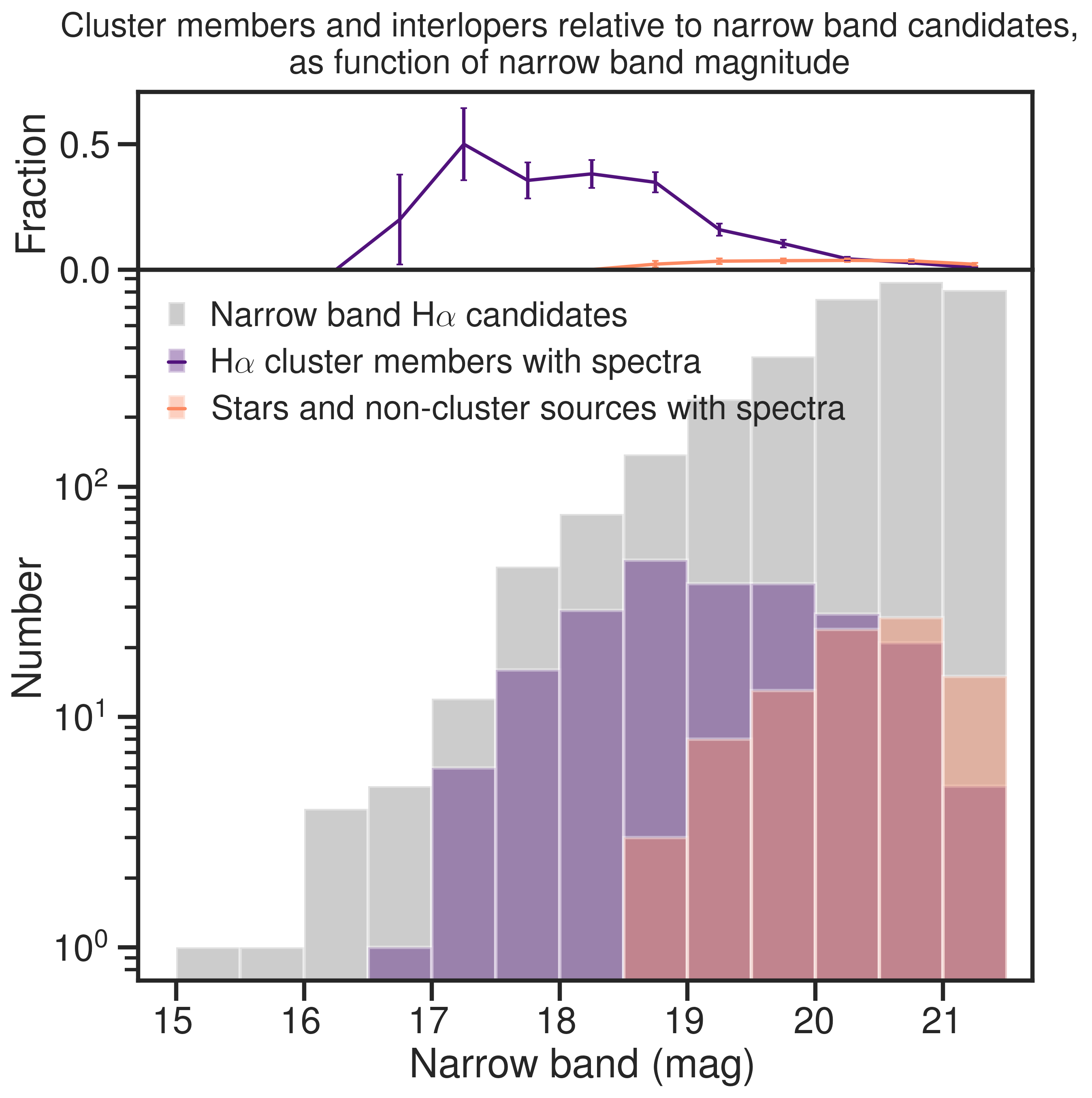}
  \caption{Distribution with respect to NB magnitude and \Ha luminosity derived from the NB. We show NB candidates and sources with spectroscopy, including confirmed \Ha emitters at the cluster redshift, foreground and background sources, and stars. The targets selected for spectroscopy are preferentially bright in the continuum and have bright emission lines. On average, the fraction of NB candidates followed-up with spectroscopy and confirmed as \Ha cluster members increases for bright sources and those with large \Ha luminosities.}
  \label{fig:mag_L_distrib}
\end{figure*}

To maximize the chance of spectroscopic confirmation, we preferentially followed-up bright sources, with large NB \Ha fluxes (typically $>4\times10^{-16}$\,erg\,s\,cm$^{-2}$. Figure~\ref{fig:mag_L_distrib} highlights this effect. Of note is the high fraction of sources confirmed as \Ha emitters at the cluster redshift, particularly those with bright NB magnitudes ($<18.5$\,mag) and large \Ha luminosities. Not unexpectedly, we find the most contamination from non-cluster sources, i.e. other line emitters at lower or higher redshift (see also Figure~\ref{fig:redshift_distrib}) and stars, happens in faint sources, especially those with small expected \Ha luminosities. Overall, as shown in Figure~\ref{fig:NB_LUM}, the spectroscopic follow-up samples well the underlying NB population.

\section{Galaxy Properties from Spectroscopy} \label{sec:measurements}

The spectroscopic component of the ENISALA project involves hundreds of cluster members of interest, which all required systematic measurements of emission and absorption lines. To this end, we used \gleam\footnote{\url{https://github.com/multiwavelength/gleam}} \citep[\textbf{G}alaxy \textbf{L}ine \textbf{E}mission \& \textbf{A}bsorption \textbf{M}odeling,][]{stroe_gleam, stroe_savu}, a Python package for fitting Gaussian models to emission and absorption lines in large samples of 1D galaxy spectra. For more details on the use and capabilities of \gleam, we refer the reader to \citet{stroe_savu}, \citet{stroe_gleam} and \url{https://github.com/multiwavelength/gleam}.

We use \gleam to measure the continuum and emission-line fluxes, EWs, and luminosities from the spectroscopy. The data was processed, as much as possible, in a uniform manner and without human interaction.

For the bulk of the sources, we fit the same set of emission lines, including \OII, \Hb, \OIII, \Ha, \NII and the \SII doublet. The fitting is done using the Levenberg–Marquardt algorithm, as implemented by LMFIT \citep{lmfit}, which has the benefit of being well-behaved, fast, and enabling an easy estimation of uncertainties on the fit parameters (see LMFIT documentation for more details). We included the uncertainties in the measured spectrum (the standard deviation) as weights in solving the minimization problem.

We fit a Gaussian model plus a constant continuum for each emission line. A total range of 70\,\AA\footnote{This range was 60\,\AA{} for the VIMOS and 40\,\AA{} for Keck, as this was found to produce more reliable results.} on either side of the emission line is used for fitting the model. This range is large enough to encompass enough line free spectrum to securely estimate the continuum without being affected by galaxy colors. The constant for the continuum, as well as all Gaussian parameters (central wavelength, amplitude, and line width), are all free parameters in the fit. However, the central wavelength is bound to $\pm3.0$\,\AA{} from the expected position. Individual lines are considered detections if the S/N of the amplitude of the Gaussian is greater than 2.

We jointly fit emission lines closer than 26\,\AA{} in rest-frame, meaning \Ha and \NII and the \SII doublet, respectively, are fit jointly. In a few sources with broad \Ha and strong \NII ($\lambda$\,6718\,\AA), a two-component \Ha plus the \NII doublet are jointly fit. Any other nearby lines outside of the 26\,\AA{} range are masked ($\pm20$\,\AA{} across the expected position) to not bias the continuum emission estimation. Any emission lines outside the spectral coverage are skipped. If necessary, we also mask the A ($\sim7600$\,\AA) and B ($\sim6900$\,\AA) sky absorption bands. For the Hectospec/MMT observations, the sky correction is very good and it is not necessary to mask the sky band.

\gleam first attempts to fit the full model with the constant and as many Gaussian components specified. If the fit does not converge, it simplifies the model by removing combinations of one or more Gaussian components. The final model includes a constant for the continuum and all emission lines with significant detections, with reported upper limits for non-detections. The emission line and continuum measurements are used to derive galaxy properties as described below in our results section. As a testament to the quality of data and calibration, we find good agreement between the spectroscopically derived \Ha luminosities and those derived from the NB (see Figure~\ref{fig:lum_lum}).

\section{Final Sample} \label{sec:sample}

The entire sample of sources with spectroscopy amounts to over 4200 sources, of which 381 we obtained through our targeted VLT/VIMOS, Keck/DEIMOS, and WHT/AF2 follow-up. The rest of the sample, from the ACReS program, included many stars, foreground and background sources, as well as passive cluster members that do not lie at the focus of our current project.

While our VLT/VIMOS, Keck/DEIMOS, and WHT/AF2 follow-up represents a strict sub-sample of the \Ha NB candidates, the nature of the ACReS selection (effectively a mass-limited selection) included many sources that were not identified as \Ha candidates in our NB survey. Such sources didn't pass our original NB selection criteria, either because they were too faint in the BB (i.e. faint continuum) or because their NB-derived EW did not pass the 3 sigma threshold (i.e. faint emission line). For our study, we only include sources from Hectospec/MMT that pass the same selection criteria as our VLT/VIMOS, Keck/DEIMOS, and WHT/AF2 follow-up sources, i.e. have significant detections of emission lines and have NB measurements of the \Ha luminosity (both ensuring bright emission lines).

We include sources in the sample if their redshift corresponds to the cluster redshift within a narrow range. We thus select sources whose redshift falls within the redshift range covered by the 2 times the full-width-half-maximum (FWHM) of the NB filter used for each cluster, transformed into the rest-frame of \Ha. For example, the central wavelength of the NB236 NB filter is 7839\,\AA{} with a FWHM of 110\,\AA. Within 2 FWHMs, the filter traces \Ha at redshifts between $0.1777<z<0.2112$. A source is included in the sample if its redshift falls within this range. For uniformity, we apply this selection to the ACReS Hectospec/MMT dataset as well.

\begin{figure*}[ht!]
  \includegraphics[width=\linewidth]{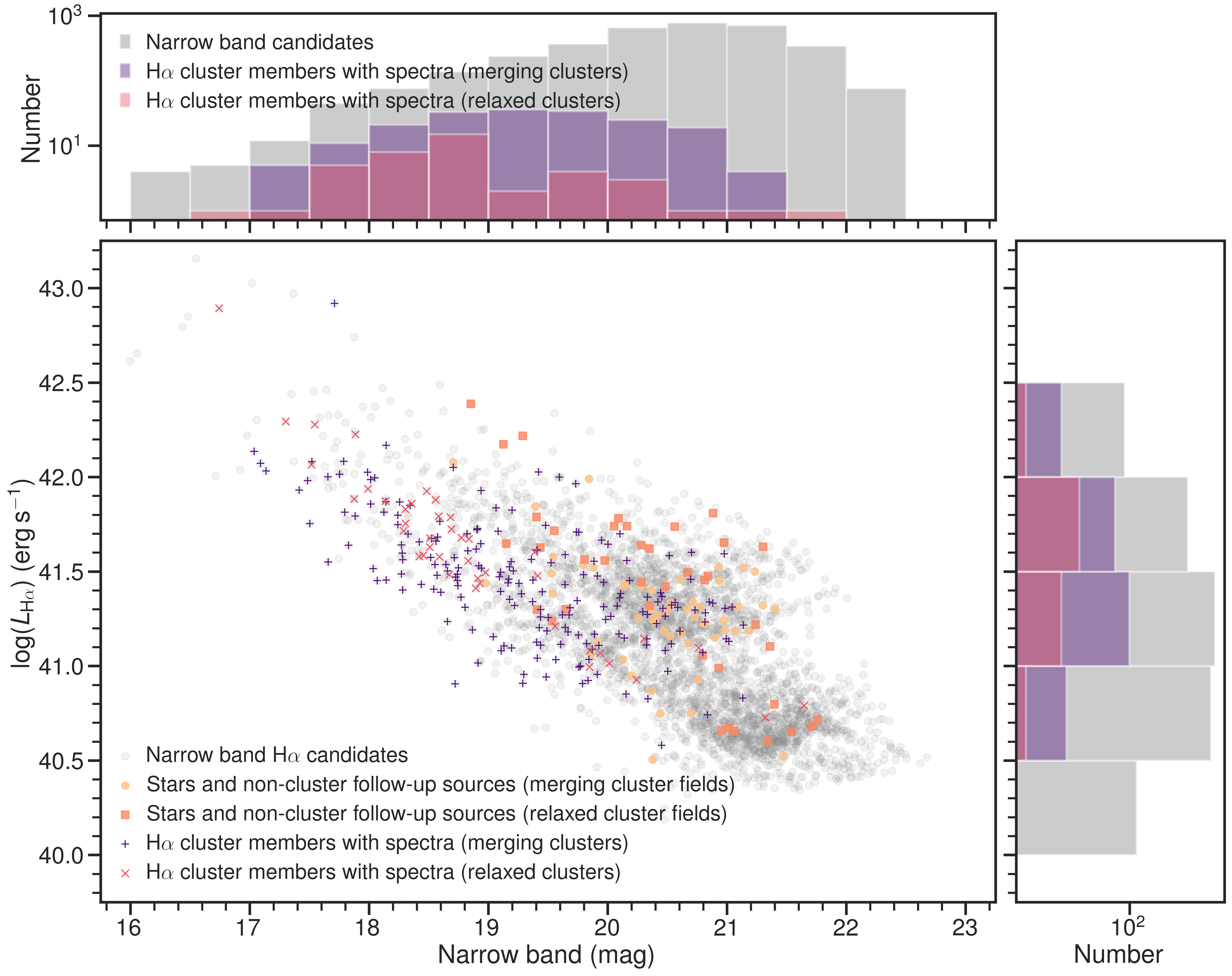}
  \caption{The magnitude-\Ha luminosity distribution of confirmed cluster \Ha and non-cluster sources, compared to the underlying NB candidate pool. We note that our spectroscopy samples well the underlying population.}
  \label{fig:NB_LUM}
\end{figure*}

Since we are interested in star-forming and active galaxies, sources at the cluster redshift must also meet a second criterion of having strong emission lines, which ensures the removal of passive ellipticals and post-starburst galaxies from the final sample. The criteria for selecting the final sample of sources are guided by detections or upper and lower limits in \Ha, \NII, \Hb and \OIII which enable the classification of a source using the BPT diagram. The selection criteria are detailed in Appendix~\ref{app:criteria}.

To ensure comparable samples between our merging and relaxed clusters, we also focus our analysis on sources with NB \Ha luminosities between $10^{40}$ and $10^{42}$ erg\,s$^{-1}$, located within 4500\,kpc from the cluster center (see Section~\ref{sec:lumdist} and Figure~\ref{fig:radial_distrib_type}).

In summary, a source must pass a set of criteria to be included in the final ENISALA sample used in our analysis. The criteria are as follows:
\begin{itemize}
  \item Source must be located at cluster redshift,
  \item Source must have emission line detections,
  \item Source must be classifiable in the BPT diagram,
  \item Source must have NB \Ha luminosity in the $10^{40}-10^{42}$
        \,erg\,s$^{-1}$ range,
  \item Source must be located within 4500\,kpc of the cluster center.
\end{itemize}

There are 818 sources with detections in at least one emission line besides \Ha. Much of our analysis relies on classifying sources based on their ionization source, which restricts the sample to 640 galaxies (see Section~\ref{sec:BPT}). The total number of sources classified in the BPT diagram with luminosities $>10^{40}$\,erg\,s$^{-1}$ is 451. The full breakdown of the sample can be found in Table~\ref{tab:sample}.

\section{Results} \label{sec:results}

Our spectroscopic observations enable us to measure emission-line properties, which, in turn, unveil the physical properties of the cluster galaxies, including their SF and ionization sources. In combination with the spectroscopic observations, we will use, where relevant, NB magnitudes and \Ha luminosities derived from the NB data (see also Figure~\ref{fig:lum_lum}) to interpret our observations.

While we focus our analysis on cluster members with \Ha luminosities in the $10^{40}-10^{42}$\,erg\,s$^{-1}$ range, located within 4500\,kpc of the cluster center, and with BPT classification in the BPT diagram, we tested the effect of using different selection cuts on our sample, such as imposing tighter restrictions to the BPT classification (see Section~\ref{sec:BPT}), including using a range of \Ha luminosity cuts (e.g. Section~\ref{sec:lumdist}) and imposing minimum EWs (e.g. Section~\ref{sec:EW}), and found that our results are robust against different selection criteria.

\begin{deluxetable}{ccccc}[tb!]
  \tablecaption{The number of emission-line galaxies in the sample, including those classified as star-forming, AGN, and composite in the BPT diagram. We display the number of sources with NB \Ha luminosities greater than $10^{40}$\,erg\,s$^{-1}$, with total numbers of sources listed in parentheses. We also show the samples divided by cluster type and by the telescope of origin. \label{tab:sample}}
  \tablewidth{0pt}
  \tablehead{
    \colhead{Sample} & \colhead{All} & \colhead{SF}  & \colhead{AGN} &  \colhead{Composite} }
  \startdata
  All & 451 (818) & 275 (414) &  43 (85) & 87 (181) \\
  \hline
  Merging clusters   & 233 (311)  & 143 (168) & 23 (32) & 43 (62)    \\
  Relaxed clusters   & 218 (507)  & 132 (246) & 20 (53) & 44 (119)   \\ \hline
  VLT/VIMOS       & 77 (84)   & 52 (55)  & 5 (6)   & 11 (12) \\
  WHT/AF2         & 65 (67)   & 42 (44)  & 7 (7)   & 9 (9) \\
  Keck/DEIMOS     & 34 (44)   & 20 (25)  & 4 (6)   & 8 (8) \\
  MMT/Hectospec   & 275 (623) & 161 (290) & 27 (66)  & 59 (152) \\
  \enddata
\end{deluxetable}

\subsection{Distribution with Magnitude and \Ha Luminosity}\label{sec:maglum}

By focusing on our NB selected sample, we unveil the distribution of sources in the magnitude-\Ha luminosity plane (Figure~\ref{fig:NB_LUM}). We find that our NB selection identified \Ha candidates on two main tracks. The first represents a sequence in which brighter galaxies have, on average, brighter emission lines. Simply, this can be interpreted as a result of the stellar mass - SF relation. Above this relation, there is a second cloud of NB candidates with continuum magnitudes between 19 and 22, which have, on average, NB-derived \Ha luminosities 3 times larger than galaxies on the sequence. If located at the target redshifts, galaxies in the cloud would display large EWs, as a result of brighter \Ha luminosities on top of a relatively faint continuum.

The vast majority of sources located on the sequence are confirmed \Ha galaxies at the cluster redshift. By contrast, half of the sources in the cloud are interlopers, either stars or other line emitters as lower or higher redshift. The other half of the sources are \Ha emitters, almost exclusively in merging galaxy clusters.

Therefore, the spectroscopic follow-up reveals a fascinating dichotomy in the distribution of confirmed \Ha line emitters. \textbf{Merging galaxy clusters host a population of \Ha emitters with large EWs, that is virtually absent in relaxed clusters.}

\begin{figure}[hbt!]
  \includegraphics[width=\linewidth]{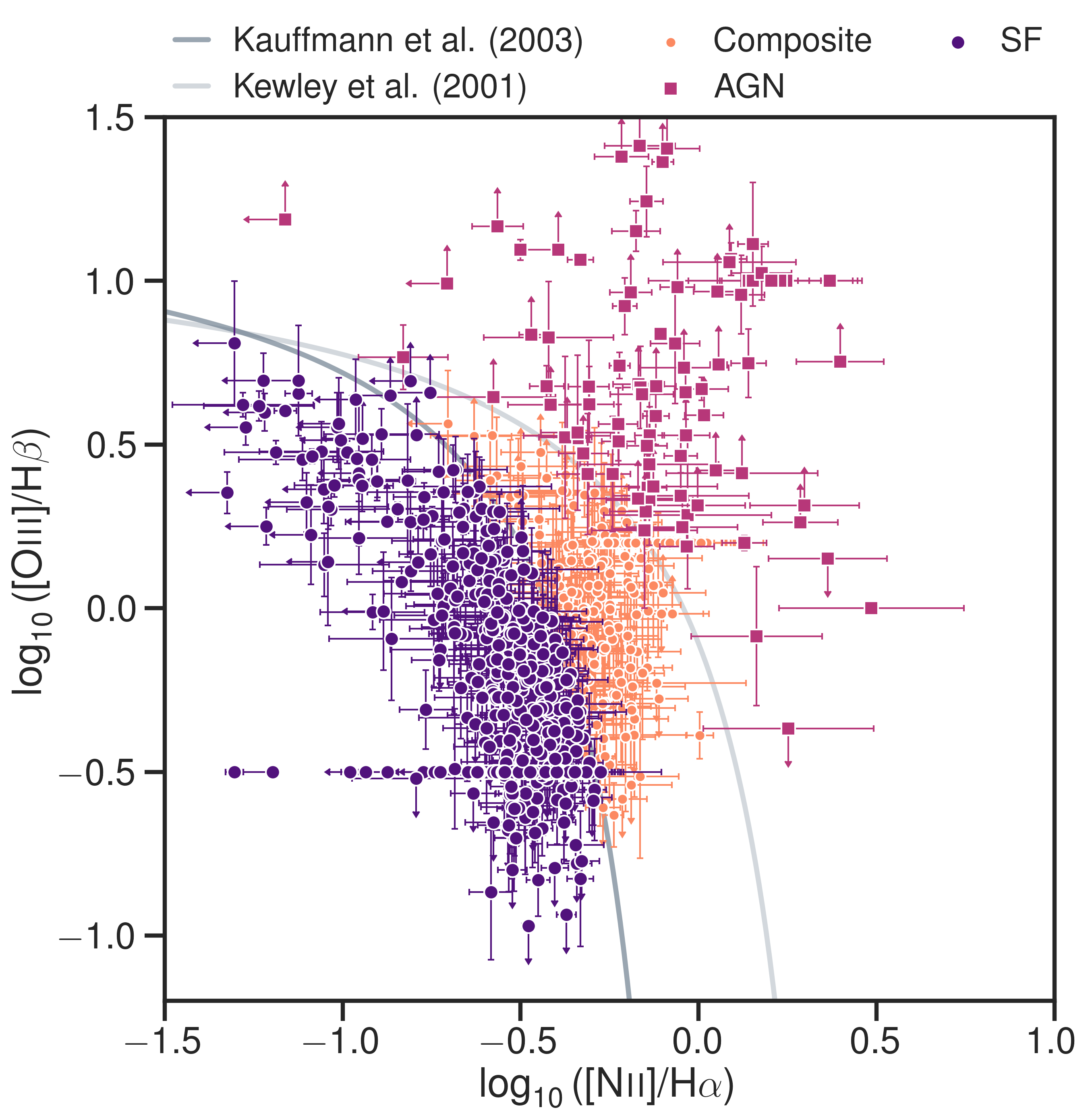}
  \caption{We classify our galaxies as dominated by SF (big purple circle), dominated by AGN emission (pink squares), or composite with contributions from H{\sc ii} regions and AGN (small orange circles), by using the \citet{Kauffmann2003} (dark gray line) and \citet{Kewley2001} (light gray line) relations. We classify 680 sources, of which pure star-forming galaxies represent 61\% of the sample, 26\% are composite, and 13\% are dominated by AGN emission.}
  \label{fig:BPT}
\end{figure}

\subsection{SF versus AGN}\label{sec:BPT}

We use the \NII/\Ha and \OIII/\Hb ratios as diagnostics for placing \Ha emitting galaxies in the BPT diagram \citep{Baldwin1981}. We distinguish between different ionization mechanisms for nebular gas and classify sources as dominated by SF, AGN activity, or as composite, through the widely adopted \citet{Kauffmann2003} and \citet{Kewley2001} dividing lines. \citet{Kauffmann2003} encompasses typical SF galaxies, while the bulk of sources above \citet{Kewley2001} are Seyfert galaxies with strong AGN contributions. Galaxies located between the two lines are composite, with contributions from both H{\sc ii} regions and AGN.

No dust extinction was applied to the line ratios, which has the effect of slightly raising the \OIII/\Hb ratio \citep[by $\sim0.06$\,dex,][]{Sobral2015} and increasing the chance galaxies cross the \citet{Kewley2001} line into the AGN regime. For two reasons, the lack of absolute flux calibration in MMT observations does not significantly affect line ratios. Firstly, the line pairs entering ratios are near each other in wavelength. Secondly, we double-checked for any systematic biases by comparing the line ratio measurements for the six galaxies with MMT and VIMOS observations and found them to be consistent with each other. Any offset in line ratio does not change the classification of the galaxies in the BPT diagram.

Overall, the BPT classification enables us to include galaxies with detections in all emission lines and use significant upper and lower limits on the line ratios where they enable the secure classification of sources. Based on these criteria alone, we were able to classify a total of 524 galaxies. For some sources, the classification could not be secured based on the \NII/\Ha and \OIII/\Hb alone. Examples include sources where the \OIII/\Hb line ratio cannot be constrained (e.g. lack of coverage of \Hb, \Hb and \OIII not detected) or sources where the upper or lower limits do not securely place the galaxy in the AGN or SF quadrant of the BPT diagram. For such sources, we also explore the \textsc{cloudy} modeling presented in \citet{Sobral2018} and \citet{Sobral2019}, to use the \OIII/\Ha ratio in combination with the \NII/\Ha ratio to classify the galaxies as SF, AGN, or composite (see Figure~\ref{fig:BPT_OIII}). While the demarcation between AGN, composite and star-forming sources is not as clear as in the BPT diagram, the \OIII/\Ha ratio depends on the ionization potential, and, as such, sources with low \OIII/\Ha and low \NII/\Ha are typically powered by SF, while a hard AGN ionizing spectrum would lead to high \OIII/\Ha and high \NII/\Ha ratios. Adding the \OIII criterion particularly helps in classifying star-forming galaxies with very low \OIII/\Ha ratio, where \OIII and \Hb are not detected. Including the \OIII/\Ha criterion enables us to classify an additional 156 sources. We note that the results presented in the paper are not altered, but strengthened with the smaller error bars provided by the increased number statistics.

\begin{figure}[ht!]
  \includegraphics[width=\linewidth]{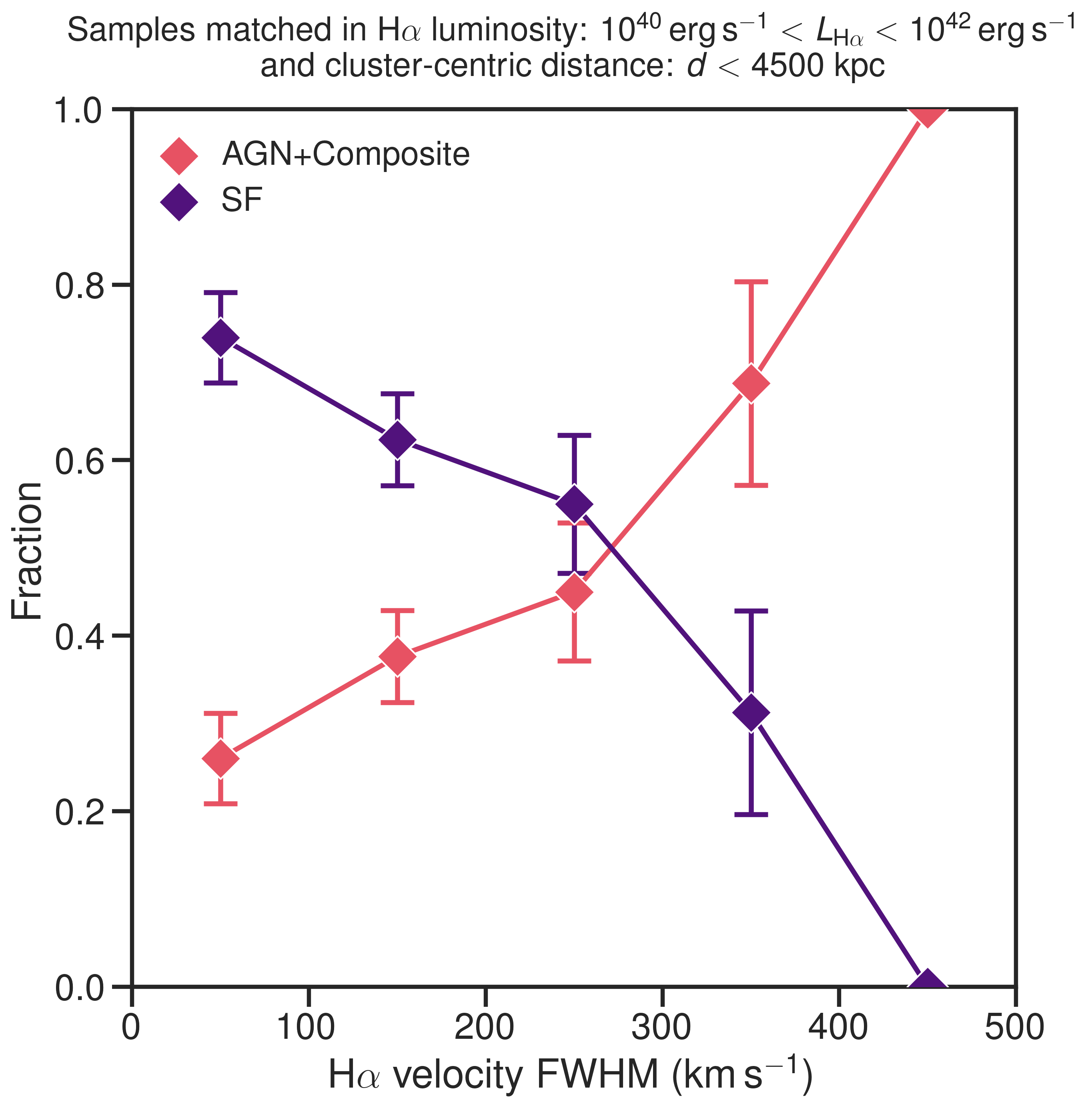}
  \caption{The fraction of AGN increases, while the SF fraction decreases with velocity FWHM of the \Ha narrow component.}
  \label{fig:frac_fwhm}
\end{figure}

\begin{figure}[ht!]
  \includegraphics[width=\linewidth]{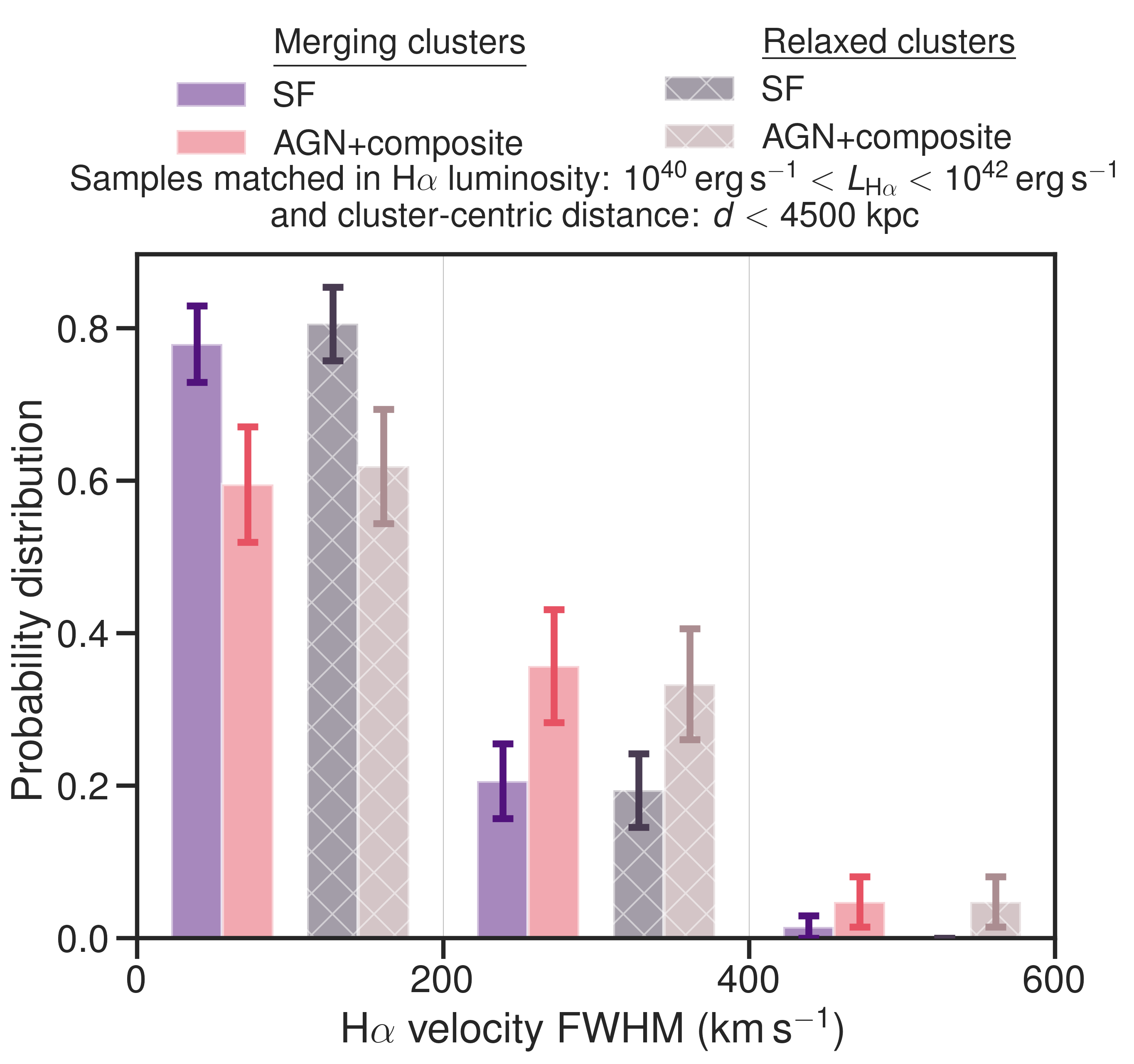}
  \caption{Distribution with \Ha velocity FWHM, separated by ionization source and galaxy cluster dynamical state. Samples are matched in \Ha luminosity and cluster-centric distance. The velocity width distribution of star-formers is statistically distinct from that of AGN. The bulk of star-forming galaxies, in both mergers and relaxed clusters, have narrow lines. On average, AGN have broader velocity widths, extending above 400\,km\,s$^{-1}$.}
  \label{fig:distrib_FWHM}
\end{figure}

Figure~\ref{fig:BPT} shows the distribution of our sample within the BPT diagram. With a total of 680 galaxies, 414 are dominated by SF (61\%), 85 are AGN (13\%), and the rest (181, 26\%) have composite spectra. As we will focus mostly on sources with \Ha luminosities greater than $10^{40}$\,erg\,s$^{-1}$, the distribution is as follows: 275 dominated by SF (68\%), 43 AGN (11\%), and 87 composite (21\%), for a total of 405 sources. Beyond the line ratios, the AGN contribution can also be clearly seen in the velocity FWHM of the narrow \Ha component of our sources. The fraction of sources with AGN contribution (including AGN-dominated and composite sources) increases with \Ha velocity width (Figures~\ref{fig:frac_fwhm} and \ref{fig:distrib_FWHM}). This is in line with theoretical expectations and large surveys, which find that the narrow lines for Seyfert 2 sources range between 200 and 900\,km\,s$^{-1}$, peaking at $350-400$\,km\,s$^{-1}$ \citep[e.g.][]{1986ARA&A..24..171O}. We test for differences in the star-forming galaxy and AGN distributions with velocity width using a two-sample Kolmogorov-Smirnov (KS) test. For the merging and relaxed cluster samples independently, we reject the null hypothesis that there is no difference between the AGN and SF-dominated samples at the 95\% confidence level\footnote{Provides the same confidence as a $2\sigma$ significance level for a normal distribution.}. When combining data from all 14 clusters, we reject the null hypothesis at the 99.6\% level\footnote{Equivalent to a $2.9\sigma$ significance level for a normal distribution.}. The significance increases when we focus on the $0-600$\,km\,s$^{-1}$ range, where the bulk of the sample is located. The distribution of star-formers and AGN differ at the 99.994\% level for relaxed clusters, at the 97.0\% level for mergers, and at the 99.9974\% level\footnote{Equivalent to a $4.7\sigma$ significance level for a normal distribution.} for the combined samples. Within the $0-600$\,km\,s$^{-1}$ range, the distribution of star-forming galaxies in merging clusters is statistically different from relaxed clusters at the 99.7\% confidence level\footnote{Equivalent to a $2.1\sigma$ significance level for a normal distribution.}.

\begin{figure}[t!]
  \includegraphics[width=\linewidth]{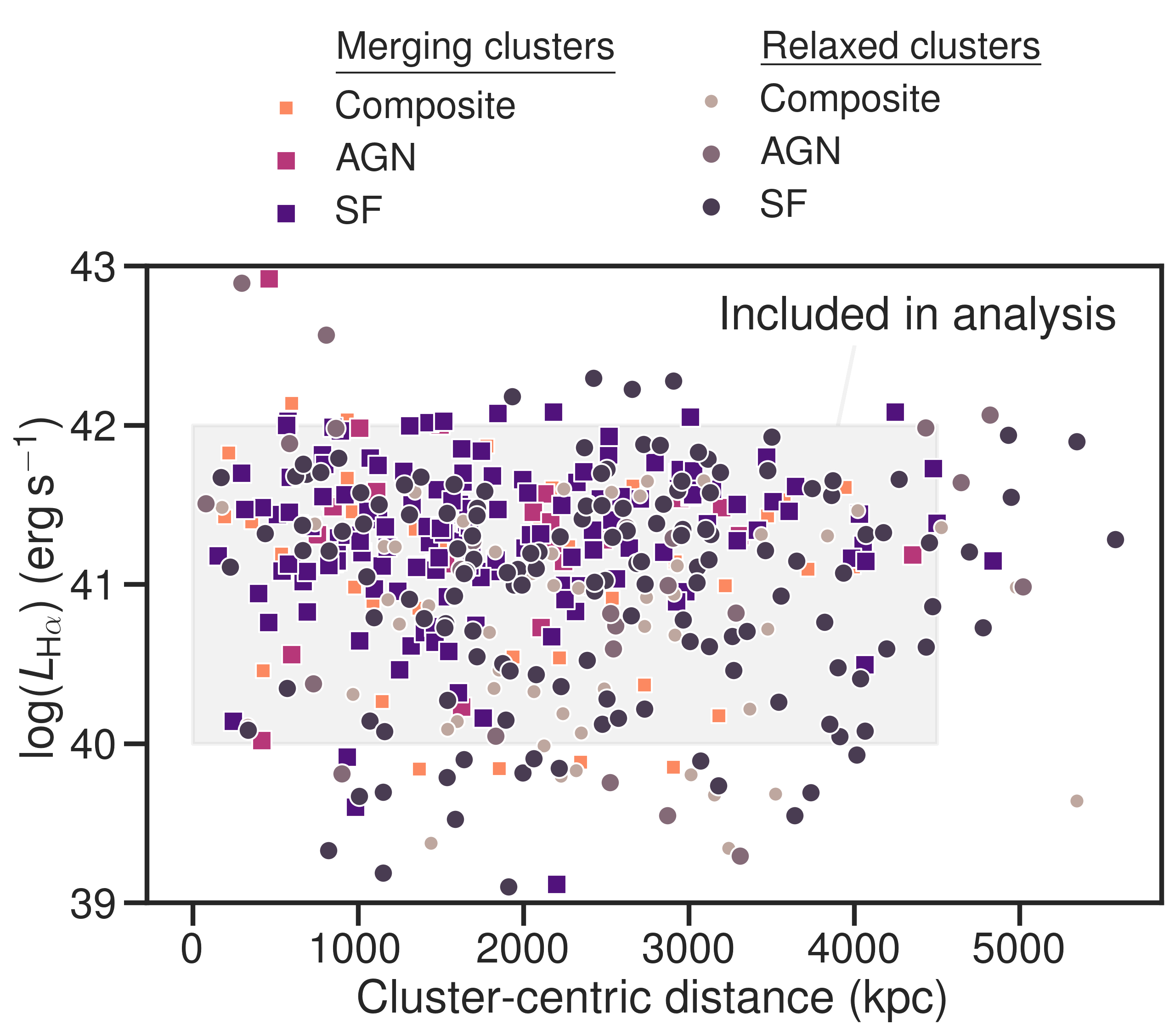}
  \caption{Distribution of sources powered by SF or AGN, as a function of NB \Ha luminosity and cluster-centric radius. Squares mark sources in merging clusters, and circles represent sources in relaxed clusters. Star-forming galaxies are marked with purple, AGN with pink and composite sources with orange. The luminosity and radius distribution for different ionization types reflect the broad selection of sources in both merging and relaxed clusters. We fully match the samples for the analysis by selecting only sources within 4500\,kpc and with \Ha luminosities in the $10^{40}-10^{42}$\,erg\,s$^{-1}$ range.}
  \label{fig:radial_distrib_type}
\end{figure}

\subsection{Radial Distribution with \Ha Luminosity}\label{sec:lumdist}

We show the distribution of our ENISALA sample with \Ha luminosity and cluster-centric distance in Figure~\ref{fig:radial_distrib_type}. Our survey covers four orders of magnitude in \Ha luminosities ($10^{39}-10^{42}$\,erg\,s$^{-1}$) and stretches out to cluster-centric radii of over 5\,Mpc. The bulk of sources have \Ha luminosities above $10^{40}$\,erg\,s$^{-1}$, with SF galaxies, AGN, and composite sources spanning the entire range of luminosities shown. The AGN dominated sources towards the cluster cores have the brightest emission lines, above $10^{42}$\,erg\,s$^{-1}$. Our sample includes AGN, star-forming galaxies, and composite sources at all luminosities, and cluster-centric radii, irrespective of cluster dynamical state (merging or relaxed).

\begin{figure}[htb!]
  \includegraphics[width=\linewidth]{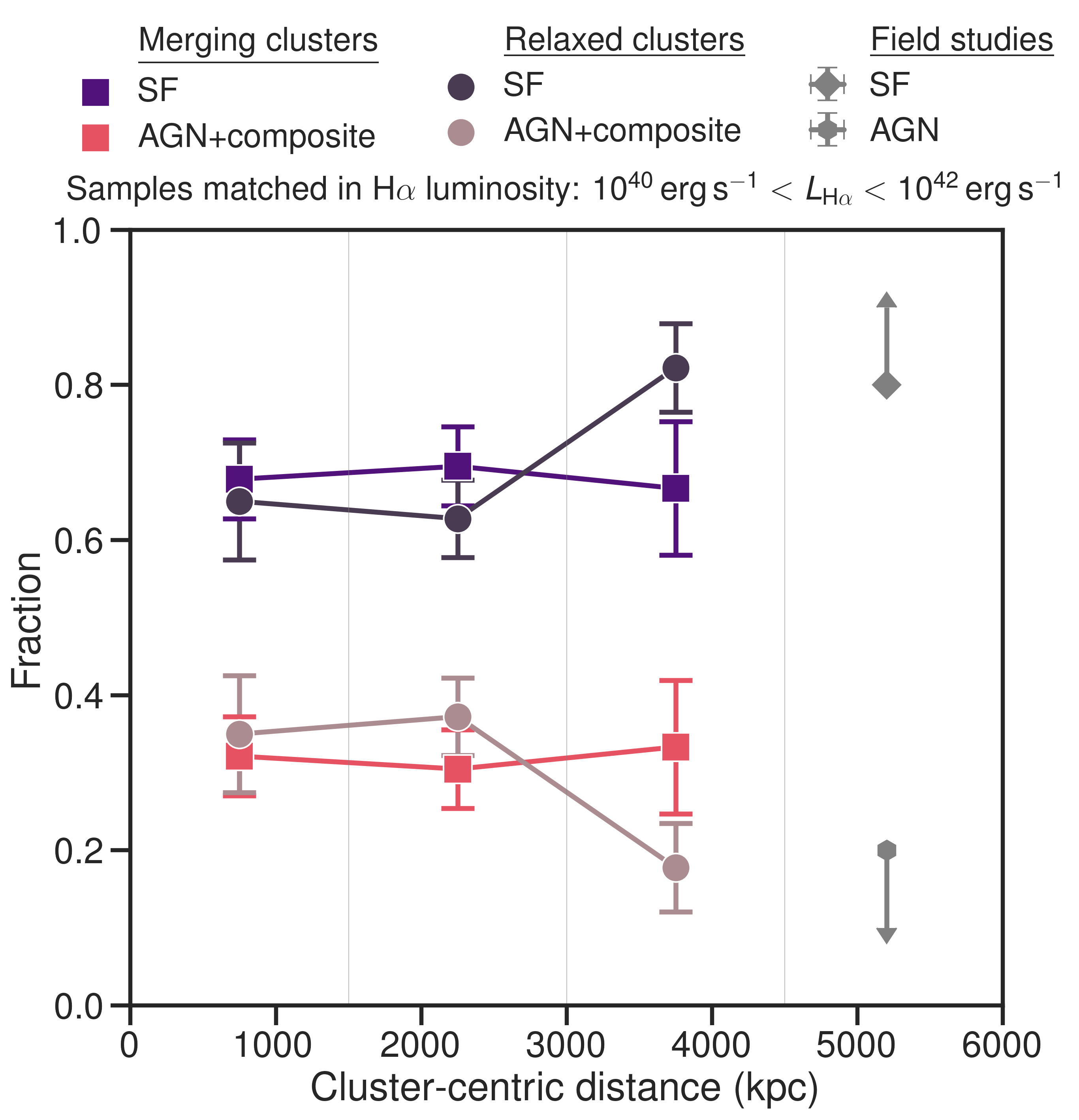}
  \caption{The fraction of pure star-forming galaxies and sources with AGN contribution, as a function of cluster-centric distance. Samples are matched in \Ha luminosity between $10^{40}$ and $10^{42}$\,erg\,s$^{-1}$. In merging clusters, the fraction of line-emitters powered by SF is constant with radial distance. In relaxed clusters, the fraction of purely star-forming galaxies drops within 3\,Mpc of the cluster core. The AGN and SF fractions at relaxed cluster outskirts match those in the field \citep[e.g.][]{Kauffmann2003, Shioya2008, Hayashi2018}.
  }
  \label{fig:radius_LUM}
\end{figure}

The fraction of sources dominated by each ionization type strongly depends on the radial distance from the cluster core and the \Ha luminosity of the source (Figure~\ref{fig:radius_LUM}). In building this plot, we focus on $10^{40}-10^{42}$\,erg\,s$^{-1}$ range and on radii $<4500$\,kpc, where the samples are comparable between merging and relaxed clusters, but note that our results are robust against other choices of luminosity ranges.

Overall, the number of bright line-emitters within merging clusters is greater than in relaxed clusters \citep{Stroe2017}. Out of those line-emitters which survive the infall into the cluster ($<3$\,Mpc), a slightly higher fraction is powered by SF in merging than in relaxed clusters. However, at only $1\sigma$, the difference is not statistically significant. Interestingly, bright emission lines are more often powered by SF at the outskirts of relaxed clusters ($3-4.5$\,Mpc). A one-tailed Z-test indicates that the fraction of star-forming galaxies in the outermost radial bin is higher than the innermost and middle bin at the 98\% ($2.3\sigma$) and 93\% ($1.8\sigma$) confidence levels, respectively. By combining the bins with sources located at $<3$\,Mpc using Fischer's method, the significance reaches $2.7\sigma$. While a linear model might not provide the best description of the data, it is still instructive in comparing the radial dependence of the star-forming fractions in mergers and relaxed clusters. In merging clusters, the relationship is consistent with no radial dependence of the star-forming fraction, with a slope of $-0.0004\pm0.0099$\,Mpc$^{-1}$ and an intercept of $0.68\pm0.02$. We find that the star-forming fraction mildly correlates with radial distance in relaxed clusters, with a slope of $0.0667\pm0.0489$\,Mpc$^{-1}$ and an intercept of $0.53\pm0.13$. A Pearson correlation analysis also indicates that relaxed clusters (Pearson $r=0.81$) present a slightly stronger linear correlation between the fraction of star-forming galaxies and radial distance than mergers (Pearson $r=-0.47$). Adopting a value of 15\% AGN for field samples at similar redshifts \citep[e.g., as per][]{Shioya2008, Hayashi2018}, we compare the fraction of star-forming galaxies in our sample to the field. By means of a Z-test, we find that the fraction of star-forming galaxies is lower than field values at a level of more than $2.5\sigma$ in all but the outermost bin in relaxed clusters. Therefore, the outskirts of relaxed clusters have AGN and SF fractions similar to average cosmic fields \citep[$80:20$, as per][]{Kauffmann2003, Shioya2008}. For mergers, the chance of a galaxy being powered by SF is constant at all cluster-centric radii. The star-forming fraction drops in relaxed clusters, from a value similar to an average field beyond 3\,Mpc to about 65\% (or factor of 1.2) within 3\,Mpc.

\begin{figure}[ht!]
  \includegraphics[width=\linewidth]{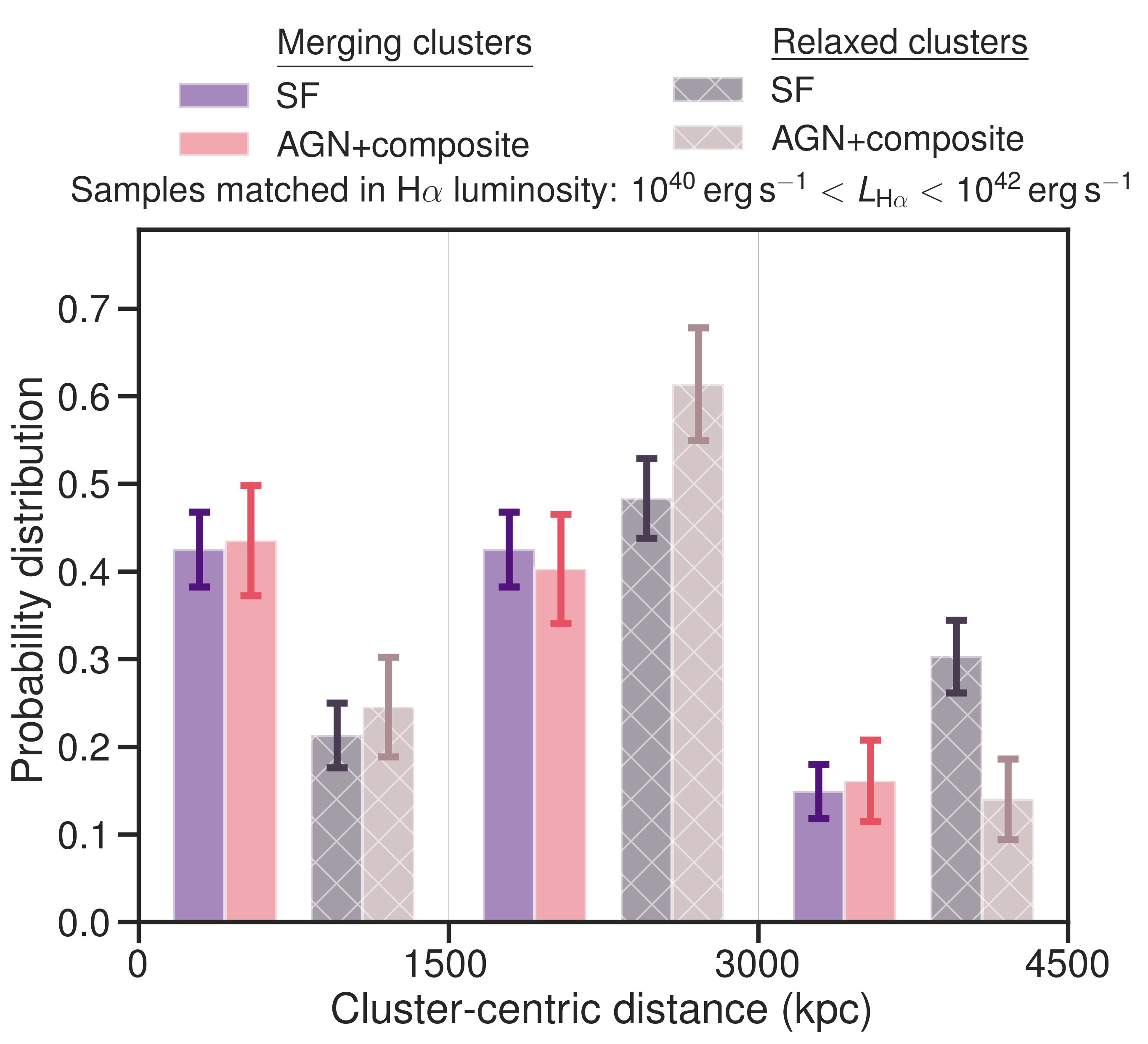}
  \caption{Normalized probability distribution with cluster-centric distance. Only sources with \Ha luminosities between $10^{40}$ and $10^{42}$\,erg\,s$^{-1}$ are included. We separate sources into four classes, based on their parent cluster type and the powering source of the \Ha emission (SF or AGN). \textbf{The radial distribution of star-forming galaxies and AGN is different between relaxed and merging clusters.} The bulk of SF and AGN activity in merging clusters is happening within 3\,Mpc of the cluster core. By contrast, the bulk of emission-line galaxies in relaxed clusters are located outside the core ($>1.5$\,Mpc).}
  \label{fig:radius_hist}
\end{figure}

Another way to look at the radial distribution of ionization types is to study the probability distribution dependence on cluster-centric radius. We split the sample into four categories as a function of ionization source (pure SF versus AGN contribution) and galaxy cluster dynamical state (merging versus relaxed). In Figure~\ref{fig:radius_hist}, we show the distribution of each population as a function of cluster-centric radius (i.e. what fraction of each population is located in each radial bin). We employ a KS and a two-sample Z-score methodology to test whether the radial distribution of sources is different as a function of ionization source and cluster relaxation. With a KS test, we reject the null hypothesis that the radial distribution of star-forming galaxies is the same between relaxed and merging clusters at the 99.99\% confidence level\footnote{Equivalent to a significance of $3.9\sigma$ for a normal distribution.}. A two-sample Z-test on the binned data gives a similar statistical significance of $3.9\sigma$ when combining the significance of each pair of radial bins using Fisher's method. We also find mild evidence that the AGN distribution differs between mergers and relaxed clusters, at the 93.2\% confidence level with a KS test\footnote{Equivalent to a significance of $1.8\sigma$ for a normal distribution.} and $2.1\sigma$ level using the Z-test in combination with Fisher's method.

For mergers, about 40\% of star-forming galaxies are located within 1.5\,Mpc of the cluster core, another 40\% are located between $1.5-3$\,Mpc, with only 20\% located beyond 3\,Mpc. Similar distributions with radius are observed for AGN in the fields of merging galaxy clusters, with only 10\% of AGN located between 3 and 4.5\,Mpc. In agreement, the KS test indicates that the AGN and SF-dominated sources have consistent radial distributions for merging clusters. Using the one-sample Z-test, we find that the core and outskirts host a statistically consistent fraction (at $<0.3\sigma$) of the star-forming population in merging clusters. The lower fraction of AGN and star-forming galaxies beyond 3\,Mpc is statistically different from the two innermost bins at a significance level of $2.5$ and $4.2\sigma$, for AGN and star-forming galaxies, respectively.

Our data paint a different picture for emission-line galaxies in relaxed clusters. The radial distribution of star-forming galaxies and AGN in relaxed clusters is different, albeit with a relatively low confidence level of 91.3\%\footnote{Equivalent to $1.7\sigma$}. Only 20\% of purely star-forming galaxies in relaxed clusters are located within 1.5\,Mpc of the core, and over 30\% are outside 3\,Mpc. Using a one-sample Z-test, the increased fraction of star-forming galaxies between $1.5-3$\,Mpc compared to the other two areas studied is statistically significant at the $3.6$ and $2.2\sigma$ level. Over 60\% of line-emitters powered, at least in part, by AGN are located at the outskirts of relaxed clusters (between 1.5 and 3\,Mpc), a fraction that is statistically higher than other cluster regions (at 3 and $4.1\sigma$, respectively). \textbf{The bulk of star-forming galaxies and AGN in the fields of merging clusters are distributed towards the core of merging clusters (within 3\,Mpc), while SF and AGN activity most likely occurs at relaxed cluster outskirts ($1.5-3$\,Mpc).}

\subsection{Distribution with Galaxy Color}

\begin{figure}[hbt!]
  \includegraphics[width=\linewidth]{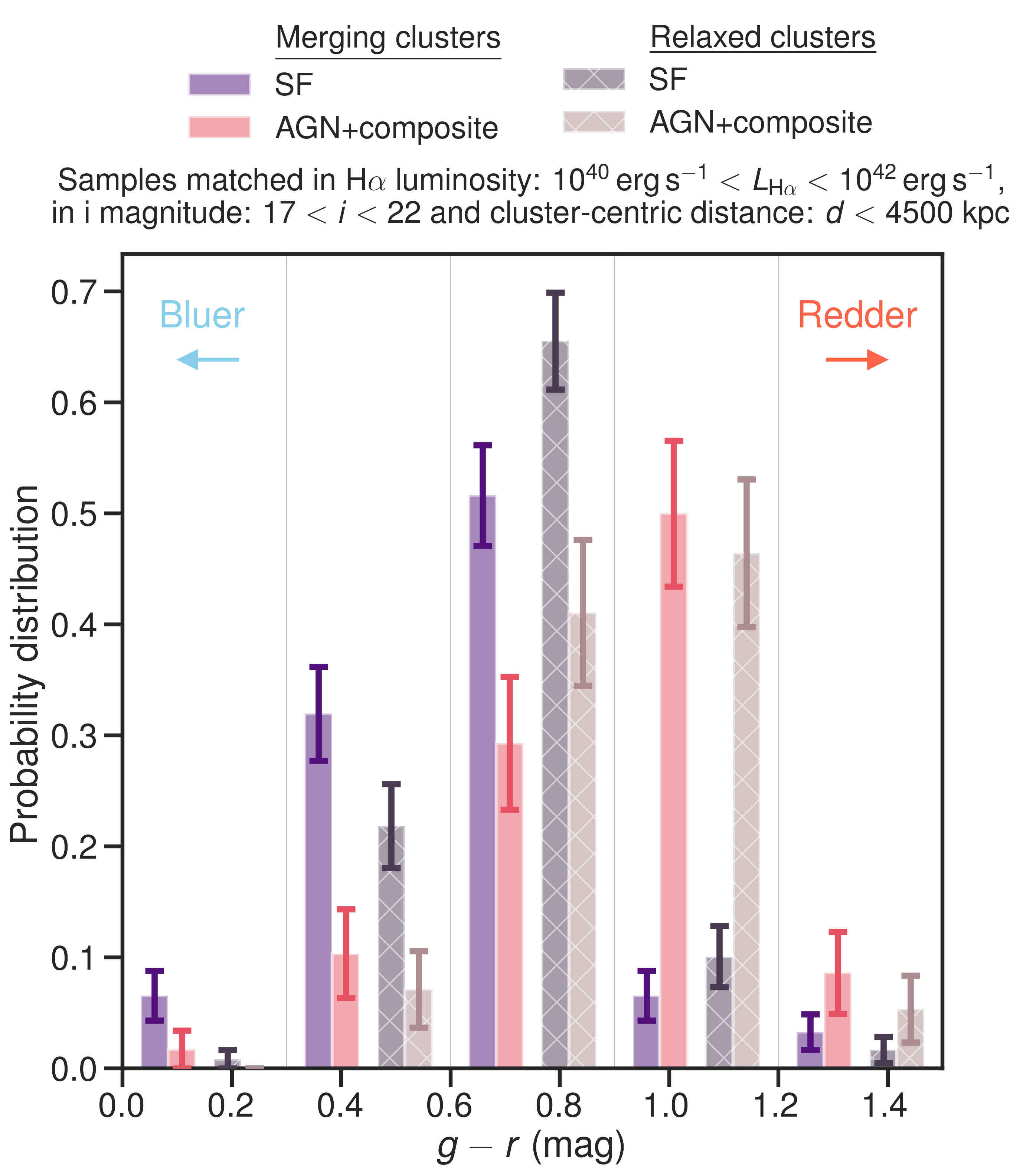}
  \caption{Normalized distribution with $g-r$ galaxy color. We compare the distribution of sources in our sample, separated by ionization source (purple for SF versus pink for AGN) and galaxy cluster type (filled histogram for mergers versus hatched histograms for relaxed clusters). Samples are matched in magnitude, luminosity, and cluster-centric distance. \textbf{AGN hosts have redder colors, while star-forming galaxies are on average bluer, especially in merging clusters.}}
  \label{fig:color}
\end{figure}

We investigate how the ionization source of the \Ha emission relates to the properties of the host galaxy (see Figure~\ref{fig:color}). Specifically, we employ the observed $g-r$ color, which traces the rest-frame UV--optical color for the redshift of our sources. This range encompasses the Balmer and 4000\,\AA{} breaks and enables a broad separation of star-forming and passive galaxies. By focusing on galaxies with $i$-band magnitudes between 17 and 22 and matching samples in \Ha luminosity ($10^{40}-10^{42}$\,erg\,s$^{-1}$) and cluster-centric distance ($<4500$\,kpc), we compare the underlying distribution of sources in our sample.

We split our sample by cluster dynamical state (merger versus relaxed) and ionization type (pure SF versus AGN-dominated or composite). A KS test confirms that the color distributions of AGN and star-forming galaxies are different at a high significance level. We reject the null hypothesis at a very high significance for the relaxed, merging, and combined sample (greater than 99.99999\%\footnote{Equivalent to a 5.7, 6.2, and $7.9\sigma$ significance level for the merging, relaxed and entire sample, respectively, when using a normal distribution.}). On average, line emitters powered by SF are by, an overwhelming margin, blue: over 80\% of star-forming galaxies in both relaxed and merging clusters have colors bluer than 0.9. By contrast, sources with AGN and composite spectral features are redder than SF galaxies.

Overall, according to a KS test, the color distribution of star-forming galaxies in merging clusters is different from relaxed counterparts at the 96.3\% confidence level\footnote{Equivalent to $2.1\sigma$ for a normal distribution.}. Further, a higher fraction ($40.0\pm5$\%) of star-forming galaxies in merging clusters have very blue colors ($<$0.6) compared to relaxed counterparts ($23\pm4$\%). A two-sample Z-test places the difference between the two fractions as significant at the $2.8\sigma$ level. By contrast, the color distribution of AGN is consistent between relaxed and merging clusters.

\subsection{\Ha Equivalent Widths}\label{sec:EW}

As mentioned in Section~\ref{sec:data}, we cannot measure \Ha luminosities, and hence SFRs, directly from our spectroscopy for the entire sample. We can, however, measure the rest-frame EW of the \Ha line. The EW traces the strength of the emission line on top of the continuum and thus broadly relates to the sSFR of the host galaxy, as long as the \Ha emission is powered by SF and dust extinction does not vary significantly between galaxies.

\begin{figure}[ht!]
  \includegraphics[width=\linewidth]{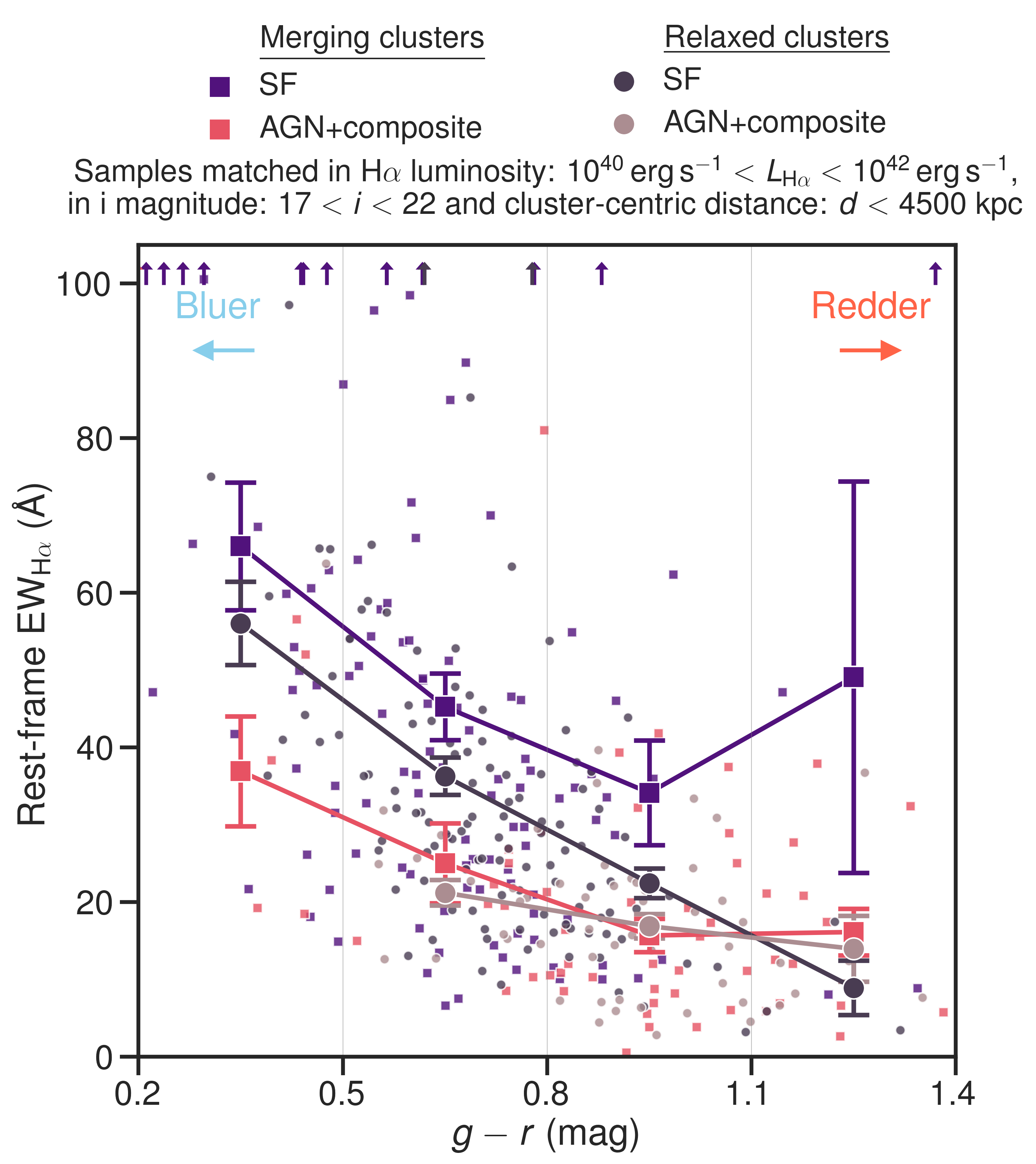}
  \caption{Distribution of sources with EW and observed $g-r$ color (tracing approximately UV--optical rest-frame), separated by ionization source and cluster dynamical state. The samples are matched in $i$-band magnitude and \Ha luminosity. Star-forming galaxies have higher EW than AGN, irrespective of galaxy color. \textbf{Star-forming galaxies in merging clusters have higher EW, or sSFRs, compared to counterparts in relaxed clusters.}}
  \label{fig:distrib_EW_color}
\end{figure}

The average EW drops with increasing (redder) rest-frame UV--optical color for all ionization sources, including pure star-forming galaxies, those with AGN contributions, and those whose \Ha is dominated by AGN. This effect is exhibited in Figure~\ref{fig:distrib_EW_color}, where we match the samples in broad-band magnitude and \Ha luminosity. On average, at all rest-frame colors, the EW of star-forming galaxies is larger than AGN and composite sources by a factor of $1.5-2$. We tested this hypothesis using a two-sample KS test, a two-dimensional two-sample KS test \citep[also known as Peacock's test,][]{Peacock1983}\footnote{Using the 2DKS Python implemention at \url{https://github.com/Gabinou/2DKS}.}, and a two-sample Student's \textit{t}-test. The rest-frame EWs distribution of SF and AGN sources is different for both the relaxed and the merging cluster sample. A KS test indicates that the null hypothesis can be rejected at $>99.999$\%\footnote{Equivalent to $>5\sigma$ for a normal distribution.} confidence level for both relaxed and merging clusters. The two-dimensional KS test in the rest-frame EW--color space yields the same conclusion at a similar confidence level. The mean EW is greater for SF-dominated sources compared to those with AGN contributions, for both relaxed clusters and merging clusters, with a \textit{t}-score of $6.5$ and $5.3$, respectively, and a high significance level of $>99.999$\%. At all rest-frame UV--optical colors, star-forming galaxies in merging clusters have higher rest-frame EWs than counterparts in relaxed clusters (see Figure~\ref{fig:distrib_EW_color}). We find that the EW distribution of star-forming galaxies differs between relaxed and merging clusters, rejecting the null hypothesis through a KS test at a $>99.66$\% confidence level\footnote{Equivalent to $>3.3\sigma$ for a normal distribution.}. The two-dimensional distribution of star-forming galaxies in the EW--color space is statistically different between relaxed and merging clusters at a 99.61\% confidence level\footnote{Equivalent to $2.9\sigma$ for a normal distribution.}. A \textit{t}-test confirms this finding and indicates that the mean EW for star-forming galaxies in merging clusters is higher than in relaxed clusters (\textit{t}-score of 3.3). By contrast, in the case of AGN, a one-dimensional KS test on EW distribution, a two-dimensional KS test in the EW--color space, and a \textit{t}-test confirm that the distribution of EWs, as well as the mean EW, do not differ between relaxed and merging clusters.

The ENISALA sample has a wide diversity of rest-frame \Ha EW properties. We note that the depth of each observation, the strength of the emission line in combination with the strength of the continuum, determine the limiting EW measurement. EW inversely correlates with stellar mass, and, in field samples at redshifts similar to our cluster $z\sim0.15-0.3$, star-forming galaxies with stellar masses $>10^{10}$\,M$_\odot$ have rest-frame \Ha EWs of $10-40$\,\AA{} \citep{Fumagalli2012}. In line with massive field galaxies, the bulk of our sources have rest-frame EWs under 50\,\AA, with AGN measuring the smallest EWs. The distribution of pure star-forming galaxies extends to higher EW, over 100\,\AA, and up to 250\,\AA, indicative of galaxies with masses of $\sim10^{9}$\,M$_\odot$ and $<10^8$\,M$_\odot$, respectively.

\begin{figure}[ht!]
  \includegraphics[width=\linewidth]{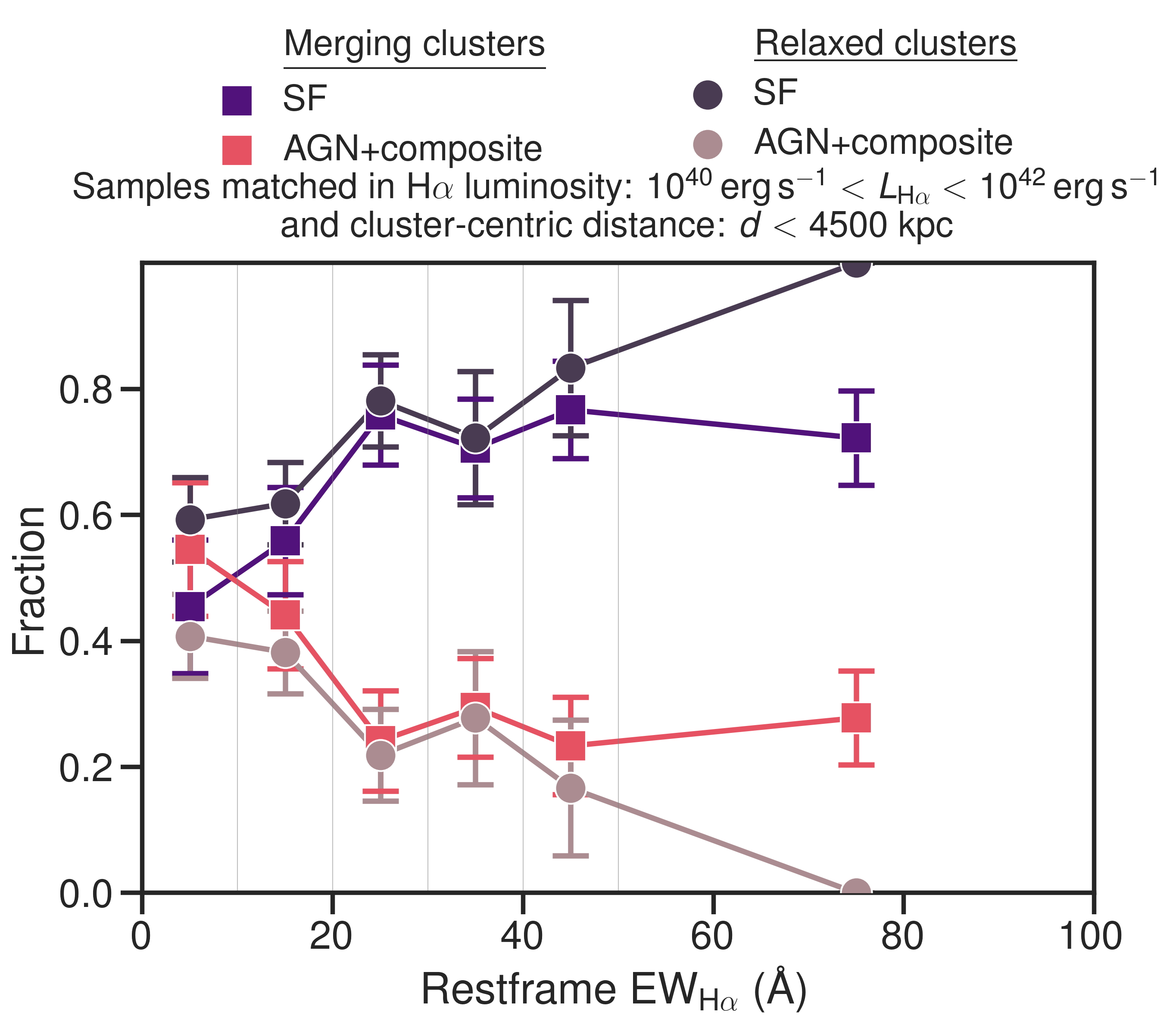}
  \caption{Fraction of targets powered by each ionization source, binned by EW. The samples are matched in \Ha luminosity. At the lowest EWs $<$10\,\AA, the emission is powered by AGN, at a higher rate in merging clusters compared to relaxed clusters. Beyond 10\,\AA, the emission is powered by pure SF in $70-80$\% of cases.}
  \label{fig:distrib_EW_frac}
\end{figure}

By matching the samples in NB \Ha luminosity, we can observe that the fraction of source powered by SF increases with increasing EW, especially for the sources in relaxed cluster fields (see Figure~\ref{fig:distrib_EW_frac}). In merging clusters, over 60\% of sources with low EW below 10\,\AA{} are powered in part by AGN. This fraction drops sharply to $20-30$\% for EW$>$10\,\AA. Conversely, for relaxed clusters, the fraction of AGN-powered sources drops from $\sim$50\% to 0 with increasing EW. In bins of EW, the fraction of sources powered by SF or AGN as a function is consistent between cluster dynamical states. However, it is noteworthy that the distribution of EWs in the two samples are vastly different in our luminosity-matched samples. Relaxed clusters in our ENISALA projects contain more low EW sources. By contrast, merging clusters have an overabundance of sources with large EWs, reflecting the preponderance of highly star-forming galaxies.

\begin{figure*}[ht!]
  \centering
  \includegraphics[width=0.7\linewidth]{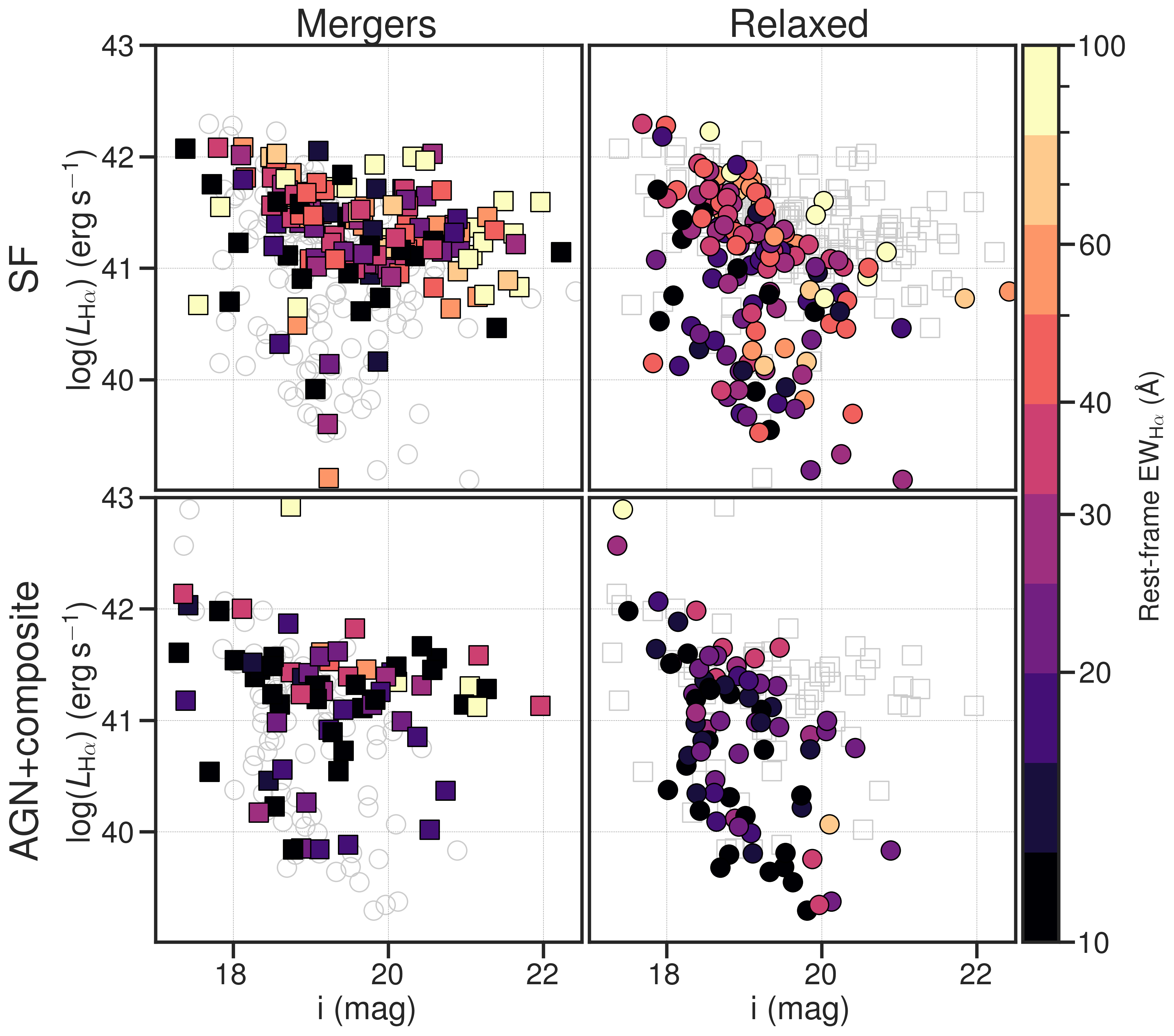}
  \caption{Distribution of the ENISALA sample with \Ha luminosity and $i$-band magnitude, highlighting in color the spectroscopic \Ha EW. Note the logarithmic scaling of the color bar. \textbf{Merging clusters host a population of continuum faint star-forming galaxies with bright \Ha emission lines, confirmed with large spectroscopic EW or sSFRs, which does not exist in relaxed clusters.}}
  \label{fig:EW_panel}
\end{figure*}

To further identify the nature of galaxies with large \Ha EW, we plot the distribution of sources with $i$-band magnitude versus the \Ha luminosity measured from the NB (Figure~\ref{fig:EW_panel}). Naively, sources with large EW likely have bright emission lines (i.e. large NB \Ha luminosity) on top of a faint continuum (i.e. large $i$ magnitude). We thus expect large EW sources to reside in the top-right quadrant of the plot.

\begin{figure}[ht!]
  \includegraphics[width=\linewidth]{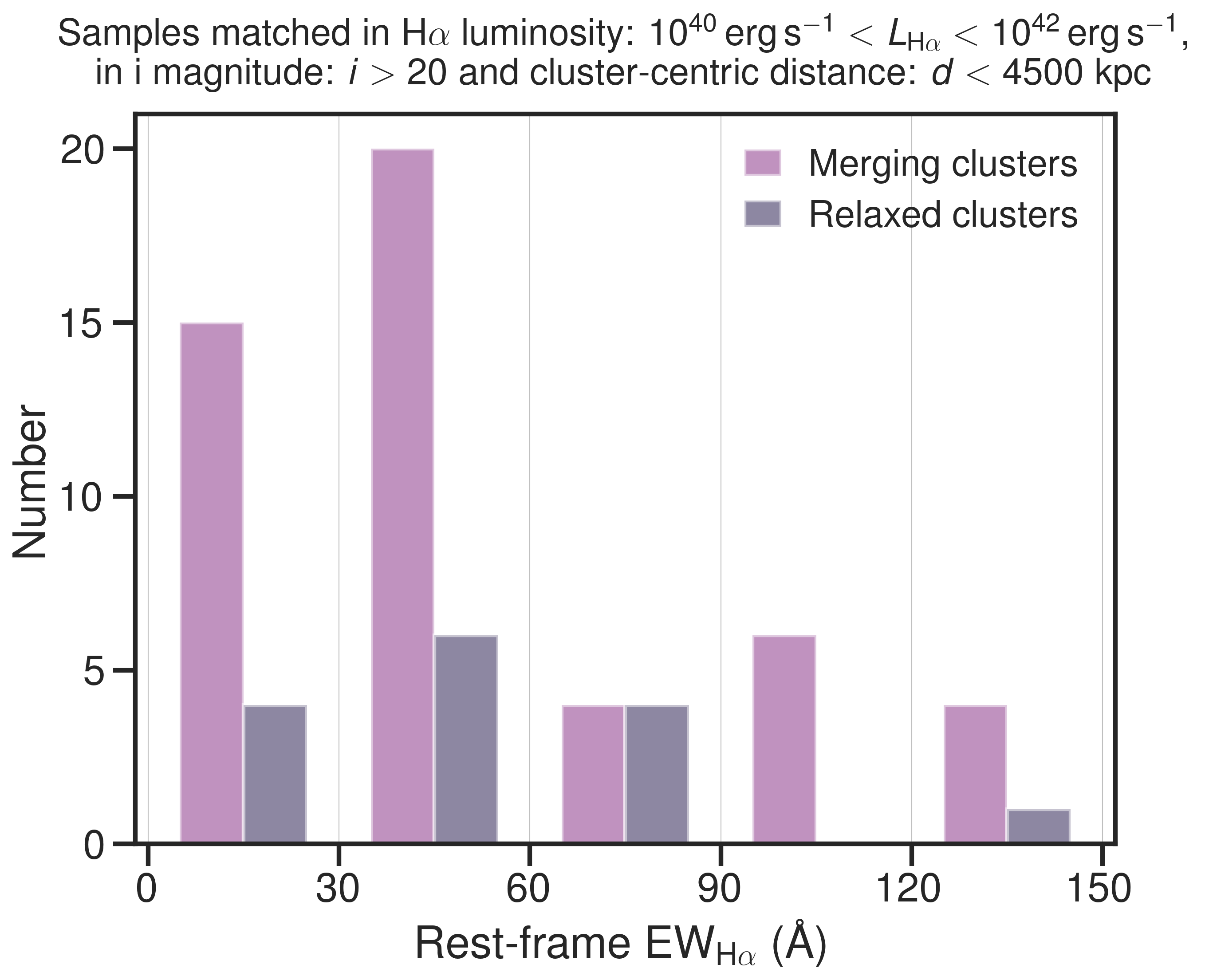}
  \caption{EW distribution of faint-continuum line-emitters. We show sources with \Ha luminosities between $10^{40}$ and $10^{42}$\,erg\,s$^{-1}$, $i$-band magnitudes $>20$, located within 4500\,kpc from the cluster center. Merging clusters host faint-continuum line-emitters, especially those with bright emission lines, at a higher rate than relaxed clusters.}
  \label{fig:faint_EW}
\end{figure}

In Figure~\ref{fig:faint_EW}, we plot the rest-frame \Ha EW distribution of emisison-line sources with faint continuum (i.e. we select sources with $i$-band magnitudes fainter than 20 mag, with $10^{40}--10^{42}$\,erg\,s$^{-1}$ \Ha luminosities, which are located within 4.5\,Mpc from the cluster center). Overall, for a similar total number of emission line galaxies in each sample, merging clusters are $>3$ times more numerous in faint-continuum emitters than relaxed counterparts (66 versus 21 sources). This effect is caused by a genuine paucity of sources with large EWs in relaxed clusters and not because of a lack of spectroscopic follow-up in parts of the magnitude-luminosity plane. As shown in Figure~\ref{fig:NB_LUM}, many galaxies were selected for follow-up from the space populated by optically faint, \Ha bright sources. However, spectroscopy confirmed that many of these sources were not at the cluster redshift, including a majority of sources hosted by relaxed clusters.

We confirm that a large fraction of the faint-continuum population consists of star-forming galaxies and a few AGN with large \Ha EWs measured from the spectroscopic observations, exceeding $80-90$\,\AA{} (Figure~\ref{fig:faint_EW}). Within the population of faint-continuum sources, merging clusters have 13 sources with EWs $>90$\,\AA{}, while we find a single such source in relaxed counterparts. An E-test confirms that the occurence rate of high EW emission-line sources is higher in merging clusters than in relaxed clusters at the 99.93\% confidence level\footnote{Equivalent to $3.4\sigma$ for a normal distribution.}. \textbf{Galaxy clusters undergoing massive mergers contain a population of highly star-forming galaxies with high sSFR, which is absent from relaxed clusters.}

\section{Discussion} \label{sec:discussion}

We specifically designed the ENISALA project to unveil the physical mechanisms through which cluster growth drives galaxy and BH evolution. With an extensive sample of over 800 high S/N spectra of \Ha-selected galaxies, of which we securely classify 680 as SF, AGN, or composite, we find striking differences between star-forming galaxy populations in merging and relaxed clusters.

The distribution of emission-line properties with cluster-centric distance reveals some interesting trends, which differ between relaxed and merging clusters. In a sample of emission-line galaxies matched in \Ha luminosity, the bulk of emission-line galaxies are located outside of the 1.5\,Mpc radius from the cores of relaxed clusters. The fraction of AGN drops outside 3\,Mpc for relaxed clusters, and, conversely, the fraction of sources powered by SF drops inside the cluster, encoding tell-tale signs of environmental quenching of SF. Overall, as shown in Paper I \citep{Stroe2017}, the density of \Ha line emitters for the relaxed clusters in the ENISALA sample is lower, resulting in an overall smaller number of star-forming galaxies towards their cores, compared to merging clusters. This effect could be reproduced if bursts of SF are triggered in gas-rich galaxies by ram pressure at the infall region of relaxed clusters, which, in turn, power bright emission lines, which rapidly fade as the galaxy further approaches the cluster core. Any bright emission lines close to the cluster core would then only be sustained through AGN activity, hinting at an almost complete lack of vigorous SF in relaxed cluster cores. In terms of relative abundance, AGN are most abundant between $1.5-3$\,Mpc in relaxed clusters, matching literature results. For example, in low redshift clusters, jellyfish, galaxies with gas and star-forming tails, have a high incidence of AGN activity, which can be attributed to ram pressure disrupting gas which flows to the core of the galaxy to trigger and fuel AGN activity \citep{Poggianti2017, Ricarte2020}. The lack of AGN, and specifically Seyfert 2 sources, has been attributed to the truncation of gas infall towards the BH, as well as a lack of fresh cold gas supply in the cores of relaxed clusters \citep[e.g.][]{Pimbblet2013, deSouza2016}.

By contrast, merging clusters are rich in both SF and AGN powering bright emission lines, echoing earlier results on individual clusters. For example, in radio studies of the merging clusters Abell 2255, A2125, and 2645, \citet{Owen1999} and \citet{Miller2003} found an excess of AGN, as well as a striking excess of faint star-forming galaxies. The authors attribute this elevated activity to cluster-wide merger shocks that cross galaxies, increase the interstellar medium pressure and thus trigger SF, or to increased galaxy-galaxy interactions in infalling groups \citep{Miller2003}. In optical analyses, \citet{Sobral2015} and \citet{Hwang2009} find an enhancement of star-forming galaxies and AGN, which they attribute to the cluster mergers funneling gas into the BH, thus promoting AGN. The findings in $z<0.5$ disturbed clusters are similar to increased AGN and SF activity in clusters beyond $z\sim1$ \citep[e.g.][]{Alberts2016, Moravec2020}.

Studies of field galaxies \citep[e.g.][]{Kinney1996, Delgado1998} established that Seyfert 2 type AGN have flat UV--optical continua, significantly redder than star-forming galaxies, caused by nuclear starburst contributions on top of old stellar population and minimal broad-line region emission from the AGN. We confirm this trend in cluster galaxies: the rest-frame UV--optical colors of AGN in the ENISALA project are redder than galaxies dominated by SF.

As expected, star-forming galaxies are bluer than the average. The tail of the color distribution does include red galaxies with observed $g-r$ redder than 1.2\,mag. While this present study does not focus on morphology, we speculate this population might be related to the population of red spirals found at intermediate local densities nearby galaxy groups and clusters \citep[e.g.][]{Bamford2009}. In explaining their origin, \citet{Bamford2009} suggests that red spirals might be low-mass sources, recently accreted onto the cluster, which after a fast removal of their gas supply will eventually evolve in S0s.

Surprisingly, a higher fraction of star-forming galaxies in merging clusters have very blue colors compared to counterparts in relaxed clusters. Invoking older stellar populations for star-forming galaxies in relaxed clusters would translate into redder average colors. An alternative interpretation comes from \citet{Sobral2016}, who find that star-forming galaxies in a $z\sim0.4$ cluster are significantly dustier than galaxies in lower-density environments. Another important piece of the puzzle comes from the EWs: on average, star-forming galaxies in merging clusters also have larger rest-frame \Ha EWs, irrespective of galaxy color. More generally, the bulk of star-forming galaxies with large EW ($>$50\,\AA) are found in merging clusters, surprisingly, within 3\,Mpc of the cluster center and likely embedded in hot ICM plasma. This readily supports a scenario in which star-forming galaxies in merging clusters have large sSFR. Other studies have unsuccessfully searched for variations in sSFRs across the shock fronts in merging clusters \citep{Chung2009, Shim2010}. These studies targeted individual clusters, did not benefit from spectroscopic observations, and relied on mid-infrared colors to estimate sSFR, which are more susceptible to contamination from foreground and background interlopers, and AGN. By contrast, the ENISALA project benefits from a large sample and contains hundreds of confirmed cluster members in different clusters, with EWs (and thus sSFRs) measured from spectroscopic observations, facilitating more conclusive results.

\begin{figure*}[ht!]
  \centering
  \includegraphics[width=0.7\linewidth]{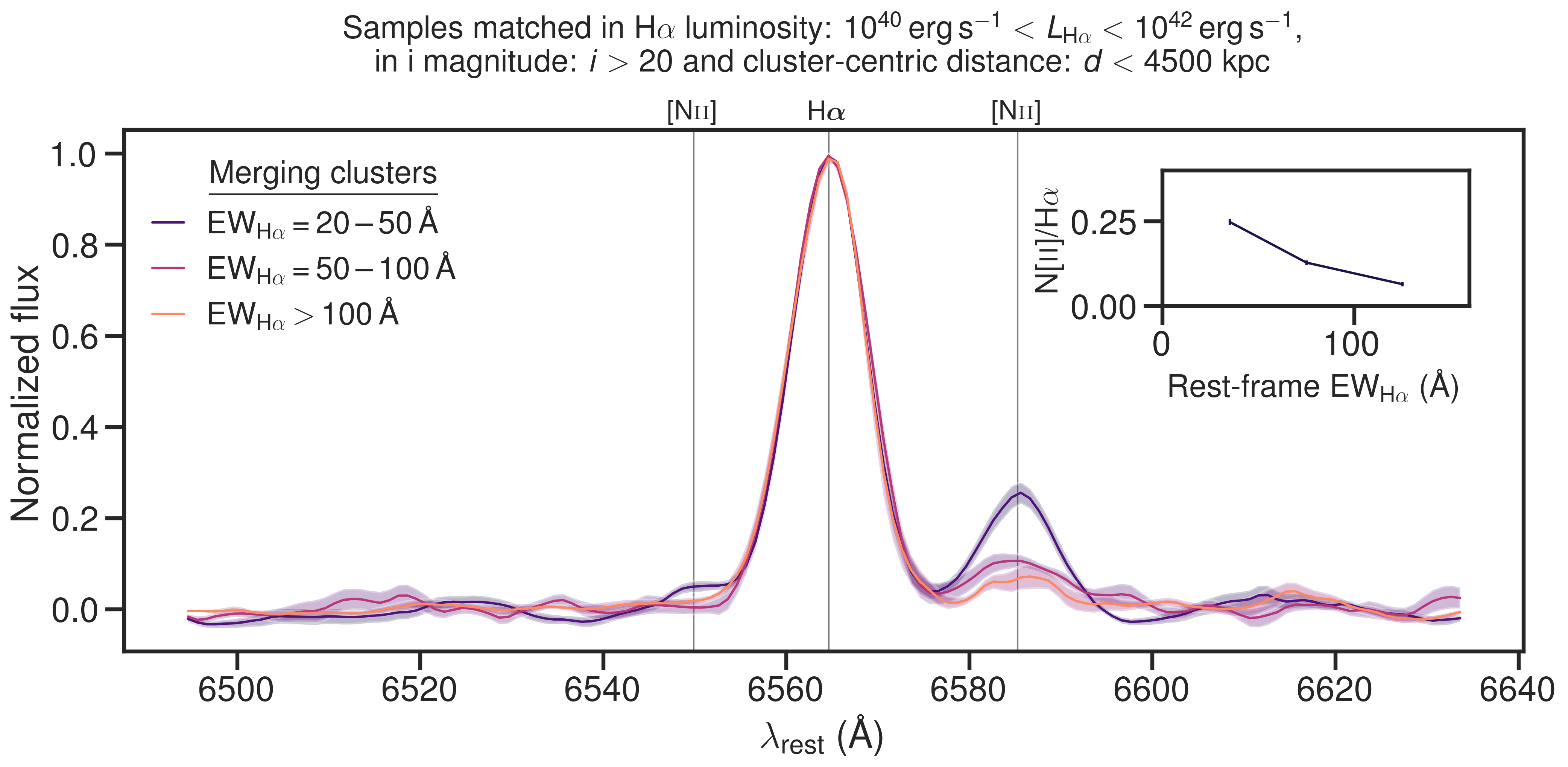}
  \caption{
    Stacked average spectra for faint star-forming galaxies in merging clusters, binned by rest-frame \Ha EW. The 68\% confidence intervals are plotted. We only show sources with $i$-band magnitudes $>20$ and \Ha luminosities between $10^{40}$ and 10$^{42}$\,erg\,s$^{-1}$. The average \NII/\Ha ratio decreases with increasing \Ha EW. From the low \NII/\Ha ratios, we can infer that the unique population of faint galaxies with high EW (sSFRs) in merging clusters have very low metallicities.
  }
  \label{fig:EW_stack}
\end{figure*}

The most consequential finding of this paper might be the existence of a star-forming population unique to merging clusters. We discover star-forming galaxies with large sSFR (EW$\sim90-150$\AA) and faint continuum magnitudes ($i\sim20-22$), which are absent in relaxed clusters. Using the relation between sSFRs and \Ha EW from \citet{Belfiore2018}, the \Ha EWs imply sSFRs ranging between $0.5-2.0$\,Gyr$^{-1}$. Assuming stellar masses anywhere between 10$^7$ and 10$^{11}$\,M$_\odot$ would place these high EW galaxies $0.7-1.7$\,dex above the main-sequence at $z\sim0.2$ \citep{Shioya2008, Stroe2015a}. Since the dispersion of the main-sequence is $0.3-0.4$\,dex in average density environments as well as in relaxed clusters \citep[e.g.][]{Erfanianfar2016}, the high EW population in merging clusters is securely located above the typical relation for field galaxies ($2-4$ times the dispersion). In stark contrast, the main-sequence in relaxed clusters at $0.15<z<0.5$ is suppressed compared to the field relation by $-0.2$ below masses of 10$^{10}$\,M$_\odot$, up to $-0.8$\,dex at masses above 10$^{11}$\,M$_\odot$ \citep{Erfanianfar2016}. Furthermore, the faint star-forming galaxies with large EWs also display lower \NII/\Ha ratios ($0.06\pm0.01$) compared to lower EW galaxies with similar $i$-band and \Ha luminosity sources, implying metal-poor gas consistent with low-mass, highly star-forming galaxies. This effect is illustrated in Figure~\ref{fig:EW_stack}, where we show the average stacked spectrum for different EW bins. The 68\% confidence intervals plotted in Figure~\ref{fig:EW_stack} are obtained through a bootstrapping method.

To sustain this high level of SF and AGN activity, a large supply of gas is necessary. Even in relaxed clusters, infalling jellyfish star-forming galaxies have higher SFRs fueled by large molecular gas reservoirs \citep{Moretti2020}. The authors attribute the increased molecular gas fractions to the efficient conversion of neutral into molecular gas under ram pressure. Unlike relaxed clusters, where cluster star-forming galaxies become increasing deficient in neutral hydrogen towards the cluster core \citep[e.g.][]{Chung2009a}, in \citet{Stroe2015}, we discovered that star-forming galaxies in the massive, binary merging `Sausage' cluster (part of the ENISALA sample) have large reservoirs of neutral hydrogen, comparable to counterparts in the field around the cluster. \citet{Jaffe2012, Jaffe2016} find a strong correlation between substructure and the presence of neutral gas-rich galaxies, supporting a post-processing scenario in which ram pressure, possibly increased by shock waves, can trigger SF. \citet{Cairns2019} uncover a large population of galaxies rich in molecular gas in a disturbed low-redshift cluster, reminiscent of gas-rich galaxies in young clusters at $z\sim1.5$ \citep{Noble2017}. We cannot make any definitive claims in the absence of the necessary observations. Neutral or molecular gas measurements for the bulk of the ENISALA sample would prove very useful in understanding how the large sSFRs in merging clusters are fueled.

Every merging cluster is an ecosystem where the ICM, shocks, and turbulence help drive the evolution of the galaxies and AGN through different pathways than relaxed clusters. In Paper I of the series \citep{Stroe2017}, we have shown that, on average, merging galaxy clusters have a higher density of \Ha line emitters, compared to relaxed counterparts. The results are even more pronounced in individual clusters. In a pilot study using the NB technique to uniformly select star-forming galaxies inside and around merging clusters through their strong \Ha emission, \citet{Stroe2014, Stroe2015a} found a spectacular increase of a factor of 20 in the density of star-forming galaxies a massive merging galaxy cluster (which is part of the ENISALA project), attributed to cluster-wide shock-induced SF or collapsed filaments and groups rich in star-forming galaxies. The cluster underwent a very recent massive merger about 0.5\,Gyr ago, which induced a cluster-wide, large-scale shock-wave, which passed through the cluster galaxies and possibly triggered SF, thus effectively elevating their SF efficiency in the last 0.5\,Gyr. In a study of 105 clusters, \citet{Yoon2020} find that the fraction of star-forming galaxies is enhanced by a factor of 1.2 in interacting clusters compared to relaxed clusters, with the most prominent effect happening in possibly gas-rich galaxies with stellar masses $<10^{10.4}$\,M$_\odot$. The harsh ICM in relaxed clusters affects all infalling galaxies to a certain degree, resulting in a complete shutdown on SF within one crossing of the cluster. In a study of over 100 SDSS nearby clusters, \citet{Cohen2014} corroborate these results. \citet{Cohen2014} attribute the weak correlation between cluster substructure and SF fraction to either cluster mergers enhancing the SF in cluster galaxies or to the less evolved nature of mergers compared to relaxed clusters. This interpretation is echoed by other authors, who find a clear overabundance of blue and star-forming galaxies in clusters exhibiting substructure, with most sources tracing infalling sub-clusters \citep{Wang1997, Cortese2004, Hou2012, Cava2017}. These studies might support a scenario in which only a fraction of galaxies lose their gas supply upon infalling into the ICM of a merging cluster, while a number manage to hold onto their gas reservoirs. The results from the spectroscopic analysis of the ENISALA sample tell a different story, which suggests SF is triggered in merging clusters, with sSFRs in few dozen galaxies exceeding values expected from the main sequence relation. Several scenarios that partly explain the observations from the literature:
\begin{itemize}
  \item \textbf{Scenario 1}: Cluster shocks/turbulence trigger activity in gas-rich cluster galaxies, which would imply cluster-wide effects and marked differences between merging and relaxed clusters.
  \item \textbf{Scenario 2}: Active accretion of groups/filaments rich in star-forming galaxies would show an increase in SF activity in merging/young clusters, as they are located in active cosmic web nodes.
  \item \textbf{Scenario 3}: Fast and localized processes such as ram pressure, with pronounced effects at the outskirts of all clusters, irrespective of merger state.
  \item \textbf{Scenario 4}: Slow, cluster-wide process, where SF slowly quenches because of a lack of new gas supply, with the strongest effects in relaxed clusters.
\end{itemize}
Overall, the detailed spectroscopic analysis of the ENISALA survey enables us to break degeneracies between the models and lend support for Scenarios 1 and 2. We unveil that the overwhelming majority of emission-line galaxies with \Ha luminosities above 10$^40$\,erg\,s$^{-1}$ powered by either pure SF or AGN activity populate the central 3\,Mpc region of merging clusters. Merging clusters also host highly star-forming galaxies deep within their hot ICM plasma. Our results are readily explained by the model proposed by \citet{Ebeling2019}, in which gas-rich galaxies infall along filaments in the rich cosmic web around merging clusters, followed by triggered SF. Merging clusters are permeated by cluster-wide shocks and turbulence traveling at speeds of $1000-2500$\,km\,s$^{-1}$ \citep[][]{Stroe2014,vanWeeren2019}, which can act as a catalyst for triggering SF in gas clouds, as well as funnel gas into the galaxy core, causing accretion onto the BH and promoting AGN activity \citep[as per][]{Poggianti2017, Ricarte2020}. \citet{Stroe2015} and \citet{Roediger2014}, for example, posit that a shock passing through a gas-rich galaxy should lead to a sharp rise of SF for up to 100-500\,Myr, which is perfectly compatible with the high sSFR galaxies we find exclusively in merging galaxy clusters. By contrast, we find evidence for mild enhancement of AGN and SF at the outskirts of relaxed clusters, followed by a dearth of SF towards their core. Our observations corroborate the extensive literature in the field, which invokes mild ram pressure triggering activity at large radii and rapid quenching taking over as the infalling galaxy crosses the cluster core.

\section{Summary} \label{sec:conclusions}

The ENISALA project is a multiwavelength observing campaign exploring the evolution of galaxies in merging and relaxed clusters. Here, in Paper II of the series, we introduce the spectroscopic follow-up survey of star-forming galaxies and AGN in 14 relaxed and merging massive ($0.5-3.5\times$10$^{15}$\,M$_{\odot}$) galaxy clusters at $0.178<z<0.308$, drawn from our homogeneously, NB-selected sample \citep[Paper I,][]{Stroe2017}. We leverage deep spectroscopy of over 800 emission-line galaxies to contrast the properties of \Ha line-emitters in merging and relaxed cluster environments and constrain evolutionary pathways from the perspective of large scale structure growth. Our main findings are:
\begin{itemize}
  \item Our cluster emission-line sample comprises about 16\% AGN-dominated sources, the majority of which dominated by narrow-line (hence classifying them as Seyfert 2 sources), with an additional 20\% of source with composite spectra, in line with studies of \Ha selected emission-line galaxies in the field. The fraction of emitters powered by AGN increases sharply with the velocity FWHM of \Ha, with over 80\% of star-forming galaxies measuring profiles under 200\,km\,s$^{-1}$.
  \item Pure star-forming galaxies in merging clusters, have on average, bluer UV--optical colors than counterparts in relaxed clusters. The bulk of the AGN have flat UV--optical colors, firmly classifying them as Seyfert 2 type source.
  \item \Ha line emitters in merging clusters are powered by SF at a higher rate than in relaxed clusters. Further, the bulk of emission-line galaxies in merging cluster fields are located within 3\,Mpc from the center. By contrast, AGN peak at the outskirts ($\sim1.5-3$\,Mpc) of relaxed clusters, and the fraction of pure star formers drops sharply inside the 3\,Mpc cluster-centric radius.
  \item Galaxies powered by SF have larger EWs than those with AGN contribution. The star-forming population in merging clusters have higher \Ha EW, or sSFR, than relaxed clusters.
  \item We measure \Ha EWs exceeding 90\,\AA, powered almost exclusively by SF. The sources with the largest EWs are found within 3\,Mpc of the centers of merging clusters.
  \item We discover a population of continuum-faint \Ha emitters with bright emission lines, with large EW, or large sSFR, exclusively found in merging clusters. These galaxies are likely located well above the field main-sequence.
  \item We introduced \textsc{redshifts}, which enables the user to obtain all spectroscopic redshifts from the literature, publicly available on VizieR and NED.
\end{itemize}

In conclusion, we find significant populations of star-forming galaxies and Seyfert 2 type AGN in merging galaxy clusters. The emission-line galaxy population in merging clusters permeates the inner parts of the ICM, which suggests that these galaxies are surviving the strong environmental effects typically seen in relaxed clusters. Their blue colors, in combination with high sSFRs, imply that they are undergoing episodes of vigorous star formation, contrary to expectations from models of galaxy evolution in relaxed clusters, which predict an exponential decline of SFRs over $0.5-2$\,Gyr through the removal of gas through ram pressure or a lack of new gas supply. Our results lend support to a scenario in which gas-rich galaxies in merging clusters, likely accreted along filamentary pathways, undergo SF and BH activity triggered by a cluster-wide process, such as a merger-induced shock wave.

Forthcoming papers will expand and build upon the results presented here: we will deepen our analysis of the star-forming main sequence, complete it by exploring the full fundamental plane of SF, mass, and metallicity, study electron densities in detail to constrain nebular chemical abundances, investigate how AGN and SF-driven outflows interact to drive the evolution of galaxies and understand the role of cluster mergers have in shaping galaxy morphology.

\acknowledgments

We are grateful to the referee for their comments, which improved the clarity, presentation, and interpretation of the results. We thank Sergio Santos for his support with the data reduction. We thank Tessa Vernstrom and Jeremy Harwood for insightful discussions, and Gabriel Taillon for making his two-dimensional KS code public. Andra Stroe gratefully acknowledges the support of a Clay Fellowship. Some of this work is based on observations made with ESO Telescopes at the La Silla Paranal Observatory under program ID 099.A-0562. Based in part on observations made with the William Herschel Telescope, operated on the island of La Palma by the Isaac Newton Group in the Spanish Observatorio del Roque de los Muchachos of the Instituto de Astrof{\'i}sica de Canarias. Some of the data presented herein were obtained at the W.M. Keck Observatory, which is operated as a scientific partnership among the California Institute of Technology, the University of California and the National Aeronautics and Space Administration. The Observatory was made possible by the generous financial support of the W.M. Keck Foundation. We would like to thank the ACReS team for making their data publicly available. Some of the observations reported here were obtained at the MMT Observatory, a joint facility of the Smithsonian Institution and the University of Arizona. We acknowledge the Smithsonian Astrophysical Observatory Optical/Infrared Telescope Data Center archive for data access to reduced MMT observations. Some of the data used to make the figures are based on archival observations obtained with XMM-Newton, an ESA science mission with instruments and contributions directly funded by ESA Member States and NASA, and  data obtained from the Chandra Data Archive. Some of the photometric data used in this paper were obtained from the Mikulski Archive for Space Telescopes (MAST). This research has made use of the NASA/IPAC Extragalactic Database, which is funded by the National Aeronautics and Space Administration and operated by the California Institute of Technology. This paper has benefited from using the VizieR database \citep{Ochsenbein2000}.

\vspace{5mm}
\facilities{VLT:Melipal (VIMOS spectroscopy), ING:Herschel (AF2 spectroscopy), Keck:II (DEIMOS spectroscopy), MMT (Hectospec spectroscopy), ING:Newton (WFT imaging, photometry), Max Planck:2.2m (WFI imaging, photometry), Subaru (Suprimecam imaging, photometry), CFHT (Megacam imaging, photometry), Sloan (SDSS survey imaging, photometry), PS1 (PS1 survey imaging, photometry), CXO (ACIS-I imaging), XMM (EPIC imaging), VLA (NVSS survey imaging)}

\software{
  \textbf{gleam} \citep{stroe_gleam},
  \textbf{redshifts} \citep{stroe_redshifts},
  Matplotlib \citep{matplotlib},
  Astropy \citep{2013A&A...558A..33A},
  Astroquery \citep{Ginsburg2019},
  APLpy \citep{aplpy},
  LMFIT \citep{lmfit},
  TOPCAT \citep{topcat},
  STILTS \citep{stilts},
  DS9 \citep{ds9},
  AstrOmatic Software \citep{1996A&AS..117..393B},
  EsoReflex \citep{2013A&A...559A..96F},
  2DKS \citep{Taillon2019}
}

\appendix
\restartappendixnumbering

\section{Selection Criteria for the Final Sample}\label{app:criteria}

Sources located at the cluster redshift were included in the final sample if they meet any of the following criteria:
\begin{itemize}
  \item Detection in \NII and \Ha, or
  \item Upper limit in, lower limit in, or unconstrained \NII/\Ha and detection in \OIII/\Hb, or
  \item No coverage in \NII/\Ha or \OIII/\Hb, as long as the other ratio is detected securely, or
  \item Lower limit in \NII/\Ha and upper limit in \OIII/\Hb, or
  \item Upper limit in \NII/\Ha and lower limit in \OIII/\Hb, or
  \item Upper limit in \NII/\Ha and upper limit in \OIII/\Hb, or
  \item Unconstrained \NII/\Ha and upper limit in \OIII/\Hb, or
  \item Upper limit \NII/\Ha and unconstrained in \OIII/\Hb.
\end{itemize}

\section{\OIII/\Ha ratios to help AGN vs. SF classification}

\begin{figure}[ht!]
  \includegraphics[width=\linewidth]{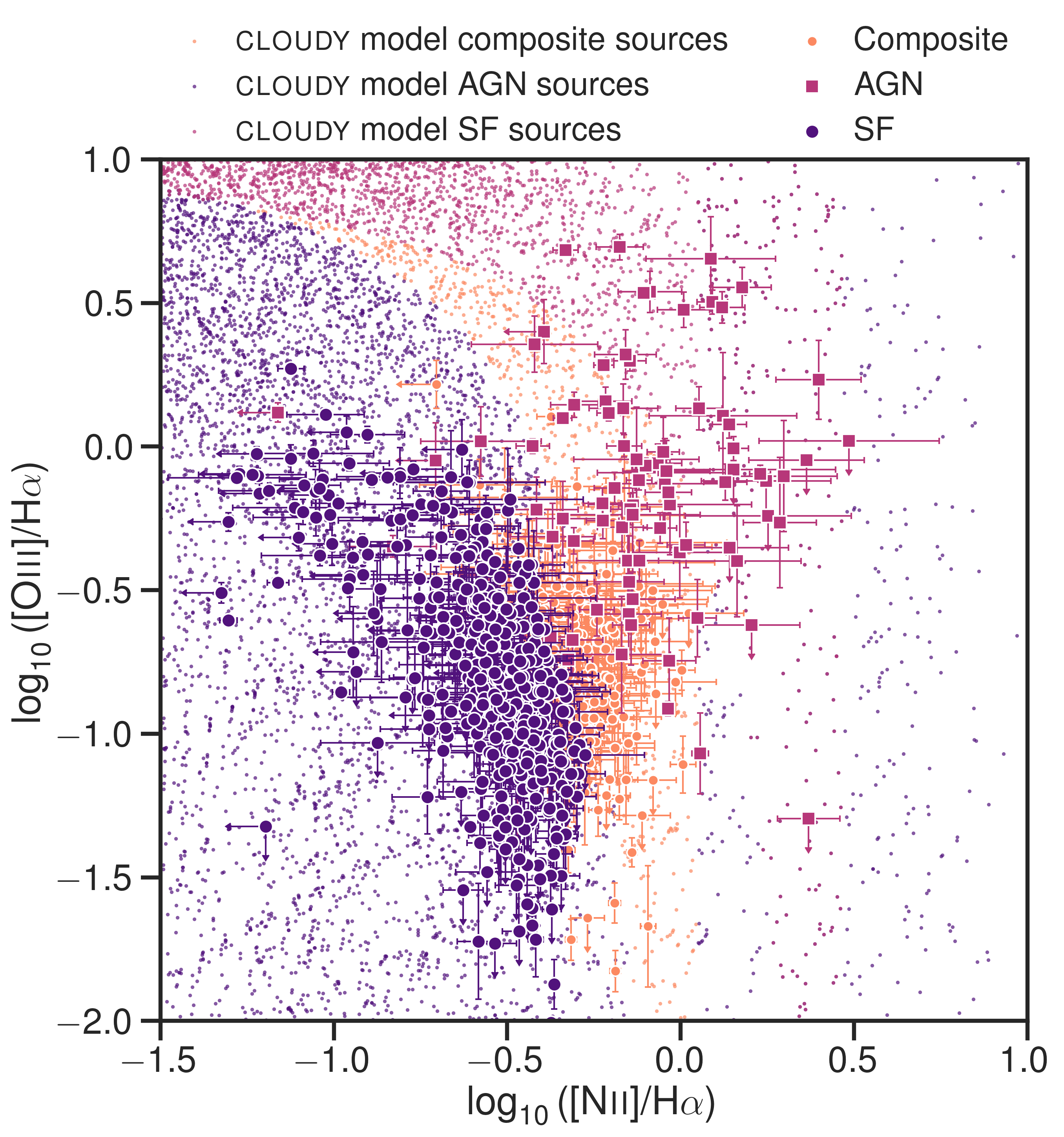}
  \caption{Distribution of our galaxies in the \OIII/\Ha versus \NII/\Ha space, color-coded by AGN, composite, and SF emission. We show sources securely classified by using the \citet{Kauffmann2003} and \citet{Kewley2001} criteria, as well as sources classified with the help of the \OIII/\Ha ratio. We show the distribution of sources from \textsc{cloudy} modeling in the background \citep[from][]{Sobral2018, Sobral2019}.}
  \label{fig:BPT_OIII}
\end{figure}

For sources that could not be securely classified based on the BPT diagram alone, we employed the \OIII/\Ha ratio in relation to the \NII/\Ha and the \OIII/\Hb ratios to classify the sources, only when the classification was unambiguous using the combination of the three ratios. We classified sources based on the spaces occupied by model sources from \citet{Sobral2018} and \citep{Sobral2019}, classified as AGN, composite, or SF dominated using the BPT criteria, as well as sources from our sample, which we securely classified in the BPT diagram. We note that our \OIII/\Ha ratios are on average lower than those predicted by the models because we they are not corrected for dust extinction. This effect was taken into account when classifying new sources based on the \OIII/\Ha ratio. For sources securely classified based on the BPT diagram alone, we do not alter the classification. We show the distribution of sources in the \OIII/\Ha vs. \NII/\Ha space in Figure~\ref{fig:BPT_OIII}.

\bibliography{AStroe_ENISALA_II}{}
\bibliographystyle{aasjournal}

\end{document}